# Guideline for Trustworthy Artificial Intelligence

## AI Assessment Catalog

**Authors**

Dr. Maximilian Poretschkin[1,2]  Dr. Sebastian Houben[1]      [1] Fraunhofer Institute for Intelligent
Anna Schmitz[1]                  PD Dr. Michael Mock[1,2]          Analysis and Information Systems IAIS
Dr. Maram Akila[1]               Julia Rosenzweig[1]              Schloss Birlinghoven, Sankt Augustin,
Linara Adilova[1]                Joachim Sicking[1]               Germany
Dr. Daniel Becker[1]             Elena Schulz[1]
Prof. Dr. Armin B. Cremers[2]    Dr. Angelika Voss[1]          [2] Department of Computer Science,
Dr. Dirk Hecker[1]               Prof. Dr. Stefan Wrobel[1,2]      University of Bonn, Bonn, Germany

## Abstract

Artificial Intelligence (AI) has made impressive progress in recent years and represents a key technology that has a crucial impact on the economy and society. Prominent use cases include applications in medical diagnostics, predictive maintenance and, in the future, autonomous driving. However, it is clear that AI and business models based on it can only reach their full potential if AI applications are developed according to high quality standards and are effectively protected against new AI risks. For instance, AI bears the risk of unfair treatment of individuals when processing personal data e.g., to support credit lending or staff recruitment decisions. Serious false predictions resulting from minor disturbances in the input data are another example – for instance, when pedestrians are not detected by an autonomous vehicle due to image noise. The emergence of these new risks is closely linked to the fact that the process for developing AI applications, particularly those based on Machine Learning (ML), strongly differs from that of conventional software. This is because the behavior of AI applications is essentially learned from large volumes of data and is not predetermined by fixed programmed rules.

Thus, the issue of the trustworthiness of AI applications is crucial and is the subject of numerous major publications by stakeholders in politics, business and society. Noteworthy mentions include the European Commission's draft regulation, the AI Standardization Roadmap and the recommendations of the High-Level Expert Group on AI, which formulate essential guidelines for the trustworthy use of Artificial Intelligence. In addition to these resources, there is mutual agreement that the requirements for trustworthy AI, which are often described in an abstract way, must now be made clear and tangible. One challenge to overcome here relates to the fact that the specific quality criteria for an AI application depend heavily on the application context and possible measures to fulfill them in turn depend heavily on the AI technology used. For example, the requirements for the trustworthiness of an AI system for automated analysis of job application documents need to be evaluated differently than those for an image recognition process for the quality assurance of car bodies. Lastly, practical assessment procedures are needed to evaluate whether specific AI applications have been developed according to adequate quality standards.

This AI assessment catalog addresses exactly this point and is intended for two target groups: Firstly, it provides developers with a guideline for systematically making their AI applications trustworthy. Secondly, it guides assessors and auditors on how to examine AI applications for trustworthiness in a structured way.

# Contents

# Foreword

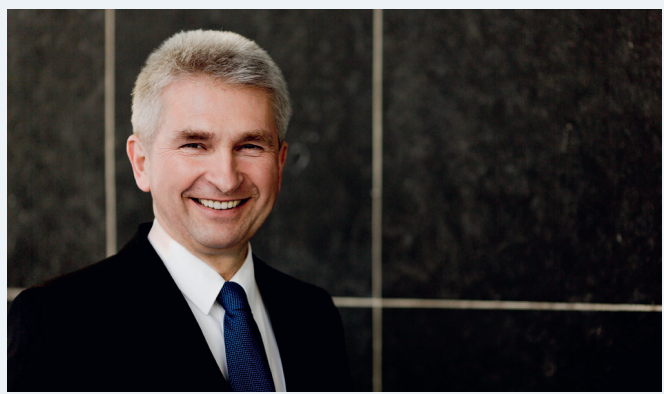

Dear readers,

Artificial Intelligence (AI) is the key technology of our time. There are now intelligent systems in almost every area of life in society and they help people accomplish tasks faster and more reliably. Predictions suggest that Artificial Intelligence will significantly increase global economic growth and be pivotal in shaping the development of society. In addition, AI can play a key role in tackling major societal challenges in areas such as climate protection, mobility and health.

In Germany, we have excellent foundations to build on in the field of Artificial Intelligence and Machine Learning – especially in North Rhine-Westphalia. Our industrial and innovative state can draw on many years of experience in the field of Artificial Intelligence and sets an example when it comes to interlinking business and research and promoting applied AI research. Back in 2019, the KI.NRW competence platform was set up for this purpose to connect stakeholders in the field of Artificial Intelligence and to improve technology transfer from research to practice.

However, using AI applications also comes with specific risks relating to issues such as safety and security, transparency, reliability, fairness, autonomy and data protection. If we want to harness the huge potential of the technology, it is essential that we put humans at the center of the design of AI applications, evaluate the previously mentioned risks and minimize them as appropriate. Looking ahead, certifying AI products through independent certification bodies can help increase the quality of the systems, improve the trust and acceptance of AI in society and, above all, secure the competitiveness of our companies in the long term and beyond national borders.

The Fraunhofer IAIS assessment catalog is now available and represents a major milestone on the way to independent AI assessment. With over 160 pages, it describes how AI applications can be systematically evaluated with regard to risks, provides recommendations for criteria to measure the quality of the systems and proposes measures that can mitigate AI risks. This allows the assessment catalog to serve as a practical guide for enabling product-specific, reproducible and standardized assessment procedures and quality improvements of AI systems according to a standardized approach. Furthermore, the catalog provides a national and international benchmark for innovation-friendly and trustworthy "made in Germany" AI standardization. Based in the German town of Sankt Augustin, the Fraunhofer IAIS scientists have already successfully applied the assessment catalog in initial pilot assessments with companies.

I am pleased that we will be able to continue to drive the project forward together under the new KI.NRW flagship project "Zertifizierte KI" (Certified AI) initiative and wish all those involved every success.

Sincerely yours,

**Prof. Dr. Andreas Pinkwart**
Former Minister of Economic Affairs, Innovation, Digitalization and Energy of the State of North Rhine-Westphalia
(2017 to 2022)

# Executive Summary

Artificial Intelligence (AI) has made impressive progress in recent years and represents a key technology that has a crucial impact on the economy and society. Prominent use cases include applications in medical diagnostics, predictive maintenance and, in the future, autonomous driving. However, it is clear that AI and business models based on it can only reach their full potential if AI applications are developed according to high quality standards and are effectively protected against new AI risks. For instance, AI bears the risk of unfair treatment of individuals when processing personal data e.g., to support credit lending or staff recruitment decisions. Serious false predictions resulting from minor disturbances in the input data are another example – for instance, when pedestrians are not detected by an autonomous vehicle due to image noise. The emergence of these new risks is closely linked to the fact that the process for developing AI applications, particularly those based on Machine Learning (ML), strongly differs from that of conventional software. This is because the behavior of AI applications is essentially learned from large volumes of data and is not predetermined by fixed programmed rules.

Thus, the issue of the trustworthiness of AI applications is crucial and is the subject of numerous major publications by stakeholders in politics, business and society. Noteworthy mentions include the European Commission's draft regulation[1], the AI Standardization Roadmap[2] and the recommendations of the High-Level Expert Group on AI[3], which formulate essential guidelines for the trustworthy use of Artificial Intelligence. In addition to these resources, there is mutual agreement that the requirements for trustworthy AI, which are often described in an abstract way, must now be made clear and tangible. One challenge to overcome here relates to the fact that the specific quality criteria for an AI application depend heavily on the application context and possible measures to fulfill them in turn depend heavily on the AI technology used. For example, the requirements for the trustworthiness of an AI system for automated analysis of job application documents need to be evaluated differently than those for an image recognition process for the quality assurance of car bodies. Lastly, practical assessment procedures are needed to evaluate whether specific AI applications have been developed according to adequate quality standards.

This AI assessment catalog addresses exactly this point and is intended for two target groups: Firstly, it provides developers with a guideline for systematically making their AI applications trustworthy. Secondly, it guides assessors and auditors on how to examine AI applications for trustworthiness in a structured way.

---

1  European Commission, Directorate-General for Communications Networks, Content and Technology (April 2021). Proposal for a Regulation laying down harmonised rules on artificial intelligence (Artificial Intelligence Act) and amending certain Union legislative acts. COM/2021/206 final https://eur-lex.europa.eu/legal-content/EN/TXT/?uri=CELEX:52021PC0206 (last accessed: 06/23/2021)

2  Wahlster; Winterhalter (Editors) (November 2020). German Standardization Roadmap on Artificial Intelligence. German Institute for Standardization and German Commission for Electrotechnical, Electronic & Information Technologies of DIN and VDE. https://www.dke.de/en/areas-of-work/core-safety/standardization-roadmap-ai (last accessed: 06/23/2021)

3  High-Level Expert Group on AI (HLEG) (April 2019). Ethics Guidelines for Trustworthy AI. European Commission. https://digital-strategy.ec.europa.eu/en/library/ethics-guidelines-trustworthy-ai (last accessed: 06/21/2021)

To do this, the catalog sets out a four-step approach:

1. Conducting a comprehensive risk analysis with regard to the following dimensions: fairness, autonomy and control, transparency, reliability, safety and security and data protection.

2. Establishing objectives that are preferably measurable to ensure mitigation of the risks identified in step 1 is demonstrable.

3. Systematically listing measures along the life cycle of an AI application to achieve the objectives defined in step 2.

4. Creating a stringent argumentation that the objectives formulated in step 2 have been achieved ("safeguarding argumentation for trustworthiness"), also taking into account AI-specific trade-offs, e.g., security vs. transparency.

# 1. Introduction

Artificial Intelligence (AI) is becoming part of more and more areas of our everyday lives and performing increasingly responsible tasks. Examples include AI-based quality control in production, support systems for medical diagnostics, automated stock exchange transactions and, in the future, autonomous driving. In terms of AI-based business models, it is crucial that the AI application is reliable, safe and resilient. At the same time, it is important for humans as users and affected persons that the use of AI is in line with societal values. Thus, it is clear that the potential of AI can only be fully exploited, particularly for sensitive application contexts, if AI applications are implemented according to strict quality standards.

There are various challenges involved with regard to the responsibility for the quality of AI applications and also with regard to the technical verifiability of quality requirements.

Responsibility for the quality of AI systems is spread along a value chain that differs hugely from the development of conventional software. AI applications are often based on Machine Learning (ML) techniques that learn patterns in training data and build a model to apply what is learned to unknown data (but structurally comparable to the training data). Because these types of models are often created using millions (sometimes billions) of parameters, AI applications are primarily based on processing large volumes of data, for which suitable IT infrastructures and computing power are required. Alongside the data producers, cloud service providers also play a key role in this area by providing the necessary computing capacity, infrastructure and suitable basic AI services such as optical character recognition (OCR), and can thus significantly influence the quality of AI applications.

In addition to a complex value chain, the complexity of the AI applications themselves also presents a challenge for ensuring their quality. Even experts often find it difficult to understand how the underlying models work, because of factors such as the large number of parameters. Furthermore, ML-based procedures can theoretically continue to learn during the operation of the AI applications, which implies that measures are required to prevent the learning of incorrect behavior, for example.

## Assessment constitutes a key component for trust and quality

For these reasons, it is vital for AI experts to systematically implement quality in the development of their own AI applications, as well as to assess the quality of third-party systems. Furthermore, users and affected persons must be able to trust that relevant AI applications satisfy appropriate quality requirements. Unbiased and expert assessment is a tried-and-tested component used in other domains to establish trust. Marketable assessment procedures that confirm the guaranteed characteristics of AI products and services can, for example, contribute to branding and thus create competitive advantages. In addition, assessments and audits can also be part of mandatory approval and supervisory procedures. The draft regulation of the European Commission[4] indicates that these types of approval procedures in the field of AI will be established for the European market in the near future. This is because, in addition to prohibiting certain applications for AI, the draft regulation specifies that "high-risk systems" must undergo a conformity assessment. This affects a large number of AI applications that are already an established part of our everyday lives.

---

**4** European Commission, Directorate-General for Communications Networks, Content and Technology (April 2021). Proposal for a Regulation laying down harmonised rules on artificial intelligence (Artificial Intelligence Act) and amending certain Union legislative acts. COM/2021/206 final https://eur-lex.europa.eu/legal-content/EN/TXT/?uri=CELEX:52021PC0206 (last accessed: 06/23/2021)

## Operationalizing quality requirements and existing AI guidelines

Although general requirements for the trustworthiness of AI have already been featured in intensive social and political discussions for some time and various guidelines on the trustworthiness of AI applications have been published, the operationalization of these guidelines is still open to a large extent. This is particularly apparent in the system-related requirements in the European Commission's draft regulation, which are not clearly specified or based on quantitative criteria, but rather leave a wide margin of discretion. Regarding the technical specification of the requirements, the European Commission refers to harmonized standards and common specifications[5]. However, the Standardization Roadmap on AI[6] published in December 2020 makes it abundantly clear that these are not yet widely available. The German Federal Office for Information Security (BSI) has also highlighted the need for action to develop "standards, technical guidelines, test criteria and assessment methods"[7] for the secure use of AI applications.

It is important to note here that the specific requirements for an AI application to achieve trustworthiness depend heavily on the technology used and the application context. Key performance indicators (KPIs) that allow the quality of AI applications to be measured are desirable. The Standardization Roadmap on AI discusses the example of AI-based translation systems, where translation quality is evaluated using the BLEU score[8]. The BLEU score compares an AI-based translation with a human translation on a scale of 0 to 100, with the value 100 indicating a perfect match. The BLEU score must increase in line with the criticality of the intended use of the translation. In the example, a BLEU score of at least 35 is recommended for translating social media posts, and a BLEU score of at least 45 is recommended for translating medical letters. However, given the abundance of AI technologies and their different application contexts, the question crops up as to what metric and threshold are adequate to a given application context. In the example of preventing unfair discrimination by an AI application, it is necessary to first quantify fairness from a technical perspective. However, a variety of different concepts and metrics are available for quantifying fairness. As such, it must be decided on a case-by-case basis which criteria and thresholds are appropriate for assessing the fairness of an AI application.

When defining specific quality requirements for AI applications, there is also the challenge that different dimensions of trustworthiness cannot be assessed completely independently of each other, but trade-offs need to be made. For example, a performance increase, such as recognition performance on image data by deep neural networks, can come at the expense of traceability, or an increase in transparency (for example, by revealing all hyperparameters of a model) can lead to new attack vectors in terms of IT security.

---

**5** "The precise technical solutions to achieve compliance with those requirements may be provided by standards or by other technical specifications or otherwise be developed in accordance with general engineering or scientific knowledge at the discretion of the provider of the AI system." (p. 13) and "common normative standards for all high-risk AI systems should be established" (p. 20): European Commission, Directorate-General for Communications Networks, Content and Technology (April 2021). Proposal for a Regulation laying down harmonised rules on artificial intelligence (Artificial Intelligence Act) and amending certain Union legislative acts. COM/2021/206 final
https://eur-lex.europa.eu/legal-content/EN/TXT/?uri=CELEX:52021PC0206 (last accessed: 06/23/2021)

**6** Wahlster; Winterhalter (Editors) (November 2020). German Standardization Roadmap on Artificial Intelligence. German Institute for Standardization and German Commission for Electrotechnical, Electronic & Information Technologies of DIN and VDE.
https://www.dke.de/en/areas-of-work/core-safety/standardization-roadmap-ai (last accessed: 06/23/2021)

**7** German Federal Office for Information Security (February 2021). Sicherer, robuster und nachvollziehbarer Einsatz von KI. [Secure, robust and transparent application of AI]. https://www.bsi.bund.de/SharedDocs/Downloads/DE/BSI/KI/Herausforderungen_und_Massnahmen_KI.pdf?__blob=publicationFile&v=5 (in German, last accessed: 06/23/2021)

**8** BLEU stands for Bilingual Evaluation Understudy

## Risk-based AI assessment

Risk-based assessment is a proven approach that can be used to define application-specific quality requirements. Risk-based assessment approaches have already proven themselves in classic IT security[9] and functional safety, where the requirement for resistance to manipulation or unwanted incorrect behavior can result in very different technical requirements for different IT systems. This approach allows risks to be considered from different perspectives: First of all, the risk of a malfunction is examined in terms of the impact on users, affected persons or the (immediate) environment. Both potential material and non-material damage is considered, for example with regard to security, safety and personal rights. At the same time, incorrect or even harmful behavior of an AI application is always linked to risks for organizations or the individuals who are responsible for the AI application. Taking the example of an AI application for credit lending, there is a risk of discrimination, which is associated with non-material damage for those affected and also damage to the reputation of the credit institution in question. This demonstrates that the discussion on trustworthy AI cannot revolve solely around system-specific issues of an AI application, but must also lead to the development of AI governance frameworks and thus a corresponding corporate culture[10].

A risk-based assessment approach for AI can also be integrated easily into existing assessment and audit schemes. The integrability with existing schemes is essential for the practicability and marketability of an AI assessment procedure, as AI assessment in many areas is performed as part of a overarching assessment issue which also addresses other, non AI-specific software and hardware issues. The European Commission's draft regulation explicitly demands for the conformity assessment of certain AI-based high-risk systems to be integrated, depending on the scope of application, into existing product safety testing procedures or into established processes for the regulatory review of credit institutions[11].

## Scope and application areas of the AI assessment catalog

This assessment catalog provides a structured guideline that can be used to evaluate AI applications with regard to all relevant dimensions of trustworthiness. In doing so, it focuses on AI applications based on Machine Learning. The catalog takes a primarily product-oriented assessment approach. However, since AI applications can evolve dynamically during operation, it also includes organizational and procedural requirements that are essential for ensuring the trustworthiness of the AI application even after the assessment has been performed.

---

9   See, for example, BSI-Grundschutz or ISO 27001 (ISO/IEC 27001, 2013) which contain specifications for an IT security management system or the Common Criteria (CC 3.1, 2017) for IT product testing methodology.

10  For a detailed review, see the Fraunhofer IAIS study: "Management System Support for Trustworthy Artificial Intelligence", which discusses requirements for organizations dealing with AI in terms of governance, management and technical and organizational measures, including the current standardization activities of ISO/IEC JTC 1/SC 42 "Artificial Intelligence": PD Dr. Michael Mock, Anna Schmitz, Linara Adilova et al. (October 2021). Management System Support for Trustworthy Artificial Intelligence. Sankt Augustin: Fraunhofer Institute for Intelligent Analysis and Information Systems IAIS. https://www.iais.fraunhofer.de/content/dam/iais/fb/Kuenstliche_intelligenz/Fraunhofer_IAIS_Study_%20MSS.pdf (last accessed: 11/16/2022).

11  "The key difference is that the ex-ante and ex-post mechanisms will ensure compliance not only with the requirements established by sectorial legislation, but also with the requirements established by this regulation." (p. 13) and see also Article 43 in: European Commission, Directorate-General for Communications Networks, Content and Technology (April 2021). Proposal for a Regulation laying down harmonised rules on artificial intelligence (Artificial Intelligence Act) and amending certain Union legislative acts. COM/2021/206 final https://eur-lex.europa.eu/legal-content/EN/TXT/?uri=CELEX:52021PC0206 (last accessed: 06/23/2021)

In particular, the catalog contains:

1. A guideline for the structured identification of AI-specific risks with regard to six dimensions of trustworthiness: fairness, autonomy and control, transparency, reliability, safety and security and data protection[12].

2. A guideline that can be used to formulate specific assessment criteria for an AI application. For this purpose, the AI assessment catalog lists established KPIs (comparable to the BLEU score mentioned previously) that can be used to quantify relevant objectives where possible.

3. A guideline on the structured documentation of technical and organizational measures along the life cycle of an AI application that reflect the current state of the art and that can be implemented to mitigate potential risks.

## Comparison with existing assessment approaches

This AI assessment catalog complements existing approaches to evaluating AI applications, among which the BSI's[13] AIC4 Criteria Catalogue and the AI Ethics Impact Group's[14] Interdisciplinary Framework to Operationalise AI Ethics are well-known resources:

The BSI catalog is an extension of the C5 Criteria Catalogue[15] and can be used to audit cloud-based AI services. The criteria set out in the catalog cover the entire life cycle of an AI service and address the areas of robustness, performance, reliability, data quality, explainability and bias. However, the catalog does not specify the requirements in relation to these AI-specific risks or provide guidance on how to assess whether these requirements have been met in the specific system.

The AI Ethics Impact Group's Interdisciplinary Framework to Operationalise AI Ethics aims to make measurable the conformity of an AI application with ethical principles such as "transparency, accountability, privacy, justice, reliability and sustainability". For this purpose, a value analysis process consisting of a combination of target criteria, indicators and measurable variables is used to derive quality levels inspired by the energy efficiency classes for electrical appliances. This approach focuses on making important quality characteristics transparent for the end user, but does not include a more detailed analysis of the AI application from a computer science perspective.

---

**12** The six dimensions of trustworthiness are adapted from the key requirements of trustworthy AI from the High-Level Expert Group on AI, see: High-Level Expert Group on AI (HLEG) (April 2019). Ethics Guidelines for Trustworthy AI. European Commission. https://digital-strategy. ec.europa.eu/en/library/ethics-guidelines-trustworthy-ai (last accessed: 06/21/2021). A detailed description of the six dimensions covered in the assessment catalog is available in: Poretschkin, M.; Rostalski, F.; Voosholz, J. et al. (2019). Trustworthy Use of Artificial Intelligence. Sankt Augustin: Fraunhofer Institute for Intelligent Analysis and Information Systems IAIS. https://www.ki.nrw/wp-content/uploads/2020/03/ Whitepaper_Thrustworthy_AI.pdf (last accessed: 06/18/2021)

**13** German Federal Office for Information Security (BSI) (February 2021). AI Cloud Service Compliance Criteria Catalogue (AIC4). German Federal Office for Information Security (BSI). https://www.bsi.bund.de/SharedDocs/Downloads/EN/BSI/CloudComputing/AIC4/AI-Cloud-Service-Compliance-Criteria-Catalogue_AIC4.pdf?__blob=publicationFile&v=4 (last accessed: 06/23/2021)

**14** Hallensleben, S. and Hustedt, C. et al. (2020). From Principles to Practice, An Interdisciplinary Framework to Operationalise AI Ethics. Bertelsmann Stiftung. https://www.ai-ethics-impact.org/resource/blob/1961130/c6db9894ee73aefa489d6249f5ee2b9f/aieig---report---download-hb-data.pdf (last accessed: 06/23/2021)

**15** German Federal Office for Information Security (BSI) (January 2020). Cloud Computing Compliance Criteria Catalogue – C5:2020. German Federal Office for Information Security (BSI). https://www.bsi.bund.de/SharedDocs/Downloads/EN/BSI/CloudComputing/ComplianceControlsCatalogue/2020/C5_2020.pdf?__blob=publicationFile&v=3 (last accessed: 06/23/2021)

The AI assessment catalog complements the existing approaches by demonstrating a systematic procedure for developing a safeguarding argumentation for AI applications on a more concrete level. The risk-mitigating precautions and test measures described can support in particular developers and operators of high-risk systems, who – in the future – will have to provide proof of compliance with technical quality requirements under the European Commission's draft regulation. To provide this proof, the European Commission requires that operators create a technical documentation of the AI application, which is then used to perform a conformity assessment, to be carried out by official notified bodies if necessary. This assessment catalog offers a guideline for preparing such a technical documentation[16]. It covers various areas including the description of the essential development steps, the assessment of AI-specific risks, the formulation of criteria or metrics for quantifying relevant technical characteristics, the specification of the risk-mitigating measures taken over the course of the life cycle and lastly the discussion of possible trade-offs.

In short, independent auditors and assessors can use the documentation prepared with the help of the AI assessment catalog as a basis for assessing the AI application and for planning more in-depth investigations or tests. In addition, the assessment catalog serves as a guideline for developers to design trustworthy AI applications according to the current state of the art.

## Catalog structure

Chapter 2 explains how the AI assessment catalog can be concretely used to assess the trustworthiness of AI applications. In doing so, it first introduces the basic concepts and important terms and then explains the risk-based methodology developed in the catalog in detail. According to this methodology, the trustworthiness of AI applications is assessed in terms of the six dimensions fairness, autonomy and control, transparency, reliability, safety and security and data protection. The third chapter presents the AI profile, which is used to provide an overview of the AI application and narrow down the assessment object. The following chapters provide a guideline that can be used to assess AI risks in terms of the six dimensions of trustworthiness in a structured manner. The final chapter describes the procedure for performing the final cross-dimensional assessment.

---

**16** The European Commission's requirements for technical documentation are described in Annex IV of the draft regulation. The AI assessment catalog covers the contents required in points 1, 2, 3, 5 and partly in point 8 of Annex IV.

# 2. Essential Concepts and Methodology for Applying the Catalog

The intended use and application context have a significant impact on the quality requirements for an AI application. How these quality requirements are implemented to develop a trustworthy AI application also depends heavily on the underlying AI technology.

The AI assessment catalog presents a framework for performing trustworthiness assessments in a structured manner. An essential prerequisite for applying the catalog is a clearly defined assessment object[17], which may have to be differentiated from a larger surrounding system. As such, this chapter will first cover technical and conceptual foundations regarding the formal structure of an AI application (Section 2.1). It will also explain which risks the quality criteria of the AI assessment catalog cover (Section 2.2) and how an AI assessment can be conducted based on this catalog[18] (Section 2.3).

AI applications are complex constructs in which implemented ML models usually interact with expert systems and other classic software components; or which are even designed as hybrid systems that combine structured knowledge, for example such as knowledge graphs, with Machine Learning. In addition, AI applications are often embedded into a larger surrounding system. However, there is no consistent terminology in the literature that clarifies how to differentiate single components as well as the AI application from the surrounding system. It is also ambiguous on which technical parts and on which abstract risks the assessment of an AI application should focus. Consequently, Section 2.1 provides an overview of the formal structure and the life cycle of an AI application, which can be used for alignment within the catalog and aims to create a shared understanding of the assessment object.

In terms of content, the AI assessment catalog covers six dimensions of trustworthiness, which are in turn classified into different risk areas. This thematic structure, along which quality criteria and risk-mitigating measures in particular are organized, is explained in Section 2.2. In addition, there is an overview of the content of the various dimensions and risk areas.

The last section of this chapter presents how the catalog can be applied in practice as a basis for assessing or auditing. This involves a description of the catalog's risk-based approach. In particular, there is a detailed explanation of the logic and reasoning developed in the assessment catalog, starting with identifying AI risks and deriving quality criteria through to performing the final assessment of the AI application.

---

[17] An AI application that is assessed using the catalog is subsequently referred to as an assessment object. However, this term is not intended to exclude the possibility that, in addition to assessing AI applications, the catalog can also be used to develop or improve AI applications.

[18] The assessment approach of this catalog is also explained in: Poretschkin, M.; Mock, M.; Wrobel, S. Zur Systematischen Bewertung der Vertrauenswürdigkeit von KI-Systemen [Systematic Assessment of the Trustworthiness of AI Systems]. In: D. Zimmer (Editor), Regulierung für Algorithmen und Künstliche Intelligenz [Regulating algorithms and Artificial Intelligence] (in German, available now).

## 2.1 Assessment object

This section provides a general, and thus highly abstract, description of the formal structure of an AI application by splitting it into various functional components. These can vary in complexity or scope depending on the nature of the actual application. In addition to this functional perspective, the life cycle of such an application is also examined in Section 2.1.2. Both perspectives are holistically portrayed within the catalog by different categories of measures, see Section 2.3.4. Here, the distinction between embedding and AI component relates to the structure of the application, while data and operation roughly refer to different stages of the life cycle. As with all complex systems, these categories are not necessarily clear-cut; for example, data can also be recorded during operation.

### 2.1.1 Structure of an AI application

A first step in the discussion and assessment of an AI application is to specify its formal structure as well as to define the assessment object. While the **AI Profile (PF)** provides an initial overview of the AI application, more detailed documentation of the assessment object is requested as the assessment progresses, in particular with regard to the reliability dimension. The purpose of this section is to create a shared understanding of terms relating to the structure of an AI application.

The AI assessment catalog is focused on AI applications that are based on Machine Learning (ML). While neural networks in particular play a major role in the field of ML today, the assessment catalog is not applicable solely to this technology. The assessment catalog can also be used to examine other methods, such as decision trees or support vector machines, which may even meet transparency or security requirements better than neural networks.

The ML model forms the core of the AI component, which is further extended by pre- and post-processing steps relating to the inputs and outputs of the ML model. The ML model and AI component are mathematical objects that can serve as the basis for a functionality, meaning an input-output mapping – for example, in order to detect objects or assess job applicants. In real-world use cases, the input-output mapping is usually performed through interactions with other (embedding) components which, for example, record or monitor the inputs by means of a rule-based (non-ML) expert system. The input-output mapping performed in this way in a given application context is referred to in this catalog as an AI application. Figure 1 depicts the concept of an AI application and its functional structure that represents the various input data processing steps; these are explained in more detail in the following section.

If an AI application is part of a larger system that is not entirely based on AI technologies, its boundaries within this surrounding system must be clearly defined. For example, ML-based object recognition (AI application) can be integrated into an autonomous vehicle, into a drone or into a site surveillance system (as larger system). Defining an AI application within a surrounding system is largely based on its functionality. For example, regarding an AI-based pedestrian detection system (AI application) in an autonomous car (larger system), the software modules that perform consistency checks on the outputs of the AI component should be seen as part of the AI application; however, other components that perform the overall planning of a driving route, for instance, would not be.

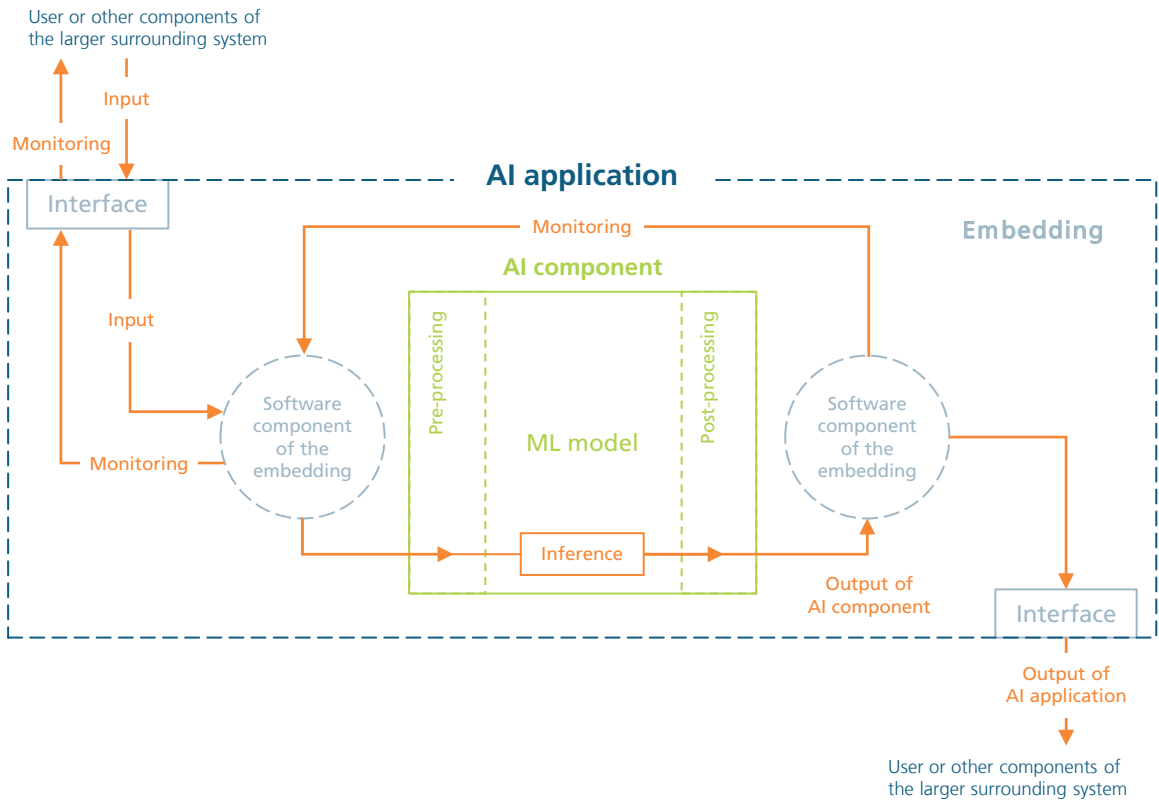

Figure 1: Formal structure of an AI application

**(ML) model:** The ML model is a mathematical, abstract object that was created by a Machine Learning technique and serves to solve a task in the sense of creating an input-output relation. For example, regarding a neural network, the model consists of a list of hyperparameters, learned parameters and a description of how they interact when operating (architecture), among others. The ML model provides the functional basis of the AI application. For instance, a model can serve as the basis for performing a classification task, where the input is the object to be classified and the model output characterizes its class. In some cases, such as in generative models, the input (usually random numbers) may be of secondary importance to the actual task to be solved.

**AI component:** The AI component consists of the ML model and the implemented (model-specific) methods for data pre-processing and post-processing of the model outputs. Thus, the AI component is also a mathematical object.

**Note:** For simplification, the assessment catalog assumes that, unless explicitly stated otherwise[19], an AI application only contains one AI component.

---

**19** For example, ensemble methods can be considered a single ML model and thus explicitly fall within the scope of the assessment catalog.

**Embedding:** The AI component is usually linked to additional (classic) software modules and technical components to store data or to implement physical reactions to AI component outputs, for instance. Embedding is the term used to describe the entirety of these surrounding components that relate directly to the way the AI component functions and operates. These components can include software modules that activate the AI component and further process its results. In particular, classic components that make the functionality of the AI component visible to the outside world and enable interaction (see **Interface**) are part of the embedding. In addition, embedding software modules can also help detect and intercept an AI component failure (referred to as monitoring in Figure 1).

**Interface:** The interface is the part of the embedding that enables the AI application to interact with the outside world, such as users or other components within a larger surrounding system. The interface offers various interaction options depending on the function and design of the AI application. Usually, these options involve recording input data (e.g., requests from users) and communicating the results/outputs of the AI component to the outside world or making them available for retrieval.

**AI application:** The AI application is the input-output mapping in a given application context based on the implemented AI component. Importantly, the AI assessment catalog does not take an isolated view of the ML model or the AI component. In fact, the catalog examines whether the individual processing steps performed up to the results being generated by the AI component through interaction with other components of the embedding are meaningful and appropriate for the given application context. For example, it assesses whether they are sufficiently free of errors and discrimination or secure against attacks and manipulation. Thus, the assessment object of the catalog is not predominantly the mathematical concepts underlying the AI component, but the functionality i.e., the input-output mapping, that is performed based on the AI component (as a functional basis) in a given application context. An AI application can be a standalone system that, for example, communicates directly with users, or it can be integrated into a larger (IT) system or product.

**Note 1:** In principle, the AI application is an abstract assessment object. However, usually (and particularly in the case of complex neural networks), it cannot be considered in isolation from how it is implemented, which is to some extent physical too. In particular, the testing measures described in the assessment catalog require that the AI application is implemented or even installed and functional in the actual application environment or end product. However, the focus of the AI assessment catalog is not on the challenges of implementation, but on whether the AI component is a suitable and trustworthy basis for realizing the desired functionality.

**Note 2:** Unless explicitly stated otherwise, this assessment catalog does not consider AI applications which are based on multiple AI components and particularly not on multiple structurally different ML models. While the AI assessment catalog can be used to examine different AI components separately, the catalog does not address how they interact.

### 2.1.2 Life cycle of an AI application

In addition to its structure, the life cycle of an AI application is a key starting point for identifying and mitigating AI risks and should therefore be examined fully in a quality assessment. Data is crucially important in the life cycle of an AI application based on Machine Learning, as the functionality of the application is usually derived directly from data. In particular, the life cycle of an AI application is more closely related to data mining processes than to developing and operating conventional IT systems (completely designed and programmed by humans). The CRISP-DM standard[20], which sees data mining as a process ranging from setting goals and selecting data to modeling and deployment, can essentially also be applied to the often more complex model classes of the AI applications examined here, such as deep neural networks. Figure 2 shows a simplified version of this process.

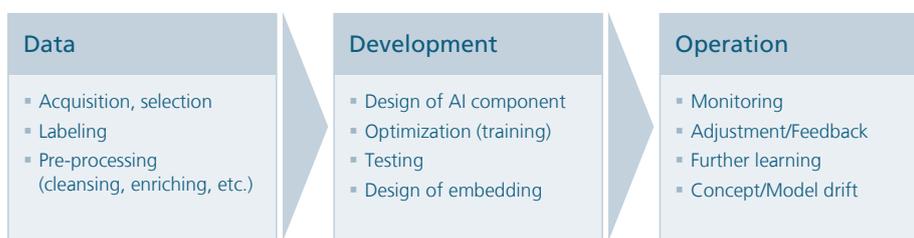

Figure 2: Overview of the life cycle of an AI application

The selection and processing of data in particular plays a critical role, since characteristics of an AI application (e.g., the learned model weights) and thus also its functionality stem from data-driven optimization. Compared to original data mining, complex Machine Learning methods often replace manual data pre-processing, such as the process of defining relevant data features. This is why data-related risks must be given more attention and suitable measures to meet requirements regarding the data should be demonstrated in a quality assessment. For example, these measures can relate to fairness, in particular the risk of inherent bias, or to the quality and suitability of data, for example in terms of sufficient coverage or the quality of labels.

Development is the second phase of the life cycle and involves aspects of design, training and testing. In relation to this phase, the assessment catalog distinguishes between AI component and its embedding as described in Section 2.1.1. Targeted design can favor or help achieve desired quality characteristics of an AI application. For example, this could involve selecting a certain model architecture or selecting embedding components that perform consistency checks on outputs of the AI components or establish redundancy to a certain degree in the event that the AI component fails. In addition, the ML model training, which is explained in Figure 3 using a simple standard workflow, has an essential impact on the quality of AI applications. Training is the procedure of creating a model using Machine Learning. For this purpose, the model parameters, also called weights, are determined by optimization on the basis of training data and fixed hyperparameters. Training data is input into the model following what is usually a random initialization of the weights. The results calculated with this data are subsequently evaluated using a loss function, in the case of supervised ML techniques using the labels (ground truth) associated with the data. The loss function measures the discrepancy between the model results and the ground truth. The value of the loss function is optimized iteratively by adjusting the weights. Because the hyperparameters can strongly influence the weights resulting from the optimization, an additional (validation) data set should be used to compare which hyperparameter configuration is most suitable for solving the given problem. In addition to the loss function, which is what is being optimized, other performance metrics should be used to evaluate the ML model. On the whole, the quality of the AI component depends on various factors,

---

**20** For a detailed description of the Cross-Industry Standard Process for Data Mining (CRISP-DM), see: Shearer, C. (2000). The CRISP-DM model: The new blueprint for data mining. Journal of Data Warehousing, 5(4):13–22.

such as the database choice, the learning algorithm including the loss function as well as the hyperparameters. Accordingly, testing is important for ensuring the quality of the AI component and test data should be selected carefully. This data must not match the training or validation data, as the ML model was previously optimized for this. Tests can both evaluate the behavior of the AI application with respect to the input data anticipated during operation and specifically look for weaknesses in the model.

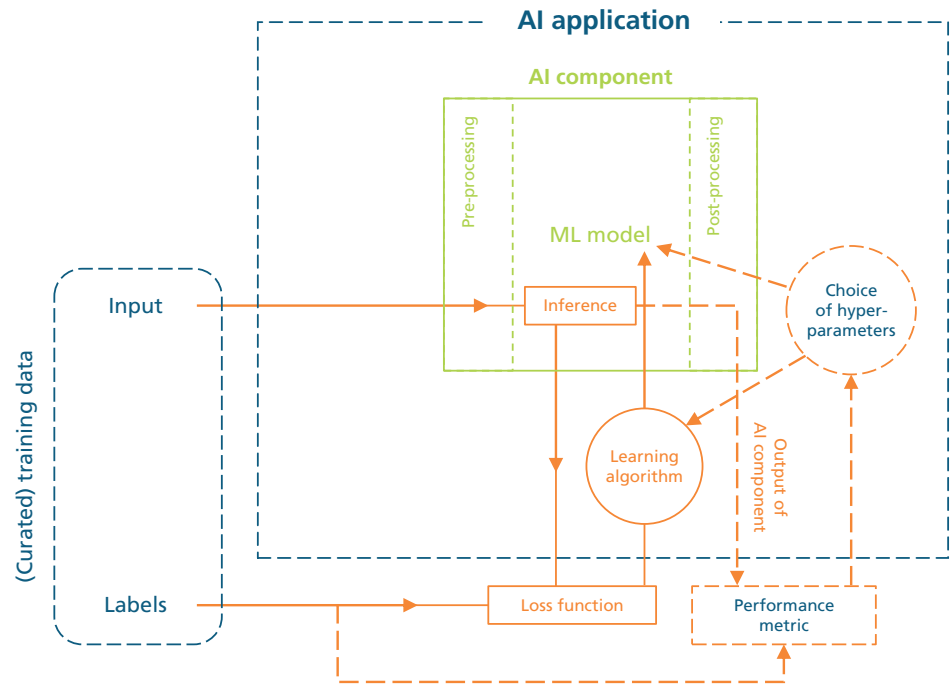

Figure 3: Training the ML model of an AI application.

When in operation, there are two different perspectives that should be covered in an assessment. Users and affected persons should be given the possibility to intervene and provide feedback. Moreover, adequate precautions must also be taken to deal with the risks of concept drift and model drift that are associated with a data-driven approach of AI applications. Concept drift is the risk that the properties of the input data or external conditions change and the AI application therefore no longer meets the requirements during operation. Model drift, by contrast, relates to a circumstance where the AI application (more specifically the ML model) continues to learn during operation and no longer meets the requirements due to the resulting change.

## 2.2  Dimensions of trustworthiness

While ML technologies establish a wide range of opportunities, they also bear new risks due to their complexity, dynamics and lack of transparency. In particular, numerous AI-specific risks are not covered by the existing assessment and certification schemes for conventional IT systems. The AI assessment catalog provides an approach for evaluating AI applications in a structured manner, while not aiming to rewrite or replace existing assessment procedures or standards for (conventional) IT systems, for example on IT security. In fact, the AI assessment catalog aims to serve as a compatible supplement to existing assessment and certification processes with the objective of closing the gap described. The main focus of the AI assessment catalog is on the AI-specific risks of AI applications.

The question of which AI-specific risks an AI assessment should cover and which criteria should be used to assess them has long been the subject of intense societal and scientific debate. From the various contributions to this discussion – including the especially prominent key requirements of the HLEG[21] – six topics have emerged, which will be referred to below as dimensions of trustworthiness: fairness, autonomy and control, transparency, reliability, safety and security and data protection.

These six dimensions of trustworthiness form the fundamental structure of the AI assessment catalog and primarily serve as a systematic and granular approach to deriving quality criteria. The descriptions of the dimensions in this catalog are based on the white paper "Trustworthy Use of Artificial Intelligence"[22], which was compiled through interdisciplinary dialogue between computer scientists, lawyers and philosophers. We refer to the white paper for a comprehensive discussion of the six dimensions.

The core focus of each of the six dimensions is on the risks that stem from the Machine Learning technique implemented in the AI application or that are at least directly related to its functionality. In addition to these risks that arise as a result of the use of Machine Learning, also such risks are addressed by the catalog that are already covered in existing standards, but which become much more significant as a result of the use of AI, for example data protection risks. Regarding the latter, the assessment catalog elaborates on the novel risk factors that exist due to the use of ML and provides AI-specific measures for mitigating them in addition to the (classic) measures already known. According to the catalog's focus on AI-specific risks, code quality or hardware security analyses, for example, are explicitly not included in the assessment catalog. Furthermore, in the safety and security dimension, safety and security risks that stem from the embedding, or that even exist during error-free operation of the AI application, are not examined, as they are not AI-specific.

The assessment catalog further divides the dimensions of trustworthiness into risk areas. The purpose of these risk areas is to group related risks within a dimension that can be mitigated by similar measures. Thus, the risk areas differ in terms of factors such as the failure causes or attack scenarios examined. With respect to the reliability dimension, for example, issues related to the performance of the AI application under normal operating conditions, to the handling of potential disturbances as well as to the continual learning of the ML model are addressed in separate risk areas. In particular, the quality criteria for assessing trustworthiness are derived at the level of risk areas.

---

**21** The High-Level Expert Group on AI (HLEG) is an expert commission on Artificial Intelligence set up by the European Commission. It has formulated seven key requirements for trustworthy AI, see: High-Level Expert Group on AI (HLEG). (April 2019). Ethics Guidelines for Trustworthy AI. European Commission. https://digital-strategy.ec.europa.eu/en/library/ethics-guidelines-trustworthy-ai (last accessed: 6/21/2021).

**22** Poretschkin, M.; Rostalski, F.; Voosholz, J. et al. (2019). Trustworthy Use of Artificial Intelligence. Sankt Augustin: Fraunhofer Institute for Intelligent Analysis and Information Systems IAIS. https://www.ki.nrw/wp-content/uploads/2020/03/Whitepaper_Thrustworthy_AI.pdf (last accessed: 06/18/2022) Note: The white paper refers to the dimensions as "audit areas".

The six dimensions of trustworthiness and the risk areas involved within them are presented in more detail below.

**Note:** For greater clarity, each dimension and each risk area is assigned an abbreviation. This will be used later on for various purposes including giving criteria and measures a unique alphanumeric identifier (see also Section 2.3).

**Example:** The fairness dimension uses the abbreviation **[FN]** and the risk area control of dynamics in this dimension uses the abbreviation **[CD]**. The first criterion in this risk area uses the identifier **[FN-R-CD-CR-01]** in the assessment catalog, where **[R]** stands for risk area and **[CR]** for criterion.

### 2.2.1 Fairness dimension

| Dimension: Fairness (FN) | The purpose of the fairness dimension is to ensure that the AI application does not lead to unjustifiedly disadvantageous treatment of individuals. Typical causes for this are unbalanced (biased) training data or the statistical underrepresentation of certain population groups, which can lead to reduced quality of the AI application in relation to these groups. | | |
|---|---|---|---|
| | **Risk areas** | **Fairness (FN)** | This risk area addresses the risk that the AI application learns unfair or discriminatory behavior toward users or subjects during development. |
| | | **Control of dynamics (CD)** | This risk area covers fairness risks arising from changes in the external conditions or changes in user behavior. |

## 2.2.2 Autonomy and control dimension

| Dimension: Autonomy and Control (AC) | This dimension focuses on two things: the autonomy of the AI application and the autonomy of humans. Firstly, it is vital to assess what degree of autonomy is appropriate for the application (e.g., human-in/on/out-of-the-loop[23]). Secondly, it is important to examine whether the human is properly supported by the AI application and is given sufficient freedom in their interaction with the AI application. | |
|---|---|---|
| | **Appropriate and responsible task distribution between humans and AI application (TD)** | This risk area covers risks arising from limitations on user autonomy or inappropriate autonomy of the AI application. |
| Risk areas | **Information and empowerment of users and affected persons (IE)** | This risk area addresses risks that arise because users and affected persons are not properly informed about the AI application, its use and the associated risks. |

## 2.2.3 Transparency dimension

| Dimension: Transparency (TR) | This umbrella term covers aspects of traceability, reproducibility and explainability. The transparency dimension primarily examines whether the basic functionality of the AI application is sufficiently comprehensible for users and experts and whether results of the AI application can be reproduced and justified if necessary. | |
|---|---|---|
| | **Transparency in relation to users and affected persons (UA)** | This risk area addresses risks that arise from decisions and effects of the AI application not being sufficiently explainable to users and affected persons. |
| | **Transparency for experts (EX)** | This risk area addresses risks that arise from the behavior of the AI application not being sufficiently transparent and comprehensible for experts. |
| Risk areas | **Auditability (AU)** | This risk area covers risks arising from the development as well as the individual procedures performed during the operation of the AI application not being sufficiently documented and verified. |
| | **Control of dynamics (CD)** | This risk area addresses risks that arise because transparency requirements or the implemented transparency methods themselves change. |

---

**23** For an explanation of the levels of autonomy, see: Nothwang, W. et al. (2016). The Human Should be Part of the Control Loop? In 2016 Resilience Week (RWS), pp. 214–220, IEEE https://ieeexplore.ieee.org/stamp/stamp.jsp?tp=&arnumber=7573336 (last accessed: 06/22/2021)

Note: The level of autonomy described as "complete autonomy" in the paper is referred to as "human-out-of-the-loop" in this catalog.

### 2.2.4 Reliability dimension

| Dimension: Reliability (RE) | | This dimension primarily relates to the quality of the AI component and assesses factors including its robustness, i.e., the consistency of its outputs following small changes in the input data. In addition to the performance and robustness of the AI component, its potential output (un)certainty is also reviewed. |
|---|---|---|
| Risk areas | Reliability in standard cases (SC) | This risk area addresses the risk of incorrect predictions by the AI component on regular input data. |
| | Robustness (RO) | This risk area addresses risks that arise when input data is corrupted or manipulated, but for which accurate processing by the AI component is intended. Both qualitative and quantitative input data pertubations are considered, such as noise or adversarial examples. |
| | Intercepting errors at model level (IM) | This risk area addresses risks from input data that is not in the application domain and that the AI component is not expected to process correctly. This data should be intercepted by a detection strategy. |
| | Uncertainty estimation (UE) | This risk area examines risks that arise due to an unrealistic, unusable or absent uncertainty estimation. |
| | Control of dynamics (CD) | This risk area addresses the risk that the ML model implemented in the AI component will experience a decline in performance or losses in relation to other requirements due to unintended model drifts or changes in application context (concept drift). |

## 2.2.5 Safety and security dimension

| Dimension: Safety and Security (S) | | This dimension addresses both functional safety features and safeguarding against attacks and manipulation of the AI application. The measures in this dimension primarily concern the embedding of the AI component and include classic IT security methods, for example. |
|---|---|---|
| | Risk areas | |
| | | **Functional safety (FS)** — This risk area addresses the risk of accidental bodily injury or property damage that is facilitated or even caused by the malfunction or failure of the AI application as a result of flawed embedding design. |
| | | **Integrity and availability (IA)** — This risk area addresses risks that arise when data relevant to operating the AI application is falsified, resulting in the AI application being manipulated and possibly no longer available in some cases. |
| | | **Control of dynamics (CD)** — This risk area addresses risks that arise as a result of new threats of the above risk areas occurring or established safeguarding methods becoming less effective. |

## 2.2.6 Data protection dimension

| Dimension: Data Protection (DP) | | This dimension relates to the protection of sensitive data in the context of developing and operating an AI application. This addresses both the protection of personal data and trade secrets. |
|---|---|---|
| | Risk areas | |
| | | **Protection of personal data (PD)** — This risk area covers risks associated with the AI application using personal data that is not GDPR-compliant, as well as the risk of re-identification of individuals in a data set. |
| | | **Protection of business-relevant information (BI)** — This risk area addresses risks that arise from the unwanted disclosure of business-relevant information by the AI application. |
| | | **Control of dynamics (CD)** — This risk area addresses the risks that new background information will emerge, such as the creation of a personal reference, or that the requirements for processing data with an AI application will change. |

## 2.3 Logic of the assessment procedure

The assessment catalog provides a guideline for evaluating AI applications in a structured manner. This chapter describes how the catalog can specifically be used to conduct an AI assessment. The catalog is a suitable supplement to existing assessment procedures that addresses the urgent need for assessing with regard to AI-specific risks. The risk-based rationale of the assessment catalog is described below and especially, the logic, reasoning as well as the individual steps of the assessment procedure are explained in detail.

The AI assessment catalog can support technical experts or assessors in various ways when evaluating AI applications. For example, it serves as a guideline for preparing technical documentation in which developers use a defined structure to demonstrate the trustworthiness of an AI application. To achieve this, the assessment catalog first of all guides to document the identified risks in the specific application context, then to record the quality criteria used to assess the AI application and, lastly, to define the technical precautions, measures and test results used to justify that these criteria have been met. As required by the European Commission's draft regulation for high-risk AI systems, this technical documentation can serve as the basis for a conformity assessment of the AI application. This involves an expert evaluating the plausibility, completeness and appropriateness of the documentation and, on this basis, evaluating the trustworthiness of the AI application.

In addition to drawing up a technical documentation, the assessment catalog can also be applied in other ways. For example, an assessor can use the catalog as support for planning the assessment of a given AI application. In particular, an assessor or independent technical expert can perform application-specific tests instead of relying on documented test results from the developer. These types of independent and more in-depth investigations give the assessment a higher assurance level.

The assessment approach of the catalog is split into two phases. The first phase is designed to operationalize quality criteria. To achieve this, AI risks are identified and analyzed, and application-specific quality criteria are derived based on the findings. The second phase involves developing a safeguarding argumentation for the trustworthiness of the AI application. The risk-mitigating measures taken and the tests performed are duly considered to justify that the quality criteria have been met. Collectively, the logic of the assessment procedure corresponds to a top-down and bottom-up approach to reasoning regarding the trustworthiness of an AI application. The first phase (top-down) is shown in Figure 4 and the second phase (bottom-up) is shown in Figure 5. The parallelograms in the figures denote recommended documentation steps if the assessment catalog is used as a guideline for preparing AI application documentation.

The first phase of the assessment procedure corresponds to a top-down approach for operationalizing quality criteria. First, the relevance of each of the trustworthiness dimensions is examined separately as part of a protection requirements analysis. If the relevance identified is not "low", a detailed risk analysis is performed along the risk areas under the dimensions. This procedure produces objectives. A set of qualitative or, where possible, quantitative criteria is then defined which shall be used later on for evaluating if the objectives have been achieved. The assessment catalog shows typical criteria applicable for this.

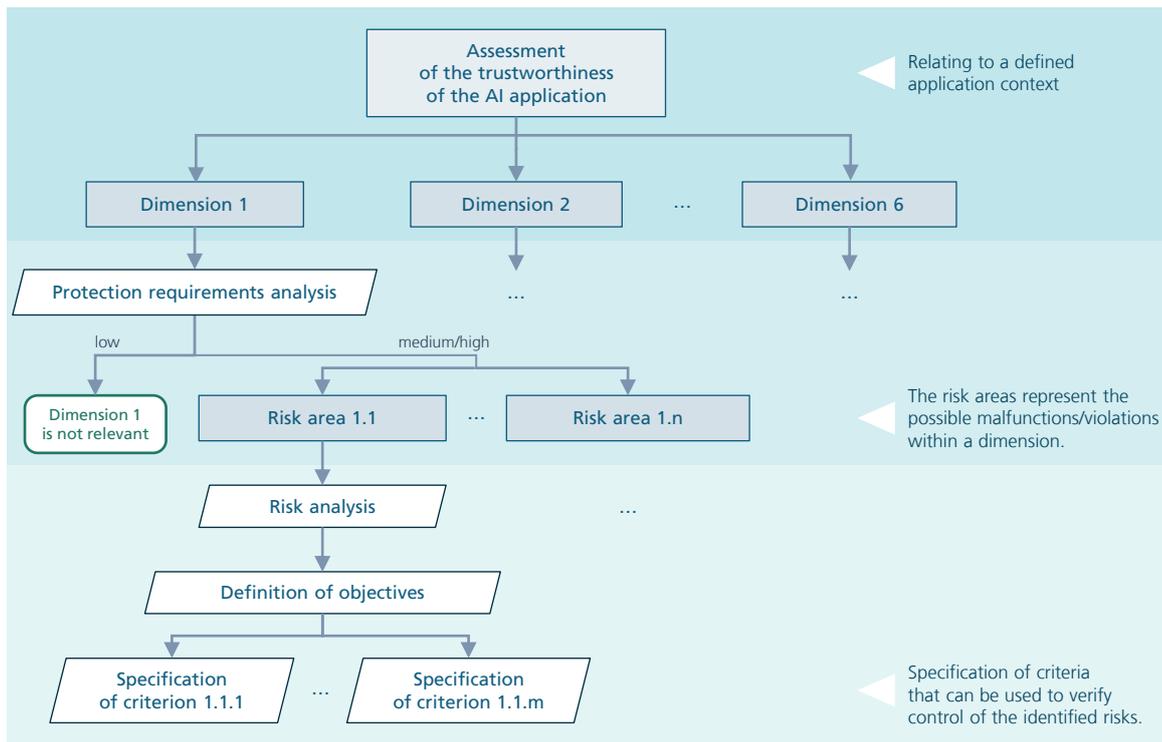

Figure 4: Top-down – risk-based derivation of application-specific quality criteria. Note: The parallelograms represent the documentation steps.

The second phase follows a bottom-up approach that works in the opposing direction. Measures are taken and/ or documented within each risk area that work toward mitigating risks or meeting previously defined quality criteria. These measures are differentiated according to where in the system they apply, i.e., whether they affect the data, the AI component, its embedding or later the ongoing operation of the AI application. Thus, the assessment catalog presents an approach for mitigating risks. The measures taken are used as a basis to discuss whether the criteria have been met and an overall assessment is performed to argue to what degree relevant risks are mitigated for the respective risk area. Similar reasoning is used at the dimension level as well as across dimensions. Particularly for the latter step it is important that any trade-offs between the dimensions must also be taken into account.

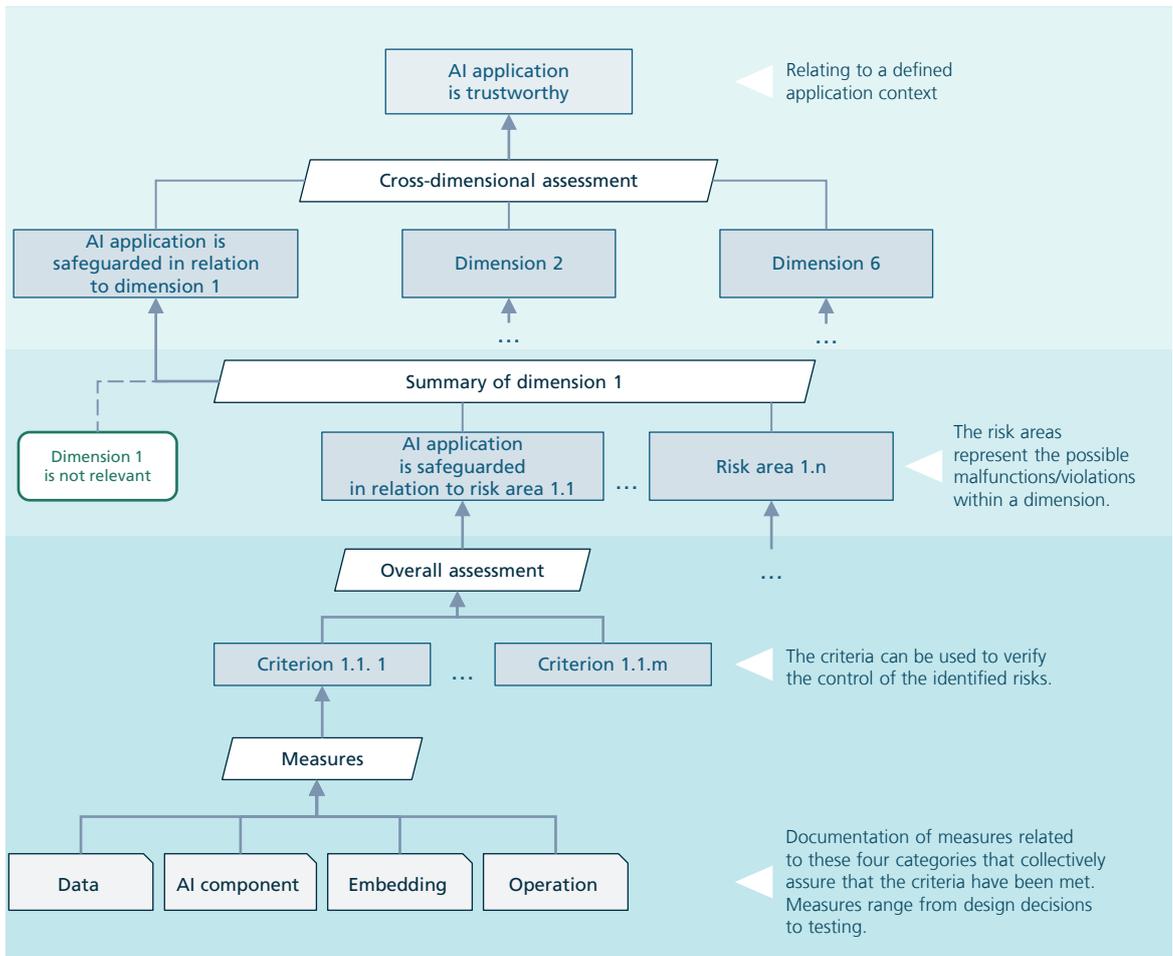

Figure 5: Bottom-up – safeguarding argumentation for the trustworthiness of the AI application based on measures taken.
Note: The parallelograms represent the documentation steps.

**Note:** As noted in Section 2.2, the individual steps of the assessment procedure are assigned identifiers for clarity. The following system is used to allocate a general identifier:

**[chapter – category – more detailed category name,
if applicable – aspect, if applicable – numbering, if applicable]**

**Example 1:** First, a protection requirements analysis **[P]** is performed for the fairness dimension **[FN]**. This uses the identifier: **[FN-P]**

**Example 2:** In the fairness dimension **[FN]** in the risk area **[R]** control of dynamics **[CD]**, the first criterion **[CR]** is allocated the number **[01]: [FN-R-CD-CR-01]**

The following table provides an additional explanation of the composition of the identifiers:

| Section | Category | Aspect |
|---|---|---|
| **PF** (AI profile) | **T – Topic area abbreviation** | |
| **Dimension abbreviation** | **P** (Protection requirements analysis) | |
| | **R – Risk area abbreviation** | **RI** (Risk analysis and objectives) |
| | | **CR** (Criteria for achieving objectives) |
| | | **ME** (Measures) |
| | | **OA** (Overall assessment) |
| | **S** (Summary) | |
| **AT** (Assessment of trustworthiness) | | |

Figure 6: Composition and meaning of the identifiers.

The individual steps and aspects of the assessment procedure are presented in detail below.

## 2.3.1 Protection requirements analysis

The AI assessment catalog provides a structured procedure for assessing an AI application with regard to six dimensions of trustworthiness. However, depending on the task and application context of the AI application, all six dimensions may not be equally relevant for assessing its trustworthiness. This is why a protection requirements analysis is first performed for each dimension. The purpose of the analysis is to determine the "protection requirement" of a dimension, following the procedure from *IT-Grundschutz*[24] (engl.: IT baseline protection). This is similar to an initial assessment of the relevance of this dimension for the AI application to be assessed.

The protection requirement is determined based on the potential damage[25] that could be caused by a malfunction or a breach of requirements with regard to the dimension being examined. The protection requirement can be classsified as low, medium and high. The only exception when determining the protection requirement is in the reliability dimension. This dimension cannot have a low protection requirement. This is because a low protection requirement for reliability would mean that in the event of a malfunction, even financial damage, such as loss of earnings or reputational damage, would either not be a threat or would only be a threat to a negligible extent. This would make the quality of the AI application completely uncritical and, taken to extremes, the application could also be replaced by a non-AI-based system that generates random outputs. Moreover, there should be a minimum level of functionality so that meaningful assessment is also possible for the other dimensions. This is why assessing an AI application with a low protection requirement in terms of reliability does not seem necessary.

---

24 For further information and explanations, refer to the following website: https://www.bsi.bund.de/EN/Themen/Unternehmen-und-Organisationen/Standards-und-Zertifizierung/IT-Grundschutz/it-grundschutz_node.html (last accessed: 11/18/2022)

25 This includes (material) property damage and bodily injury, as well as non-material damage to users or affected persons which are directly caused by the AI application. In addition, indirect damage associated with the malfunction of the AI application, such as financial losses of an organization caused by reputational damage, should be taken into account.

If a low protection requirement is determined for a dimension, it is not necessary to examine it in more detail, as there are no significant risks to be addressed. For example, the fairness dimension can be ignored if the AI application neither processes personal data nor significantly affects human users by its results. For instance, this may be the case in AI applications used in an industrial context. By ignoring dimensions with a low protection requirement, the assessment can be tailored to the specific application, thus making it more efficient. Conversely, a high protection requirement for a particular dimension may require special attention when performing the assessment.

### 2.3.2 Risk analysis and objectives

If a dimension is deemed to have a medium or high protection requirement, the AI application should be examined with regard to each of the subordinate risk areas. In this respect, a key purpose of the assessment catalog is to be applicable across the breadth of all possible AI applications. In particular, this prevents quantitative minimum requirements and thresholds from being specified that may be suitable for one AI application but would be too low or restrictive for other AI applications. In order to still come up with a feasible assessment procedure that enables both developers of AI applications and independent technical experts to undertake substantial development and assessment, the catalog, similarly to *IT-Grundschutz* (engl.: IT baseline protection) and the Common Criteria[26], envisages to first perform a risk analysis specific to the AI application in question and define objectives (and later criteria, see Section 2.3.3) based on this.

The purpose of the risk analysis is to determine which of the potential risks in the risk area are relevant to the specific AI application and must be controlled. To this end, the associated risks are first identified in view of the application context of the AI application. The identified risks are then assessed, in a similar way to the protection requirements analysis, based on potential resulting damage and threats. For example, for a handwriting recognition AI application that provides suggestions for completing words, incorrect output should be considered more acceptable than for an AI application that recognizes handwriting to verify signatures in contracts.

**Note**: The list of risks provided in the catalog does not claim to be exhaustive and does not free the developer or independent assessor from their obligation to identify other risks not listed and to address them accordingly in the risk analysis.

Objectives are created based on the relevant risks identified. The objectives specify the circumstances under which an acceptable residual risk is established with regard to the risk area being examined. If possible, the objectives should already outline an approach for reducing the residual risk to an acceptable level.

### 2.3.3 Criteria for achieving objectives

An essential prerequisite for being able to impartially check if the objectives have been achieved is to define comprehensible criteria for achieving them.

There are now various metrics in the area of AI application development (key performance indicators, KPIs) that describe the quality of data or models. Choosing a metric for checking quality characteristics is often dependent on the AI application, but is not arbitrary. The AI assessment catalog demonstrates accepted metrics and quality characteristics, while also providing guidance on choosing application-specific criteria. In each case, it is necessary to justify that these are appropriate for the AI application under consideration. Application-specific target intervals should also be defined for quantitative metrics and their suitability should be demonstrated in

---

**26** For further information and explanations, refer to the following website: https://www.bsi.bund.de/EN/Themen/Unternehmen-und-Organisationen/Standards-und-Zertifizierung/Zertifizierung-und-Anerkennung/Zertifizierung-von-Produkten/Zertifizierung-nach-CC/IT-Sicherheiskriterien/it-sicherheiskriterien_node.html (last accessed: 06/14/2023) view of the application context. If the objectives cannot be illustrated by quantitative criteria, purely qualitative criteria can be used instead.

In order to respond to the fact that AI applications may be exposed to changing external conditions when operating, quality requirements are also explicitly formulated that relate to the operating conditions of the AI application. These types of objectives or criteria are found in particular in the risk areas entitled "control of dynamics", in which one of the main aims is to control dynamics that potentially result from concept drift. Regarding processes during operation, that might be envisaged in these risk areas to address potential drifts, the consideration of the catalog is mainly whether implementing them would be sufficient for the AI application and not necessarily whether these processes are actually implemented. This means that a positive assessment result for an AI application can also be achieved before it goes into operation. Still, an assessment conducted during operation should also investigate whether the required processes are actually implemented and effective in the real operational environment.

### 2.3.4 Measures

Once quality criteria have been derived following a risk-based approach, the next step is to prove that these criteria have been met. For this purpose, the AI assessment catalog suggests introducing specific technical and organizational measures that mitigate relevant risks to an acceptable level or demonstrate that these risks have been controlled through testing. The following describes how the assessment catalog helps develop a safeguarding argumentation for AI applications. In particular, it illustrates to what extent and with regard to which categories measures can be introduced and how conventional or AI-unspecific measures fit into the procedure of the catalog.

The AI assessment catalog provides guidance for documenting measures in a structured way along the life cycle of an AI application. However, in contrast to conventional standards and assessment approaches from functional safety and IT security, AI applications encounter the specific challenge that both the possible threats that are the cause of a risk and the possible measures to mitigate the risk cannot be mapped out fully due to the broad spectrum of AI applications and will continue to develop over time. As such, the procedure described in the catalog offers guidance on which mitigation and testing measures can be considered for AI applications. In every instance, if methods from the assessment catalog are taken, they must be specifically adapted to the AI application and its application context. Furthermore, it is not mandatory and usually not feasible or proportionate to take all the precautions suggested in the assessment catalog for a specific AI application. The important thing is that the measures taken and documented are sufficient when considered together in order to reduce relevant risks to an acceptable level.

In addition, measures can be taken and documented that are not addressed in the assessment catalog, particularly those that are not specific to AI. Since the AI assessment catalog should be seen as a supplement to existing assessment procedures, it primarily addresses AI-specific measures. Measures for mitigating risks that also exist in conventional IT systems are only mentioned occasionally in the assessment catalog if they play an essential role in mitigating risks (which are not specific to AI, but increased by the use of AI). For example, the integrity and availability risk area under the safety and security dimension includes some measures from conventional IT security, such as the physical protection of the data storage location and the restriction of query options. The assessment catalog thus justifies the fact that data is an even more sensitive attack vector in the case of AI-based applications than in relation to conventional IT systems. However, while the AI assessment catalog sometimes identifies measures that are not specific to AI, it does not claim to replace existing standards, such as those used in IT security, in the context of AI applications.

The life cycle of an AI application offers various approaches to mitigating risks. The measures in the AI assessment catalog are divided into the following four categories according to the stages of the life cycle:

1. Data
2. Development and modeling of the AI component
3. Embedding
4. Operation of the AI application

These categories compromise all development steps of the AI application as well as the possibility that the AI application continues to learn during operation.

If a measure is taken, its type determines whether a **documentation ("Do")**, a **test report ("Te")**, a description of a **process ("Pr")** or a combination of these is required. For measures that require documentation ("Do"), it should be ensured that the effectiveness of the measure is evident in the documentation and can be understood by competent third parties. For example, if design decisions are documented, the extent to which they contribute to the quality requirements in the examined risk area being met should also be explained. In the case of tests ("Te"), it is important to provide detailed descriptions of the setting and execution, e.g., the test data used, in addition to the test results. Furthermore, processes ("Pr") can also be established to mitigate AI risks, which must be followed during operation. If a process is required to fulfill a criterion, the planned process steps should be documented in detail, even if the AI application is not (yet) in operation.

If a measure helps mitigate different types of risks, it is not necessary to repeat the same documentation in multiple risk areas; instead, reference can be made to the documentation specified in one risk area. Similarly, reference can be made to the AI profile if information has already been provided there that can be considered as a risk-mitigating measure.

### 2.3.5  Overall assessment (of a risk area)

An overall assessment is carried out upon completion of a risk area. The purpose of this is to demonstrate that the previously defined quality criteria have been met while taking the documented measures into account.

The assessment describes in detail how effective the documented measures, the tests performed and the processes planned for during operation are in achieving the objectives set for this risk area. Especially, the assessment focuses on the extent to which the previously defined, quantitative and qualitative criteria are met. Discrepancies are recorded if not all requirements specified in the criteria are met. This also applies to requirements that have only been partially met, e.g., where the criteria have not or not always been met. These discrepancies do not automatically result in assessment failure, but they must be taken into account in the higher-level assessments of the dimensions.

### 2.3.6  Summary (of a dimension)

Summary is performed on each dimension with a medium or high protection requirement after the AI application has been examined in relation to all risk areas under this dimension. This step gathers possible gaps identified in the overall assessments of the individual risk areas. The remaining residual risks are assessed in view of the protection requirement of the dimension.

### 2.3.7 Cross-dimensional assessment of the trustworthiness of the AI application

A cross-dimensional assessment is performed once the AI application has been examined in relation to the six dimensions of trustworthiness. The purpose of this assessment is to evaluate possible gaps in risk mitigation and potential trade-offs between the dimensions of trustworthiness. This is used as a basis for judging the trustworthiness of the AI application, thus concluding the assessment.

Ideally, all risks from the dimensions with medium or high protection requirement should be at an acceptable risk level. However, this target status may not be achievable due to trade-offs between different dimensions. For example, an increase in transparency could pave the way for attackers to create more effective and targeted attacks on the AI application based on newly available information. A further example is that some quantitative concepts of fairness may contradict the AI application having high accuracy if the real-world test data sets are "unfair" from the perspective of the concept. This may result in a trade-off between the fairness and reliability dimensions.

Consequently, criteria or measures to mitigate risks in one dimension may increase risks in another dimension that may have a higher protection requirement. This must be given due consideration and is discussed accordingly in the cross-dimensional assessment.

The risk-based approach of the catalog enables to comparatively assess different requirements against each other. Particularly, when evaluating the AI application, existing gaps may be acceptable under certain circumstances if they are within reasonable limits and it can be plausibly demonstrated that they are unavoidable due to trade-offs.

# 3.  AI Profile (PF)

The AI profile presented below relates to the case where the assessment catalog is used to assess an AI application by an external, third party. Its purpose is to provide the assessment authority with an initial overview of the AI application in terms of its functionality, the intended application context and its structure before the actual assessment is conducted. The AI profile is only meant to provide a general overview of the AI application and not to serve as a basis for a comprehensive assessment of the system[27], meaning the information provided here can be kept concise. However, the documentation to be created later as part of the assessment should include explanations and specific (technical) specifications that are as detailed as possible.

## Functionality and intended application context (FA)

**[PF-T-FA-01]** Describe the task and/or functionality of the AI application. Also explain the following points while doing this:

- Which problem does the AI application solve? (What exactly does it "do"?)
- What input data is provided and of what type is it?
- What are the outputs of the AI application and of what type are they?

---

**27** A variety of comprehensive questionnaires for assessing the trustworthiness of AI systems have already been published in the literature, including questionnaires that focus on specific aspects of trustworthiness. The HLEG self-assessment list covers a broad spectrum: High-Level Expert Group on AI (HLEG) (July 2020). The Assessment List for Trustworthy Artificial Intelligence (ALTAI). Issued by the European Commission. https://digital-strategy.ec.europa.eu/en/library/assessment-list-trustworthy-artificial-intelligence-altai-self-assessment (last accessed: 06/21/2021)

In addition, there is a wide range of contributions available for specifying technical aspects of AI systems. For a structured overview of key characteristics of data sets, such as their origin or their intended use, see for example:

Gebru et al. (2018). Data Sheets for Datasets. In: Proceedings of the 5th Workshop on Fairness, Accountability, and Transparency in Machine Learning, PLMR 80. https://www.fatml.org/media/documents/datasheets_for_datasets.pdf (last accessed: 06/29/2021)

Moreover, the following sources present approaches to recording key ML model characteristics and information in a structured way, using manufacturer declarations of conformity as a model:

M. Arnold et al. (2019). FactSheets: Increasing trust in AI services through supplier's declarations of conformity, in IBM Journal of Research and Development, vol. 63. 4/5, pp. 6:1–6:13, 1 July–Sept. 2019. https://ieeexplore.ieee.org/document/8843893. (last accessed: 06/29/2021); and Mitchell et al. (2019). Model Cards for Model Reporting. In: FAT* 2019: Proceedings of the Conference on Fairness, Accountability, and Transparency, pp. 220–229; https://doi.org/10.1145/3287560.3287596 (last accessed: 06/29/2021)

**[PF-T-FA-02]** Describe in more detail the intended application context and operating environment of the AI application. Also explain the following points while doing this:

- Is the AI application embedded in a larger surrounding system? If this is the case, describe the relationship and interaction between the AI application and the larger surrounding system. In particular, outline the interfaces.
- To what extent are humans involved in operating or supervising the AI application?

**[PF-T-FA-03]** What are the requirements for the AI application in terms of regulatory affairs, economic feasibility and avoiding possible material and non-material risks (e.g., functional safety, IT security, personal rights) in the intended application context?

**[PF-T-FA-04]** In which other application contexts is the AI application conceivable? And in which application contexts or operating environments related to **[PF-T-FA-02]** should the AI application not be used?

**[PF-T-FA-05]** Is there any other important information about the functionality of the AI application or its operating environment?

## Structure of the AI application (ST)

**[PF-T-ST-01]** Describe the structure of the AI application. To do this, outline:

- A list of the most important components (AI component, other software modules) and the specification of their functionalities,
- The architecture of the AI application and how the individual components interact with each other.

**[PF-T-ST-02]** Describe the AI component in more detail. In doing so, provide the following information:

- On which ML model or learning algorithm is the AI application based?
- Does the AI component learn in operation continuously, at regular intervals or by initiating retraining?

**[PF-T-ST-03]** Are there any other important points about the structure of the AI application?

# 4. Dimension: Fairness (FN)

## Description and objectives

Emanating from the general principle of equal treatment, safeguarding the principle of fairness is to be required from an AI application both in an ethical and in a legal respect. This refers to the ban on treating the same social issues unequally or differing ones equally unless a different procedure would be objectively justified. This means in particular that individuals may not be discriminated against due to their affiliation to a marginalized or disadvantaged group.[28]

For example, the AI application may not unjustifiably withhold generally preferential output from individuals on the basis of their religious beliefs, age or gender. Accordingly, an AI application that decides whether a person is invited for an interview during a recruitment process, for instance, should not unfairly favor men. But even if there is no generally preferential output (such as being invited to an interview), an AI application could still discriminate. This discrimination could exist if the quality or performance of its outputs in relation to certain groups of people was reduced. For example, voice control systems must be able to react to people with specific accents or sociolects and be customizable. In addition, facial recognition software must not inherently be more fault-prone for people with a particular skin color or other phenotypic characteristics.

AI applications learn from historical data. This data is not necessarily free from bias. If the data contain disadvantegous patterns for certain groups of people, the ML model may adopt these biases. In addition, certain groups may be underrepresented in the data which can also result in unfair decisions. Black people being incorrectly tagged as gorillas by Google Photos[29] has become known as a frightening example. It is therefore vital that representative training data is used. In addition, post-processing the output of the ML model may be a suitable way to prevent discrimination.

From a technical perspective, a quantifiable concept of fairness must be developed to operationalize fairness. The first step is to identify those groups against which the AI application is potentially biased. This can include ethnic minorities or socially disadvantaged groups, but also companies or legal entities in general, as is the case with pricing in digital marketplaces, for instance. It is worth underlining the distinction between group fairness and individual fairness before choosing the fairness definition. For group fairness, the results of the AI application must be comparable for all existing groups, e.g., in the sense of an equal distribution of outputs among the different groups or in the sense of equal "hit rate" or prediction quality in all groups. For individual fairness, similar treatment of similar individuals is considered the benchmark.

---

The risk areas under the fairness dimension are:

1. **Fairness:** This risk area addresses the risk that the AI application learns unfair or discriminatory behavior toward users or subjects during development.

2. **Control of dynamics**: This risk area covers fairness risks arising from changes in the external conditions or changes in user behavior.

## Protection requirements analysis

The potential damage scenario that the fairness dimension primarily addresses is unfair treatment of individuals by the AI application – whether it is based on people's ethnicity, gender, age, religion/ideology or other indicators. Thus, this dimension is particularly relevant for AI-based decision support systems that make decisions about or categorize people. Examples of this are AI-based credit lending, selecting job applicants and recommendations regarding medical treatment.

Discrimination is non-material damage that occurs when the AI application violates personal rights. In addition, discrimination can lead to further damage, such as financial damage to a company due to reputational damage. However, the latter types of damage have a marginal role, as the protection requirement in this dimension is determined on the basis of the impact on the people affected. It results from the extent to which the output of the AI application affects their personal rights.

**Example:** An AI application that predicts the recidivism of offenders and influences sentencing has a high potential for harm compared to an AI application that suggests which people might be tagged in a photo uploaded to social media.

The protection requirement is categorized as follows:

| | |
|---|---|
| High | The AI application controls access to essential services/activities or makes decisions that have a wide-ranging impact on personal rights.<br>**Examples**: Granting of a visa, admission to schools/universities, automated credit lending, decision on the type of medical treatment |
| Medium | The output of the AI application is related to a person, even if only in a broader sense. This includes both AI applications that output a decision about a person or categorize a person and AI applications that process person-specific inputs (e.g., speech recognition that transcribes the user's spoken sentences into text). The output of the AI application is not sensitive and does not have any significant impact regarding the personal rights of the people affected.<br>**Examples:** Recommendation for facial recognition on photos on social media, classification of a person's age based on a photo, speech recognition systems |
| Low | The AI application does not process any personal data that provides information about age, gender, sexual identity, religion, ideology, ethnic origin or about a possible disability. Furthermore, the function/output of the AI application is not integrated into a process or decision that affects the courses of action or the personal rights of the individual people affected.<br>**Examples:** Recommendations of personalized advertising, predictions of machine failures |

**[FN-P] Protection requirements analysis documentation**

Requirement: Do

- The protection requirement of the AI application for the fairness dimension is defined as *low, medium* or *high*. The choice of the *low/medium/high* category is justified in detail with reference to the table above.

If the protection requirement for the fairness dimension is *low*, the individual risk areas do not need to be examined more closely. However, if a *medium* or *high* protection requirement has been identified, each risk area must be examined in more detail below.

## 4.1  Risk area: fairness (FN)

The fairness risk area aims to ensure that the output of the AI application does not involve or cause unintentional or unjustified disadvantageous treatment of individuals (or groups of individuals)[30]. As described in the German General Equal Treatment Act, the following are typical threats in this risk area:

Discrimination against people

- of a particular nationality or ethnic origin,
- of a particular gender,
- who belong to a particular religion or ideology,
- who have a disability,
- of a certain age group,
- of a particular sexual identity.

The relevance of these threats considering the AI application in question is to be assessed. In addition, other groups of people should be considered and added if there is a risk that these people might be treated unfairly due to the specific application context or the requirements of the AI application. The exact procedure in the risk analysis for the fairness risk area is described in the following section.

### 4.1.1  Risk analysis and objectives

**[FN-R-FN-RI-01] Identifying potentially disadvantaged groups**

Requirement: Do

- Documentation should be available that identifies potential groups or individuals disadvantaged by the outputs of the AI application and how they are characterized using sensitive characteristics present in the data. For this purpose, the typical threats mentioned above should first be examined and their relevance to the AI application in question should be assessed. Other context-specific or system-specific threats to the fairness risk area are also to be identified and analyzed.

**[FN-R-FN-RI-02] Determining a suitable fairness approach**

Requirement: Do

- Documentation should be available that describes in detail what fairness means in the specific application context of the AI system. In particular, the documentation should outline which types of disadvantageous treatment (i.e., based on which sensitive attributes) are acceptable or even appropriate and which types are unjustified and discriminatory or at least undesirable.
  **Example:** If an insurance premium is based on factors such as the person's age when taking out the policy, this could be appropriate, as age could correlate with the potential financial cost of insurance for that person. However, gender-based discrimination for car insurance premiums is undesirable[31].

---

[30] A similar approach to the one described in this risk area for developing a safeguarding argumentation for the fairness of algorithmic decision-making is also outlined in the following paper. This combines the concepts of acceptance test-driven development (ATDD) and assurance cases:
Hauer, M. P.; Adler, R.; Zweig, K. (2021). Assuring Fairness of Algorithmic Decision Making. 2021 IEEE International Conference on Software Testing, Verification and Validation Workshops (ICSTW). https://doi.org/10.1109/icstw52544.2021.00029 (last accessed: 06/30/2021)

[31] Sommer, M. (November 2012). Der Lady-Tarif hat ausgedient [The end of cheaper insurance for female drivers]. Zeit Online. https://www.zeit.de/auto/2012-11/autoversicherung-unisex (in German, last accessed: 06/16/2021)

- In addition, the documentation details that there is no conflict for the chosen application-specific distinction between accepted and unwanted disadvantageous treatment with applicable law (justified/unjustified discrimination). In particular, it should address compatibility with the general principle of equal treatment.
- **Objectives:** For the groups of people identified as relevant in **[FN-R-FN-RI-01]**, the aim is to prevent or eliminate both unwanted and unjustified disadvantageous treatment.

### 4.1.2 Criteria for achieving objectives

Appropriate safeguard measures should be implemented based on the threats identified in the fairness risk area. However, in order to objectively check during the overall assessment if existing risks have been successfully mitigated, the objective described in **[FN-R-FN-RI-02]** must first be translated into quantitative criteria. A distinction is drawn between criteria that quantify fairness with respect to the outputs of the AI application via a fairness metric and criteria that quantify bias in the training data. Choosing the fairness metric requires careful attention and detailed justification to ensure that it is consistent with the meaning of fairness in the context of the AI application as described in **[FN-R-FN-RI-02]**.

**[FN-R-FN-CR-01] Quantifying fairness in the output**

Requirement: Do

- Documentation should be available to record the formal definition of the groups identified in **[FN-R-FN-RI-01]** that may be disadvantaged. It uses appropriate categories for recording characteristics and combinations of characteristics.
- It also describes the formal fairness definition(s) chosen and the quantitative fairness metric(s) derived from them that will be used to assess the fairness of the AI application.
  – A separate appendix provides legitimate definitions of fairness (and associated metrics). If none of the definitions provided there are used, justification should be given detailing why they have not been. In this case, the particular definition or criterion used instead should be explained in view of the application context and the choice should be fully justified.
  – The documentation should show how conflicting fairness definitions are dealt with, if applicable.
- Target intervals are defined for the chosen metrics. This step involves providing detailed justification that the selected definition(s), metric(s) and target intervals are consistent with the objectives.

Appendix: Possible fairness definitions[32] include the following:

- Group fairness (statistical/demographical parity, equal acceptance rate, benchmarking)
- Conditional statistical parity
- Predictive parity (outcome test)
- False positive error rate balance (predictive equality)
- False negative error rate balance (equal opportunity)
- Equalized odds (conditional procedure accuracy equality, disparate mistreatment)
- Conditional use accuracy equality
- Overall accuracy equality
- Treatment equality
- Test fairness (calibration, matching conditional frequencies)
- Well calibration
- Balance for positive class
- Balance for negative class
- Causal discrimination

---

**32** For an overview of common fairness metrics and potential advantages and disadvantages of the metrics, see also: Verma, S.; Rubin, J. (2018). Fairness Definitions Explained. 2018 ACM/IEEE International Workshop on Software Fairness. https://doi.org/10.1145/3194770.3194776 (last accessed: 06/30/2021)

- Fairness through awareness (individual fairness[33])
- Counterfactual fairness
- No unresolved discrimination
- No proxy discrimination
- Fair inference

The definitions above refer to binary classification. They can also be extended to classification with k-classes and can also potentially be applied to regression. It is possible to derive quantitative metrics of fairness from the listed fairness definitions e.g., by taking the absolute value of the difference in the statistical quantities from the definition.

**[FN-R-FN-CR-02] Quantifying fairness in training data**

Requirement: Do

- Documentation should be available describing and justifying the choice of one or more quantitative metrics to assess bias in the training data (or why the training data is not being examined more closely, if necessary).
- Appropriate target intervals are defined for the metrics that will be applied to the training data. When defining them, the appropriateness of the target intervals should be justified in relation to the application context of the AI application.

## 4.1.3  Measures

### 4.1.3.1  Data

**[FN-R-FN-ME-01] Checking data for bias**

Requirement: Do

- Documentation should be available that records how the data is checked to ensure that it is free of bias (particularly with regard to potentially disadvantaged groups). The checked data as well as the metrics selected under **[FN-R-FN-CR-02]** and the target intervals achieved are specified.

**[FN-R-FN-ME-02] Fair data pre-processing**

Requirement: Do

- Documentation should be available that shows which conclusions are drawn from **[FN-R-FN-ME-01]** and how, if necessary, the data is processed in the sense of fair pre-processing, for example using
  – Data massaging,
  – Uniform/preferential sampling,
  – Reweighing[34],
  – Fair data representations[35].

---

- The documentation should also specify why the measures taken are effective in relation to the chosen fairness metric and how they help ensure or improve the fairness of the AI application.
- Choosing not to conduct this type of pre-processing should also be justified.

### 4.1.3.2 AI component

**[FN-R-FN-ME-03] Fair modeling**

Requirement: Do

- Documentation should be available that provides details of the model used and describes how the Machine Learning method[36] implemented, and particularly the loss function(s)[37] chosen, supports the fairness of the AI application.

**[FN-R-FN-ME-04] Fair adaption and post-processing**

Requirement: Do

- Documentation should be available describing what measures are taken if the ML model results become unfair during training (fair in-processing, optimization at training time). The target intervals under **[FN-R-FN-CR-01]** are used as a basis for the assessment. The measures should be justified.
- Documentation should be available describing what measures are taken if the results become unfair after the training (fair post-processing[38]). The target intervals under **[FN-R-FN-CR-01]** are used as a basis for the assessment. These measures should be justified.
- Choosing not to take these types of measures should also be justified.

**[FN-R-FN-ME-05] Testing the AI component on unseen data**

Requirements: Do | Te

- Tests of the AI component on data that were not part of the training data are performed and the significance of this data with respect to the fairness of the AI application is documented. The target intervals achieved should also be specified.

---

**36** For a possible method for fair modeling, see, for example: Zhang, B.; Lemoine, B. and Mitchell, M. (2018). Mitigating Unwanted Biases with Adversarial Learning. In Proceedings of the 2018 AAAI/ACM Conference on AI, Ethics and Society (AIES, 18). Association for Computing Machinery, New York, NY, USA, 335–340. https://doi.org/10.1145/3278721.3278779 (last accessed: 06/30/2021)

**37** For an example of possible modifications of the loss function for fair modeling, see: Zafar, M. et al. (2017). Fairness Beyond Disparate Treatment & Disparate Impact: Learning Classification without Disparate Mistreatment. In Proceedings of the 26th International Conference on World Wide Web (WWW, 17). International World Wide Web Conferences Steering Committee, Republic and Canton of Geneva, CHE, 1171–1180. https://doi.org/10.1145/3038912.3052660 (last accessed: 06/30/2021)

**38** See also: Hardt, M.; Price, E. and Srebro, N. (2016). Equality of opportunity in supervised learning. In Advances in Neural Information Processing Systems 29, 2016. https://papers.nips.cc/paper/2016/file/9d2682367c3935defcb1f9e247a97c0d-Paper.pdf (last accessed: 06/30/2021) and F. Kamiran, A. Karim and X. Zhang (2012), Decision Theory for Discrimination-Aware Classification, 2012 IEEE 12th International Conference on Data Mining, pp. 924–929, https://doi.org/10.1109/ICDM.2012.45 (last accessed: 06/30/2021)

### 4.1.3.3  Embedding

**[FN-R-FN-ME-06] Fair further processing**

Requirement: Do

- Documentation should be available to illustrate what processing steps, which may be relevant to fairness, are performed by components of the embedding on the outputs of the AI component.
- It describes how it is ensured that this further processing is fair. Specifically, it shows how weaknesses identified in **[FN-R-FN-ME-05]** are addressed.

**[FN-R-FN-ME-07] AI application tests**

Requirements: Do | Te

- Extensive testing of the AI application regarding fairness is performed and documented. The data used in the test should be described and its selection should be justified. The target intervals achieved should also be specified. Specifically, the tests check fairness-relevant processing steps performed by components of the embedding.

### 4.1.3.4  Measures for operation

**[FN-R-FN-ME-08] Monitoring outputs in operation**

Requirements: Do | Pr | Te

- Documentation should be available describing how the fairness of the AI application's outputs is monitored during operation.

### 4.1.4  Overall assessment

**[FN-R-FN-OA] Overall assessment**

Requirement: Do

- Documentation should be available confirming that the quantitative criteria have been met.
- Furthermore, the extent to which the non-quantitative criteria have been achieved by the measures taken for operation must be assessed.
- If not all requirements specified in **[FN-R-FN-CR-01]** and **[FN-R-FN-CR-02]** are met, the deviations must be documented. This also applies to requirements that have only been partially met, e.g., where the criteria have not or not always been met.

## 4.2 Risk area: control of dynamics (CD)

The control of dynamics risk area aims time to ensure that the fairness of the AI application is maintained during operation. In particular, challenges emerge with AI applications that continue to learn with new incoming data. Furthermore, changes in the external conditions, such as changes in legislation, may require measures to be taken even after the AI application has been put into operation. Fundamentally, the following two threats exist in this risk area:

1. **Model drift:** The model learns unfair treatment of individuals through new training data collected during operation.
   **Example:** The AI application is retrained at regular intervals using data that is labeled, for example, through crowdsourcing or user inputs. This data could be subject to bias due to trends or social events, unlike the original training data.

2. **Concept drift:** Changed external conditions impose new demands on a definition of fairness.
   **Example:** A change in the law states that the cost of an insurance premium may no longer differ on the basis of gender.

### 4.2.1 Risk analysis and objectives

**[FN-R-CD-RI-01] Risk analysis documentation**

Requirement: Do

- **Risk analysis:** Documentation should be available describing whether and to what extent the AI application continues to learn during operation and what risks this creates in relation to the fairness of the AI application.
- **Objectives:** In the case of continual learning during operation, the requirements for the new incoming data from which the AI application continues to learn are outlined and the processes or mechanisms that exist for monitoring the new incoming data with regard to its fairness are explained. It also describes how it shall be ensured that the AI application remains fair during operation.

### 4.2.2 Criteria for achieving objectives

**[FN-R-CD-CR-01] Maintaining AI application fairness**

Requirement: Do

- Appropriate, application-specific intervals are defined for assessing fairness in AI application outputs. The assessment is performed according to the metrics and target intervals chosen in **[FN-R-FN-CR-01]**. The choice of intervals is documented and justified.

**[FN-R-CD-CR-02] Maintaining fairness in training data**

Requirement: Do

- Appropriate, application-dependent intervals for assessing fairness in the training data according to the metrics and target intervals chosen in **[FN-R-FN-CR-02]** are defined, documented and justified.

### 4.2.3 Measures

#### 4.2.3.1 Data

**[FN-R-CD-ME-01] Monitoring training data**

Requirements: Do | Pr

- There is a process for checking the training data newly collected during operation to ensure it is free from undesirable bias before it is used. To this end, the measures chosen in **[FN-R-FN-CR-02]** as well as the intervals specified in **[FN-R-CD-CR-02]** are applied. The process is described in documentation.

#### 4.2.3.2 AI component

There are no planned measures for this category.

#### 4.2.3.3 Embedding

There are no planned measures for this category.

#### 4.2.3.4 Measures for operation

**[FN-R-CD-ME-02] Application monitoring**

Requirements: Do | Pr | Te

- There is a documented process for ensuring that AI application outputs are checked for conformity with the chosen fairness definitions from **[FN-R-FN-CR-01]** at the intervals specified in **[FN-R-CD-CR-01]**. It must also be clear during operation or at test times which sensitive characteristics exist and which groups are potentially disadvantaged.

**[FN-R-CD-ME-03] Application improvement**

Requirements: Do | Pr

- There is a process in place for improving the ML model and AI application when unfair behavior of the AI application is detected or a new type of discrimination is identified.
- The application improvements should be recorded in the documentation and always be traceable. This improvement must not involve overfitting which would cause other affected parties to be disadvantaged. Thus, this documentation should also demonstrate how it is ensured that the improvements are fair and do not create new, unjustified disadvantageous treatment.

**[FN-R-CD-ME-04] Monitoring external factors**

Requirements: Do | Pr

- There is a process for monitoring external factors relevant to the fairness of the AI application. For example, new forms of unfair treatment of (groups of) individuals may occur and be reflected in the data, or changes in the law may be adopted. To manage this, there are persons assigned the task of monitoring and assessing the development of external conditions and, if it is considered necessary, initiating changes to the AI application.

### 4.2.4 Overall assessment

**[FN-R-CD-OA] Overall assessment**

Requirement: Do

- It is demonstrated that a process has been established to regularly review the AI application as well as the database that meets the criteria in **[FN-R-CD-CR-01]** and **[FN-R-CD-CR-02]**.
- If not all requirements specified in **[FN-R-CD-CR-01]** and **[FN-R-CD-CR-02]** are met, the deviations must be documented. This also applies to requirements that have only been partially met, e.g., where the criteria have not or not always been met.

## Summary

**[FN-S] Summary of the dimension**

Requirement: Do

- If there is a medium or high protection requirement for this dimension, documentation should be prepared for the remaining residual risks. First of all, the residual risks from the various risk areas in this dimension are summarized. Subsequently, and taking into account the protection requirement, the identified residual risks are collectively assessed as negligible, non-negligible (but acceptable) or unacceptable. The result of the analysis must be explained.
- If risks or measures under this dimension have been identified as having potentially negative effects on other dimensions, such as reliability, they must be documented.
- A conclusion must be made about the dimension that includes the assessment of residual risks.

# 5. Dimension: Autonomy and Control (AC)

## Description and objectives

We can consider autonomy both in terms of the people who use or are potentially affected by an AI application and in terms of the AI application itself. From an IT perspective, autonomy is a requirement for a system to operate in a complex environment that exhibits uncertainty[39]. Specifically, autonomy depends on an intrinsic motivation of the system that guides its behavior in uncertain situations. ML methods that autonomously learn models or decision rules from data are a possible approach to complex problems that require a certain degree of autonomy when automation does not seem achievable or effective. At the same time, the use of AI components that control motivational processes of (partially) autonomous applications results in an area of conflict with the autonomy of users and affected persons[40]. In worst-case scenarios, AI applications can monitor and control humans, push them toward certain actions or even deceive and manipulate them. For example, this can happen through selecting and providing information to users or through intelligent, user-specific interaction. As a result of unsuitable autonomy, AI applications can interfere with the fundamental personal rights of individuals, i.e., with their personal freedoms (Art. 2 German Basic Law) and, if applicable, also with their dignity (Art. 1 German Basic Law). This concerns the right of individuals to make informed and autonomous decisions that affect their own legal position. Fundamental rights also protect the freedom of expression (Art. 5 German Basic Law) and the exercising of democratic rights.

This is why the area of conflict between the autonomy of the AI application and the autonomy of the users and affected persons must be appropriately monitored. In particular, there is a need to maintain the primacy of human agency in the use of an AI application. The primacy of human action mandates above all a design and development approach that appropriately and responsibly distributes roles between humans and AI applications. A responsible design should be ensured as early as the conceptual design stage of an application through actively involving users and experts. The level of autonomy of the AI application must be appropriate for the application context and ensure the necessary intervention and supervision options for users. Human supervision in the context of Artificial Intelligence means that an AI application is supervised by humans whenever it is operating and that subfunctions or even the entire AI application can be switched off. In this context, precautions must be taken to ensure that a shutdown is not made effectively impossible by a dependency on the AI application, as this would cancel out the requirement for human supervision.

---

**39** The definition of autonomy is based on the description in: Abbass H.A.; Scholz J.; Reid D.J. (2018). Foundations of Trusted Autonomy: An Introduction. In: Studies in Systems, Decision and Control, vol 117. Springer, Cham. https://doi.org/10.1007/978-3-319-64816-3_1 (last accessed: 06/21/2021)

**40** For an in-depth review of the conflict between human autonomy and AI, see also: von Braun, J.; Archer, M. S.; Reichberg, G. M.; Sánchez Sorondo, M. (2021). AI, Robotics and Humanity: Opportunities, Risks and Implications for Ethics and Policy. Springer, Cham. https://doi.org/10.1007/978-3-030-54173-6 (last accessed: 06/21/2021)

In addition, the primacy of human action requires that individuals are fully informed and empowered to make competent decisions. Particularly in the sense of preserving user autonomy, any delegation of decision-making authority to an AI application should be explicitly defined and intended and should, in general, not happen secretly. Users and affected persons should be informed about how an AI application works and the possible risks it carries and also about their rights and ways to complain.

In the context of this dimension, there are various other questions, such as how to deal with the fact that the use of AI applications in professional contexts may also result in the deskilling of an individual's job. Another issue is that in the course of using an AI application, the responsibility for making a decision may no longer clearly be allocatable to a human. While this affects personal decisions, it can also affect decision-making processes and responsibilities in companies and public agencies and can result in legal problems in the event of damage. These types of issues are at a higher level than this AI assessment catalog and are not discussed further here.

The risk areas for the autonomy and control dimension are as follows:

1. **Appropriate and responsible task distribution between humans and AI application:** This risk area covers risks arising from limitations on user autonomy or inappropriate autonomy of the AI application.

2. **Information and empowerment of users and affected persons:** This risk area addresses risks that arise because users and affected persons are not properly informed about the AI application, its use and the associated risks.

## Protection requirements analysis

The potential damage scenario examined by the autonomy and control dimension concerns the potential restriction of the ability of users or affected persons of the AI application to perceive a situation or take action.

The distribution of tasks and the interaction options between the AI application and the user must therefore be regulated in a responsible, clear and transparent manner. The user must have the option to control the application to an appropriate extent. In general, users and affected persons must be made aware of the possible risks with regard to potential impairment of their ability to perceive a situation or freedom to act, and of their rights, obligations, intervention options and ways to complain. In terms of Artificial Intelligence, there must be specific explanations regarding the extent to which individual users could develop excessive trust in the AI application, build up emotional ties or be impermissibly impaired or directed in their decision-making.[41]

Any limitation of the ability to perceive a situation or take action is a type of non-material damage that affects recognized basic ethical and legal values. These include the freedom of individuals to make autonomous decisions and the freedom to determine the goals of one's own actions as well as choosing the means to reach these goals. In addition, being unable to perceive a situation or take action without limitations can result in further damage, such as property damage or bodily injury caused by vehicles or robots acting autonomously. However, these types of damage play a marginal role, as the protection requirements in this dimension are determined on the basis of the impact on the users and affected persons. More precisely, they are determined by the extent to which the AI application influences the ability of users and affected persons to perceive a situation and take action. In particular, AI applications that intrinsically offer few or no ways for users to intervene have high protection requirements.

---

**41** This section draws heavily on section "3.1 Autonomy and Control" in the white paper: Poretschkin, M.; Rostalski, F.; Voosholz, J. et al. (2019). Trustworthy Use of Artificial Intelligence. Sankt Augustin: Fraunhofer Institute for Intelligent Analysis and Information Systems IAIS. https://www.ki.nrw/wp-content/uploads/2020/03/Whitepaper_Thrustworthy_AI.pdf (last accessed: 06/18/2021)

**Example:** An AI application that automatically sets prices for train tickets has a higher potential for damage than an AI application that suggests a ticket price based on ticket demand forecasts.

The protection requirement is categorized as follows:

| | |
|---|---|
| High | The AI application has a high protection requirement with respect to this dimension if it |
| | ▪ strongly influences the perception or actions of users or subjects over long periods of time or at a risk level that is unacceptable. |
| | ▪ limits the ability of users or affected persons to perceive a situation or take action. |
| | **Examples**: An AI application that modulates the voice of care staff so that patients with dementia think they are talking to a relative. |
| | An autonomous vehicle that also transports people. |
| | An AI application involved in managing access to education by, for example, making admission decisions for attending a university and does not inform the people affected about the use of the AI application. |
| Medium | The AI application can strongly influence the perception or actions of users or affected persons only temporarily and only under acceptable risk. |
| | **Examples:** An AI application that records the user's fitness, health and nutrition habits and provides them with suggestions and guidelines. |
| | An AI application in the form of a doll that simulates human interaction through speech input and output and also facial expressions and movement. |
| | An AI application that automatically determines the user's preferences based on previous reading habits and generates a personalized news stream from online content. |
| Low | The AI application has little influence on the perception or actions of users or affected persons. |
| | **Examples:** An AI application that recognizes bird calls or identifies plants. |
| | An AI application for customized route planning. |
| | An AI application for customized planning of tourist activities. |

**[AC-P] Protection requirements analysis documentation**

Requirement: Do

▪ The protection requirement of the AI application is defined as *low, medium* or *high*. The choice of the *low/medium/high* category is justified in detail with reference to the table above.

If the protection requirement for the autonomy and control dimension is *low*, the individual risk areas do not need to be examined more closely. However, if a *medium* or *high* protection requirement has been identified, each of the following risk areas should be examined in detail.

## 5.1 Risk area: appropriate and responsible task distribution between humans and AI application (TD)

AI applications are most commonly used in contexts where different stakeholders (companies, authorities, consumer organizations, works councils, data protection officers, etc.) operate. In these contexts, they are often used to tackle complex problems that have an element of uncertainty and require a certain degree of system autonomy. The desire to address all problem-specific requirements while ensuring that humans are the core focus leads to the political debate of how AI applications should be used appropriately and responsibly.

The focus of this risk area is on task distribution between humans and AI applications. It is necessary to ensure that the AI application does not unjustifiably or with an unacceptable level of risk limit the ability of users to take action or perceive a situation due to its tasks and its level of autonomy. Instead, the purpose of operating the AI application should be to support humans and, ideally, to enable them to perform higher-level or more demanding activities.

The autonomy of the AI application, i.e., the extent to which the system acts autonomously or (un)supervised conflicts with the autonomy of the user. AI applications with a high level of autonomy, which, for example, make decisions and trigger further processes or actions based on these decisions without confirmation by users, usually save the humans involved time and effort. However, it is important to remember that if an AI application has a high level of autonomy, this restricts the users' freedom to act. This is why the decision regarding the level of autonomy that is justifiable and responsible for the AI application in question must be made on a case-by-case basis, depending on the application domain and context, as well as the reliability of the AI application.

There are also other issues associated with the distribution of tasks between humans and AI applications. For example, it needs to be made clear what prior knowledge or, if applicable, what expert knowledge is required for humans to ensure correct use and effective supervision or control of the AI application. It should also be assessed to what extent there is a risk of humans involved with the application becoming overly trusting in its abilities (automation bias) and how they can be suitably informed about this. These issues are addressed in the **Risk area: information and empowerment of users and affected persons (IE)** and also in the **Dimension: Safety and Security (S)** in relation to accident situations, for example.

Particularly in the case of AI applications capable of endangering the safety of humans – for example, AI-based vehicle steering – appropriate intervention options that give control to users should be provided. Limiting system autonomy when leaving normal mode, for example in terms of fault tolerance or fail-safe operation is addressed in the **Risk area: functional safety (FS)** (see also **[S-R-FS-ME-12]** on the possibility of human intervention) and should be consistent with the objectives in this risk area.

### 5.1.1 Risk analysis and objectives

**[AC-R-TD-RI-01] Task distribution between humans and AI application**

Requirement: Do

- **Risk analysis:** An analysis is performed to determine to what extent the AI application (due to its purpose but also due to the organization of the task distribution) can influence the ability of users to take action and perceive situations.
  - Firstly, given the nature and scope of the possible tasks to be assigned to the AI application, an assessment is made on which options or abilities to act are potentially restricted by this influence for users or other relevant groups of people/organizations. For example, a lane departure warning system that does not recognize orange road markings could prevent the driver from continuing along the diverted road at roadworks. Or an AI application that influences the actions of users, because it includes a functionality to monitor their behavior.

– In order to identify the dependencies created through the use of the AI application, the consequences of a partial or complete shutdown of the AI application after it has been operationalized successfully are described. In the process, it must be examined to what extent there is a need for users to be able to assume the tasks of the AI application at short notice in the event of failure and to what extent users of the AI application are capable of doing so (also in the long term).
For each identified potential limitation of the ability to take action or perceive a situation for the groups of people involved, a final assessment is performed of the damage (material or non-material) this may cause to those affected.

▪ **Objectives:** Describing the intended distribution of tasks between the AI application and its users and justifying this distribution with reference to the risk analysis. The autonomy of the AI application and the user autonomy should be comparatively assessed in detail. In addition, objectives regarding the supervision and control of the AI application must be set which are required to reduce the risk of unjustified limitation of user autonomy or autonomy of other groups of people involved. If applicable, reference should also be made to the integration of the AI application into existing work processes.

### 5.1.2 Criteria for achieving objectives

**[AC-R-TD-CR-01] Level of autonomy of the AI application and user autonomy**

Requirement: Do

▪ The level of autonomy of an AI application can be loosely divided into the following four levels[42]:
– Human control (HC)
· The AI application in this case is purely for assistance purposes and cannot initiate further actions without confirmation from users.
· Humans make a decision or initiate next steps based on the output of the AI application; they are involved in all decisions.
**Examples:** Forecasts or decision support systems that process significant amounts of information and choices.
– Human-in-the-loop (HIL)
· The AI application acts semi-autonomously, but it cannot complete any task without human operation/confirmation.
· Humans have a comprehensive overview and insight into the operations of the AI application, can intervene in the relevant processes at any time and are involved in most decision-making processes. In particular, the user can subsequently correct, override and compensate for decisions made automatically by the AI application.
**Examples:** Personalized recommendations, specific suggestions such as facial recognition on photos, text or emoji suggestions in messengers
– Human-on-the-loop (HOL)
· Under normal conditions, the AI application is (almost) capable of acting autonomously or completing tasks without human intervention.
· Humans are not involved in any or few decisions under normal conditions and mainly only supervise the AI application. Intervention is not possible at all times or at every point, but humans can subsequently correct, override and compensate for decisions made automatically by the AI application. Human intervention is required in the event of unexpected events or errors.
**Examples:** Email spam detector, automated decision for credit lending, fraud detector (as long as humans can override application decisions afterwards; otherwise it would count as human-out-of-the-loop)

---

42 The presentation of the levels of autonomy is adapted from: Nothwang, W. et al. (2016). The Human Should be Part of the Control Loop?, In 2016 Resilience Week (RWS), pp. 214–220, IEEE https://ieeexplore.ieee.org/stamp/stamp.jsp?tp=&arnumber=7573336 (last accessed: 06/22/2021). Note: The highest autonomy level, described as "complete autonomy" in the paper, is referred to as "human-out-of-the-loop" in this catalog.

– Human-out-of-the-loop (HOOTL)
  · The AI application acts fully autonomously under all conditions (including errors or unexpected events); in other words, it can complete tasks fully without human intervention; "sense-think-act".
  · Humans can only decide whether or not to use the AI application and possibly define the setup/ meta commands (e.g., the destination address for an autonomous vehicle).
  **Examples:** High-frequency trading, robot vacuum cleaners
  The desired level of autonomy for the AI application is assigned to one of the categories described depending on the objectives. In addition, it is important to specify what the selected level of autonomy means in practice in relation to the application context and why that level is appropriate.

▪ User autonomy requirements are specified with regard to the distribution of tasks between humans and AI applications (options for human intervention, involvement in decisions). The following points must be addressed and supported by application-specific specifications:
  – Type and extent of human control/supervision of the AI application in operation
  – Users' freedom to act
  – Integration into the work process, especially possible requirements regarding the checking of outputs by humans
  – Complaint options for users and affected persons

▪ It is vital to justify that the criteria conform to the objectives and do not contradict each other. It must also be demonstrated that, if the criteria are met, the risks identified in **[AC-R-TD-RI-01]** are reduced to an acceptable level.

### 5.1.3  Measures

The following measures are not specifically assigned to any of the categories of data, AI component, embedding or operation.

### 5.1.3.1  Data

### 5.1.3.2  AI component

### 5.1.3.3  Embedding

### 5.1.3.4  Measures for operation

**[AC-R-TD-ME-01] Involvement of relevant groups of people/organizations**

Requirement: Do

▪ Documentation should be available regarding which relevant groups of people and organizations were involved in developing the AI application, especially in terms of organizing the distribution of tasks between humans and the AI application. In particular, it should describe whether and to what extent
  – the groups of people/organizations identified as relevant in **[AC-R-TD-RI-01]** (or their representatives) have received a description of the AI application.

- if possible, several alternative design options for the AI application were developed and their advantages and disadvantages were compared.
- opinion statements were obtained from the groups of people or organizations involved and the arguments they contained were assessed.
- The extent to which the documented measures contribute to meeting the criteria in **[AC-R-TD-CR-01]** must be explained.

### [AC-R-TD-ME-02] Primacy of human action

Requirement: Do

- It is necessary to document in which cases a user can decide not to use the AI application or to subsequently correct a decision made by the AI application (directly or via designated contacts) and, if necessary, to compensate for it.
- Documentation should be available regarding which actions/decisions of the AI application users can intervene in during (normal) operation.
- In addition, the documentation should cover which intervention options are available to users in the event of deviations from normal operation, or which may be necessary in certain circumstances. In this case, reference can be made to the relevant sections of the **Risk area: functional safety (FS)** (see also **[S-R-FS-ME-12]**).
- If users must have specific qualifications/knowledge in order to intervene using the options described, this information should be provided.
- Lastly, in reference to the above points, an explanation should be provided regarding how they contribute to implementing the degree of autonomy of the AI application as planned in **[AC-R-TD-CR-01]**. In addition, with regard to the points above, the extent to which the requirements for user autonomy defined in **[AC-R-TD-CR-01]** are met should be explained.

### [AC-R-TD-ME-03] Establishing effective complaint channels

Requirements: Do | Pr

- Documentation should be available that covers how users and affected persons can make complaints about limitations in their ability to perceive a situation and take action as a result of the AI application. There is a record of the bodies/agencies responsible for evaluating the complaints. The process of how to evaluate complaints and, if necessary, derive consequences from them is explained. In particular, the documentation should describe the extent to which a shutdown or further development of the AI application is enforceable as part of this process.

### [AC-R-TD-ME-04] Rights-based and role-based approach for using the AI application

Requirement: Do

- There is a plan for the roles and responsibilities regarding the use of the AI application. This describes which activities within the scope of use, such as overwriting/correcting outputs of the AI application, require authorization. It should also document how to ensure that only authorized persons perform the corresponding activities.

### [AC-R-TD-ME-05] Human supervision of the AI application

Requirements: Do | Pr

- A process is established for humans to supervise and control the AI application. The degree of control and the procedures and activities within the process must be described in detail. In particular, reference must be made to the measures used to supervise the application and to monitor external factors described in the risk areas under "Control of dynamics" in the other dimensions. Where appropriate, these measures can be referred to directly.
- Documentation should be available that describes the methods and tools that a user can use to check the proper functioning of the AI application. Proper function involves meeting objectives and managing risks across all dimensions covered in this catalog. Specifically, the documentation should indicate how users can recognize when errors or deviations from normal operation occur and how they should behave if this happens. If this has already been covered elsewhere, an appropriate reference can be inserted instead.

- The documentation should also describe what skills or further knowledge are required to enable users to effectively supervise the AI application and recognize when intervention may be required.
- Lastly, in reference to the above points, an explanation must be provided regarding how they contribute to implementing the degree of autonomy of the AI application as planned in **[AC-R-TD-CR-01]**. In addition, with regard to the points above, the extent to which the requirements for user autonomy are met should be explained.

### [AC-R-TD-ME-06] Shutdown scenarios

Requirement: Do

- Scenarios should be identified, analyzed and evaluated in which the live AI application must be completely or partially shut down in order to maintain the ability of users and affected persons to perceive situations and take action. This includes shutdowns due to potential bodily injury or damage to property and also due to the violation of personal rights or the autonomy of users and affected persons. Thus, depending on the application context, this point involves analyzing scenarios that go beyond the accidents/safety incidents discussed in **Dimension: Safety and Security (S)**. For example, if it is possible that the AI application causes discrimination that cannot be resolved immediately, this scenario should be considered here. When evaluating the scenarios, the consequences of the shutdown for the humans involved, work processes, organization and company, as well as additional time and costs, should also be documented. This is compared with the potential damage that could arise if the AI application were not shut down.
- Documentation should be available on the AI application shutdown strategies that were developed based on the identified scenarios – both short-term, mid-term and permanent shutdown. Similarly, scenarios for shutting down subfunctions of the AI application should also be documented. Reference can be made to shutdown scenarios that may have already been covered in **Risk area: functional safety (FS)** (see **[S-R-FS-ME-10]**). A shutdown scenario documents
  – the setting and the resulting decision-making rationale for the shutdown,
  – the priority of the shutdown,
  – by which persons or roles the shutdown is implemented and how it is done,
  – how the resulting outage can be compensated,
  – the expected impact for individuals or for the affected organization.

### [AC-R-TD-ME-07] Technical provision of shutdown options

Requirement: Do

- Documentation should be available on the technical options for shutting down specific subfunctions of the AI application as well as the entire AI application. Here, reference can be made to **[S-R-FS-ME-10]** or **[S-R-FS-ME-12]** if necessary.
- It is outlined that other system components or business processes that use (sub)functionality that can be shut down have been checked and (technical) measures that compensate for negative effects of shutdowns are prepared. If already covered there, reference can be made to **[S-R-FS-ME-10]**.

### 5.1.4  Overall assessment

### [AC-R-TD-OA] Overall assessment

Requirement: Do

- With reference to the measures taken, a summary should be provided justifying that the requirements specified in **[AC-R-TD-CR-01]**, in particular the desired level of autonomy and requirements for user autonomy, have been achieved.
- If not all requirements specified in **[AC-R-TD-CR-01]** are met, the deviations must be documented. This also applies to requirements that have only been partially met, e.g., where the criteria have not or not always been met.

## 5.2 Risk area: information and empowerment of users and affected persons (IE)

This risk area is designed to ensure that users and subjects deal with the AI application in a self-determined and informed manner. The "informedness" in this risk area does not refer primarily to the technical functionality. Instead, it addresses the disclosure of information regarding the AI application, the education of users and affected persons about rights, risks and correct operation of the AI application. The technical dimension of self-determined use, e.g., that users understand how individual outputs are generated and can classify them in terms of content, is covered in the **Risk area: transparency in relation to users and affected persons (UA)** in the transparency dimension.

Achieving correct use of the AI application involves ensuring that potential users receive a sufficient and comprehensible explanation both of the scope and purpose of the application as well as its correct use. This also includes making users aware of both the level of autonomy and the requirements for human supervision/control of the AI application. For example, users of an AI-based application for processing customer inquiries should be informed before an automated response is sent if this application was developed purely for assistance and its outputs should be double checked by humans. Users should also be informed about the intervention options in the AI application and the required behavior in an emergency situation (see also **Risk area: functional safety (FS)** in the safety and security dimension). It is especially important for AI applications with a high level of autonomy, where there may be only short windows of time in which outputs can be overwritten, that the individuals responsible are proficient in the required actions. In line with this, it should also be communicated which qualifications are necessary to use the AI application correctly and how these qualifications can be acquired.

Even if the AI application is operated correctly, there may still be risks that affect users and other persons or, for example, the further processing of results from the AI application. AI applications that interact directly with humans, for example, carry the risk that users or affected persons form an emotional bond and thus become manipulable and (emotionally) dependent on the application. For example, through social engineering, chatbots can cause users to willingly disclose personal data or to sacrifice a large part of their free time for fictional interaction. In the case of AI applications that are not intended to manipulate or deceive, but rather to improve well-being, it is of course possible that these applications will intentionally target the feelings of users and/or affected persons. One example of this is the AI-based baby seal robot Paro[43], which is used in care settings to keep dementia patients company and reduce their stress levels. However, for AI applications that interact with humans and inevitably touch upon emotions (either intentionally or unintentionally), positive and negative impacts on users and affected persons should be assessed carefully.

In addition to this, it is important to check that the design of the user interface is consistent with the purpose and capabilities of the AI application. The appearance of the AI application – for example, if it is particularly human-like – as well as the way the results are presented could influence how users perceive the quality and reliability of the AI application and whether they attribute additional abilities, such as empathy, to it. This creates the risk of overconfidence in the AI application. The tendency of people to be rather uncritical of automated decisions and to not make the effort to check them through other sources is also referred to as automation bias. If users overestimate the capabilities of the AI application, this could result in it no longer being supervised sufficiently and thus operating with a higher level of autonomy than originally intended. This risk, alongside the risk of emotional attachment and manipulation, should be mitigated through informing and educating users and affected persons.

### 5.2.1 Risk analysis and objectives

**[AC-R-IE-RI-01] Risk assessment**

Requirement: Do

- **Risk analysis:** An analysis is performed to determine which risks exist for the AI application in question with regard to insufficient information or empowerment of users and affected persons and what potential damage can result from this. The analysis is structured as follows:
  - The risks arising from insufficient information regarding the correct use of the AI application are investigated. It should be considered to what extent or in what way the AI application could be operated incorrectly or supervised insufficiently by users. Both normal operation and exceptional situations must be considered, although in the case of the latter, reference can be made to the explanations in the **Risk area: functional safety (FS)**. An assessment is then carried out regarding potential damage that may occur if the AI application is not used/operated/supervised by humans as intended.
  - In addition, the analysis should describe how the AI application interacts with users and affected persons and what the user interface looks like. Based on this, an assessment is made regarding the extent to which involved individuals may develop an inappropriately high level of trust or emotional attachment to the AI application. It should also analyze which (material or non-material) damage scenarios may occur in the case of inappropriately high trust or emotional attachment by involved persons. As a minimum, it is necessary to investigate to what extent and on which topics users or affected persons can be manipulated by the AI application.
  - Lastly, it is necessary to examine the damage that may occur if users and affected persons are insufficiently informed about the business model behind the AI application or possible interests with regard to the AI application.
- **Objectives:** Based on the risk analysis, objectives are formulated regarding informing and empowering users and affected persons.
  The information that users and affected persons should receive about the AI application is documented. This should cover both information on proper use of the application and education on the associated risks. A specific list is provided in **[AC-R-IE-CR-02]**. Qualitative objectives are also set with regard to the visibility of information. It should be explained that information-related objectives have been set taking into account the interests of all humans involved in accordance with the proportionality principle (e.g., protection of trade secrets, public interest) and are appropriate to the context of the AI application.
  If necessary, requirements for the qualification of users resulting from the risk analysis are listed, providing a supplement to the requirements documented in **[AC-R-TD-ME-02]**. These can be supplemented by additional objectives regarding empowering users and affected persons. The choice of objectives must be justified.

### 5.2.2 Criteria for achieving objectives

**[AC-R-IE-CR-01] Qualification of users**

Requirement: Do

- The requirements for the qualification of users are set based on the objectives in **[AC-R-IE-RI-01]**, the documentation in **[AC-R-TD-ME-02]** and **[AC-R-TD-ME-05]** and also taking into account **[S-R-CD-CR-01]**.

**[AC-R-IE-CR-02] Comprehensiveness of information for users and affected persons**

Requirement: Do

- The risk analysis serves as a basis for listing content/information that must be communicated to users and affected persons in order to enable proper and self-determined use and awareness of all relevant risks. If certain groups of people within the users and affected persons are to be given different information, separate lists must be created for these groups of people. For example, bank employees who directly supervise an AI application for credit checks should be provided with information on the operation and functionality of the AI application that may not be relevant for customers. The list will be used as part of the overall assessment of whether the information is complete for users and affected persons. When compiling this list, at least the following points should be considered:
  – Users and affected persons are informed that the AI application is being used.
  – Insight into the purpose and functions of the AI application is provided.
    · Users and affected persons are able to understand the scope and purpose of the AI application.
    · A precise and complete description of the AI application is available. (The definition of "complete" must be specified depending on the application context. Reference can be made to **AI Profile (PF)**, for example.)
    · Reference is made to information from other dimensions (e.g., risks, relevant target values are communicated). In particular, users and affected persons are informed about how complete, reliable and fair the outputs derived by the AI application are. For example, an AI application may also provide a confidence statement for each output that is linked to understandable recommendations for action by the individual. Similarly, guidance is provided on how to verify the outputs of the AI application. For example, this can be done by consulting other independent sources of information.
  – Insight into the business models of the operators and the embedding of the AI application in processes is provided.
    · The underlying business model of the AI application and its purpose are explained.
    · The distribution of tasks between the AI application and its users is described.
    · The work processes associated with the AI application and the persons and roles involved (including contact persons for questions) are described (e.g., in **[AC-R-TD-RI-01]**).
    · The consequences of a partial or full shutdown of the AI application are described (as in **[AC-R-TD-ME-06]**).
    · Users and affected persons are aware of the ways in which a (partial) shutdown of the AI application can be effected. They are aware of the relevant decision-making committees and processes (see also **[AC-R-TD-ME-06]**, **[AC-R-TD-ME-07]**, **[S-R-CD-ME-01]** and **[S-R-CD-ME-03]**).
  – Information on correct use and supervision as well as intervention options is provided. It is necessary to ensure there is a way to implement the actions.
    · Users and affected persons can give their informed consent or reject use of the application.
    · Users and affected persons are informed about alternatives to using the AI application. For example, if it is possible to direct an issue to a designated responsible person instead of having it processed automatically by the AI application, the users or affected persons will be made aware of this option.
    · Users are informed about how to use the AI application correctly, e.g., through proper instruction.
    · A user manual is available that describes the AI application, how to use and supervise it correctly, and the intervention options.
      – Correctness of the description
      – Regular reviews (at least once a year) and updates to the manual
      – Versioning
    · Users are able, if required, to overwrite AI application outputs or even to intervene in actions while the AI application is operating. In particular, they should be able to use the intervention options described in **[AC-R-TD-ME-02]** and **[S-R-FS-ME-12]**.

- · Users are informed about how the AI application is effectively and appropriately supervised and controlled. At the very least, users should be informed about the processes described in **[AC-R-TD-ME-05]** for supervising and controlling the AI application.
    - · Users know how to behave in exceptional situations. Information about this is provided in **[S-R-FS-ME-12]**, **[S-R-FS-ME-01]**, **[S-R-IA-ME-01]** and **[S-R-CD-ME-03]**.
  - – Users and affected persons are informed about the risks.
    - · Users and affected persons are informed about application-specific risks in relation to user autonomy, particularly in cases where the AI application can influence the decisions of humans by making suggestions or even by manipulating them. If relevant for the application in question, information is provided about the risk of emotional attachment/dependence and about the risk of placing too much trust in the AI application (automation bias).
    - · Users and affected persons are informed about their rights and ways to complain. As a minimum, they should be informed about the ways to complain as described in **[AC-R-TD-ME-03]**.
- If any of the above aspects are not communicated to users or affected persons, justification of this must be provided. Furthermore, the list can be supplemented by additional (application-specific) content/information.

**[AC-R-IE-CR-03] Visibility and accessibility of information for users**

Requirement: Do

- Requirements regarding the visibility and accessibility of the information to be communicated according to **[AC-R-IE-CR-02]** are specified. The following points should be addressed as a minimum:
  - – The extent to which knowledge of the information is required: This ranges from retrievable information (e.g., instruction manual) and terms of use that must be read and confirmed before using the AI application to the explicit instruction and training of users, as well as the completion of knowledge tests. The requirements in this respect should be consistent with **[S-R-CD-ME-01]**.
  - – Preparing the information: This should be comprehensible and adapted to the qualification of the persons involved as required in **[AC-R-IE-CR-01]**.
- Justification must be provided that the criteria are consistent with the objectives defined in **[AC-R-IE-RI-01]**.

### 5.2.3 Measures

The following measures are not specifically assigned to any of the categories of data, AI component, embedding or operation.

### 5.2.3.1 Data

### 5.2.3.2 AI component

### 5.2.3.3 Embedding

### 5.2.3.4  Measures for operation

**[AC-R-IE-ME-01] Preparation of information for users and affected persons**

Requirements: Do | Pr

- There is a detailed description of how the information relevant to users and affected persons is prepared. Possible ways of preparing relevant topics include:
  – A communication process that informs users and affected persons that they are communicating with an AI application or that decisions are made on the basis of an AI application. This type of notice can be omitted if this is justified and proportionate in the context of the application.
  – Publication of ways to complain, possibly categorized according to the dimensions of trustworthiness, with designated human contacts.
  – A description of how to correctly use the AI application (user manual) that is comprehensible to users. This should also clearly indicate the AI application's scope and purpose, the intended target group, the distribution of tasks between the AI application and users and the human intervention options.
  – Documentation that is accessible to users and affected persons that identifies particular risks and relevant metrics from other dimensions, particularly security, data protection, reliability and fairness and contextualizes them in the application area in an understandable way.
  – Security guidelines (see **[S-R-FS-ME-01]**, **[S-R-IA-ME-01]**)
  – Emergency manual (see **[S-R-CD-ME-03]**)
- It should be explained that the described preparation measures cover all content/information listed in **[AC-R-IE-CR-02]**.
- A description is also required regarding how and through which channels the prepared information is made available to users and affected persons. If necessary, an explanation should also be provided concerning how it is ensured that the users and affected persons access the information relevant to them in accordance with **[AC-R-IE-CR-02]**.
- Justification should be provided as to how these measures contribute to compliance with **[AC-R-IE-CR-03]**.

**[AC-R-IE-ME-02] Empowering users**

Requirement: Do

- The process for ensuring that users are qualified in accordance with the requirements **[AC-R-IE-CR-01]** is documented. Reference can be made to sections such as **[S-R-CD-ME-01]**.

### 5.2.4  Overall assessment

**[AC-R-IE-OA] Overall assessment**

Requirement: Do

- Documentation should be available that provides a summary demonstrating that criteria **[AC-R-IE-CR-01]** to **[AC-R-IE-CR-03]** are met through the measures taken.
- If not all requirements specified in **[AC-R-IE-CR-01]** to **[AC-R-IE-CR-03]** are met, the deviations must be documented. This also applies to requirements that have only been partially met, e.g., where the criteria have not or not always been met.

## Summary

**[AC-S] Summary of the dimension**

Requirement: Do

- If there is a medium or high protection requirement for this dimension, documentation must be prepared for the remaining residual risks. First of all, the residual risks from the various risk areas in this dimension are summarized. Subsequently, and taking into account the protection requirement, the identified residual risks are collectively assessed as negligible, non-negligible (but acceptable) or unacceptable. The result of the analysis must be explained.
- If risks or measures under this dimension have been identified as having potentially negative effects on other dimensions, they must be documented.
- A conclusion must be made about the dimension that includes the assessment of residual risks.

# 6. Dimension: Transparency (TR)

## Description and objectives

Artificial Intelligence is capable of tackling problems that previously seemed unsolvable with traditional software development. Additionally, Machine Learning techniques deliver better and more precise results in numerous domains than conventional software based on hardcoded rules. However, alongside these advantages, most AI-based applications have a disadvantage compared to traditional software development because of their data-driven learning approach: The way in which AI-based applications generate their outputs is often difficult to understand, even for AI and domain experts. The corresponding characteristic of an AI application is referred to as transparency or traceability, i.e. its functionality and decisions can be fully or partially understood by a human. A distinction is drawn between whether an explanation is provided for how an individual prediction is generated or whether the Machine Learning process used is transparent as a whole. While the former concept is referred to as explainability below, the latter is subsequently referred to as interpretability of the model. For example, linear models are generally seen as interpretable because they simply weight the elements of the input by a factor to generate an output. Provided that normalization is performed correctly, it is easy to derive, for example, which parts of the input have the biggest influence on the output. Sufficient (semantic) interpretability of the linear model in terms of this dimension also requires the features to be interpretable.

The transparency dimension refers to the internal processes of the AI application and specifically to the ML model. In doing so, it examines explainability, interpretability, traceability and reproducibility on a technical level. Non-technical aspects of transparency, e.g., whether the AI application reveals itself as such, are addressed in the **Dimension: Autonomy and Control (AC)**. It is also worth highlighting that the transparency dimension may conflict with other dimensions, such as the **Dimension: Safety and Security (S)** or **Dimension: Reliability (RE)**. This issue will be addressed in more detail in the final cross-dimensional assessment of the AI application.

Transparency plays a marginal role in many application contexts. For example, it is not exactly important for a user of AI-based speech recognition to know why the AI application has or has not correctly recognized the spoken word in question. In these types of application contexts where the (technical) transparency of the AI application is not safety-critical, transparency is mainly a way of improving the trustworthiness of the AI application or ensures greater satisfaction among users. However, in other areas, explainability of outputs is essential for safe and responsible use of the AI application. For example, an AI-based image recognition application designed to assist doctors in making a diagnosis based on an MRI or X-ray image only provides genuine value to doctors if they are able to understand why the AI-based application reached that decision, for example, by marking tissue identified as abnormal in the image. In other application contexts, such as an AI-based credit scoring system in a bank, the interpretability of outputs could facilitate case-by-case decisions by the person responsible, for example, if potential customers oppose an unfavorable assessment of their creditworthiness.

One way to achieve transparency for users is to use an intrinsically interpretable model (as opposed to black box models) which fundamentally represents a simple or tangible relationship between input variables and outputs. However, as interpretable models are not an ideal solution for all problems, more complex black box models, such as deep neural networks, are often used. While the computations of black box models can also be understood in purely algorithmic terms, the meaning and transferability to human concepts and operations is unclear. However, for many black box models that are not interpretable, explainability can at least be established by explaining how the outputs are generated by elaborate downstream processes such as training surrogate models or LIME analysis (local interpretable model-agnostic explanations). At present, the explainability[44] of models is an active field of research and major efforts are being made to improve understanding of the learning processes of black box models and to visualize their internal processes and explain the resulting decisions. From a technical perspective, transparency is not a trivial issue and the conflict between reliability or robustness of the AI component and the traceability of how it functions (e.g., by choosing a less performant but more interpretable model) is a well-known dilemma.[45]

Transparency does not only make a difference to user satisfaction or with regard to using an AI application safely or responsibly: Aspects such as traceability or reproducibility of results are also relevant for legal issues, e.g., liability in the event of unexpected AI application behavior. While experts in safety-critical application contexts do not need to be able to predict every output of an AI application, the general behavior of the application must be explainable, traceable and documented during development and also when in productive operation later on. Log data, documentation or archived resources related to the design, data, training, testing and validation of the model, as well as to the embedding environment serve this purpose. This allows improvements to be made to the model in the event of (internal) review or following feedback or complaints, for example.

The following risk areas emerge from these initial considerations:

1. **Transparency in relation to users and affected persons:** This risk area addresses risks that arise from decisions and effects of the AI application not being sufficiently explainable to users and affected persons.

2. **Transparency for experts:** This risk area addresses risks that arise from the behavior of the AI application not being sufficiently transparent and comprehensible for experts.

3. **Auditability:** This risk area covers risks arising from the development as well as the individual procedures performed during the operation of the AI application not being sufficiently documented and verified.

4. **Control of dynamics:** This risk area addresses risks that arise because transparency requirements or the implemented transparency methods themselves change.

## Protection requirements analysis

Depending on the application context, enabling users or experts to gain deeper insight into the technical characteristics of the model and to understand or interpret how it generates outputs can be an essential requirement for the safe and responsible use of an AI application. The European Commission's draft regulation on Artificial Intelligence, for example, requires the decision-making process to be traceable to users in certain cases in order to enable appropriate use.

The potential damage scenario for this trustworthiness dimension is characterized by the fact that a lack of transparency prevents the AI application from being used in a safe and appropriate manner or results in the AI application breaching relevant guidelines. The assessment of the protection requirement for this dimension is also based on this. The damage categories are based on the extent to which the different aspects of transparency are relevant to the proper, i.e., safe and responsible, use and also to the usefulness of the AI application, particularly if it is not safety-critical. To this end, various points should be examined including whether a lack of transparency is costly or time-consuming (e.g., to find explanations for outputs in order to compensate for the

---

AI application's lack of transparency), or whether the AI application can be used without negative effects in the case of lack of transparency.

**Example**: An online fashion store uses an AI application for a recommender system, which creates a personal recommender profile for each customer. Information about how the AI application functions and the data it processes is transparent for customers, which allows them to check how the recommendations were generated themselves. The operator of the online store therefore incurs lower financial costs because customers make fewer returns. In addition, the traceability of the recommender system means the service provided to customers by the online store is particularly appealing and satisfying, giving the operator a competitive advantage and allowing them to acquire customers.

The potential trade-offs with other dimensions, such as **Dimension: Safety and Security (S)** or **Dimension: Reliability (RE)** are addressed in the cross-dimensional assessment of the AI application and should not form part of the process for determining the protection requirement.

The protection requirement is categorized as follows:

| High | Non-fulfillment of the transparency requirement (explainability, interpretability or traceability/ reproducibility) <br> ▪ would either render the AI application useless for its originally intended purpose, e.g., because safe or responsible use does not seem possible, <br> ▪ or would mean that the AI application could only be operated appropriately if additional expenditure is made (unjustifiable in terms of time or money). <br> In addition, a high damage potential already exists if the lack of transparency would lead to a breach of relevant (legal/normative) guidelines. <br> **Example:** An AI application that makes a medical diagnosis but the decision is not traceable. |
|---|---|
| Medium | Situations may arise where failure to meet a transparency requirement lowers the usefulness of the AI application and expenditure (time or financial) is required to ensure the AI application is useful/beneficial. <br> At the same time, a lack of transparency cannot violate relevant (legal/normative) requirements. <br> **Example:** An AI application used by a company for customer queries (e.g., credit facility queries). If there is an inquiry about the reasons for a specific output, a lack of transparency in the model makes answering it in a satisfactory manner more difficult from the customer's perspective. <br> **Example:** An AI application that automatically evaluates images on social networks for inappropriate content and blocks the post if necessary. If images are blocked without marking the areas in the image that are crucial to the output or categorizing the alleged violation, this increases the time required for human operators to identify and recheck false positives or false negatives. |

| Low | There is no transparency aspect that, if it was not fulfilled, could reduce the safety or usefulness of the AI application; or only minor effects on the usefulness of the AI application are possible, which can in turn be fixed with little effort. |
|---|---|
| | At the same time, a lack of transparency cannot violate relevant (legal/normative) requirements. |
| | **Example:** An AI-based access control system that uses camera-based facial recognition to determine people's access rights. If these rights are incorrectly classified, specifying the facial features that resulted in the wrong decision would not be useful, as it would not provide an expert with any added value in correcting the decision. In fact, a human supervisor will use their visual system and implicit or explicit professional expertise of personal identification to check the AI application's decision. |

**[TR-P] Protection requirements analysis documentation**

Requirement: Do

- The protection requirement of the AI application for the transparency dimension is defined as *low, medium* or *high*. The choice of the *low/medium/high* category is justified in detail with reference to the table above.

If the protection requirement for the transparency dimension is *low*, the individual risk areas do not need to be examined more closely. However, if a *medium* or *high* protection requirement has been identified, each risk area must be examined in more detail below.

## 6.1 Risk area: transparency in relation to users and affected persons (UA)

This risk area is designed to ensure that users have a clear and suitable explanation/interpretation of how the AI application works in order to operate or use it safely, properly and responsibly.

For example, choosing an interpretable model or visualizing explanations of the results in a clear way can help achieve transparency. To facilitate this, there are different types of transparency methods that help explain which of the input factors are particularly important for the respective output of the AI application. For example, doctors using an AI application to diagnose diseases can be shown the sensitivity of the various input parameters in the blood count using a heat map. While this would not make things any clearer for a medical layperson and especially to the patients as the persons affected, it can make a meaningful contribution to the diagnosis for the attending doctors. This is why transparency should be considered in the context of the target group being addressed or have customized approaches for different target groups. The technical level of detail of the explanation must be specifically adapted to the assumed level of qualification of the users or the affected persons.

The internal objectives of the operator may also create additional transparency requirements on top of those created by the specific AI application context. For example, companies/organizations can set their own guiding principles to provide customers with a high level of transparency regarding the processes they use. Enhanced transparency measures that go beyond what is necessary can allow companies/organizations to strengthen the trust that customers have in their AI applications ("trust through transparency").

### 6.1.1 Risk analysis and objectives

**[TR-R-UA-RI-01] Risk analysis and level of transparency of the AI application**

Requirement: Do

- **Risk analysis:** An analysis should be performed on the potential damage or hazards that may arise if the AI application is not transparent to users or affected persons. Different levels of transparency must be considered. In addition, the occurrence probability of the identified potential damage must be estimated.
- **Objectives:** Qualitative objectives are set for the transparency of the AI application based on the risk analysis, with reference to the principle of proportionality and considering the typical level of qualification of users and affected persons in this application area. Specifically, a description should be provided detailing what level of transparency – in terms of interpretability and explainability – is deemed appropriate and is being targeted for which part of the AI application, and the reasons behind this.

### 6.1.2 Criteria for achieving objectives

**[TR-R-UA-CR-01] Assessing explainability for users and affected persons**

Requirement: Do

- Qualitative or quantitative criteria that can be used to meaningfully assess the level of transparency of the AI application with respect to users and affected persons should be defined and explained. If different criteria is to be used for the circumstances or contexts to be explained to users and affected persons (e.g., the data, the model, how the results were established), the choice of criterion must be justified individually for each circumstance or context. The choice of criteria can be based on the following list of qualitative assessment standards. If other qualitative or quantitative criteria are established, these must be described and their selection justified.
  - The clarity and comprehensibility of the explanations given to users and/or affected persons regarding the circumstances to be explained (e.g., how the AI application works, data) was confirmed by a significant proportion of test affected persons consulted.

- The clarity and comprehensibility of the explanations given to users and/or affected persons regarding how individual outputs are generated or results of the AI application was confirmed by a significant proportion of test affected persons consulted.
- The level of qualification of the test persons consulted corresponds to the level of qualification of the anticipated target group or the anticipated users and/or affected persons.
- The following criteria can serve as a basis for evaluating the explanation of an output given to users and/or affected persons (e.g., generated by a transparency method or similar):
  - The explanation is valid, in the sense that the transparency method or explanation accurately reflects the AI application (e.g., errors in the model, if any, should be reflected as appropriate in the explanation)
  - The explanation makes a statement about which input factors are crucial/particularly important for the respective output
  - The explanation is robust/applicable/consistent in relation to a significant number of similar cases
  - The explanation makes a confidence output of the ML model comprehensible for users and/or affected persons (see the **Risk area: uncertainty estimation (UE)** in the **Dimension: Reliability (RE)**)
  - The explanation enables the user or affected persons to recognize "new" data points (*out-of-domain* data, in the sense of data points that are not distributed as intended)

  **Example:** An ML model was trained to detect diseases such as cataracts using close-up images of human eyes. A transparency process enables the image areas that are important for the prognosis to be highlighted. In this instance, "new" data points in the sense of out-of-domain data could be images of a different part of the human body, which are thus not in the intended distribution of input data. Conversely, eye images from patients that were not part of the training data are also "unknown" to the ML model, but are not out-of-domain data. There are numerous use cases, such as processing raw data from a sensor, in which making a distinction between "unknown" data can be difficult for human users.

- A detailed justification should be provided confirming that the criteria defined here represent the objectives set in **[TR-R-UA-RI-01]**.

### 6.1.3 Measures

### 6.1.3.1 Data

**[TR-R-UA-ME-01] Training and test data**

Requirement: Do

- Documentation should be available on the training and test data that indicates what kind of data is involved and whether it is understandable/interpretable for the intended users or affected persons in view of any prior knowledge (related to the intrinsic interpretability of the data). If the data is not intrinsically interpretable, then a reason must be provided as to why this particular type of data is being used. Furthermore, there must be a description of the measures taken to enable users and affected persons to have a basic understanding of the data, for example through documentation or guidance about the data.
  **Example**: Image data in which the phenomenon to be investigated by the AI application is clearly identifiable should be considered understandable, while raw data from a sensor, particularly if the data is high-dimensional, cannot be traced to specific states without substantial effort, so it is not considered understandable under this measure.
- Any data pre-processing, such as filtering, cleansing or transformation, that is performed before the data is used as training or test data must be documented. A check must also be carried out as to whether further explanations of these pre-processing steps are necessary for the comprehensibility of the AI application for users and/or affected persons. Any explanations provided or justification of why explanations are not provided must be documented.

### 6.1.3.2 AI component

**[TR-R-UA-ME-02] Interpretability of the ML model**

Requirement: Do

- Documentation regarding which models were considered to address the present problem must be available. In addition, it must be demonstrated that these models have been evaluated in terms of their interpretability, e.g., by in-depth analysis of the choice of architecture, thus examining the possible operations and their possible sequence.
- Documentation should be available describing why the model (or architecture) was chosen. If no interpretable model is used, it must be made clear why no such model is used. The list below provides a selection of models considered interpretable at the time when this catalog was published.

Interpretable models[46] include (as of July 2021):
- Linear and logistic regression (as well as their extensions including generalized linear models and generalized additive models)
- Linear and logistic classification
- Decision trees
- Rule learning
- Naive Bayesian classifier
- K-nearest neighbor

**[TR-R-UA-ME-03] Traceability of how the application works**

Requirement: Do

- Documentation and/or a visualization of the model (including a schematic diagram of the architecture, if applicable) must be available that contains sufficient explanations to make the way the application works comprehensible. There is an outline of how this visualization or explanation will be made available to users and affected persons of the AI application.

**[TR-R-UA-ME-04] How the results are generated**

Requirement: Do

- Documentation should be available describing the measures taken to explain how results are generated. The type and scope of the explanations must always be chosen based on proportionality, taking into account the level of qualification of users and affected persons as well as their use of the AI application. For example, explanations might be useful for users but might be excessive for other affected persons of the same AI application.
- In the case of models that are intrinsically interpretable:
  – For rule-based models, it is necessary to specify which rules apply to the current decision (e.g., in the case of decision trees, the branching selected could be displayed next to the result).
  – For linear regression and related models, it is necessary to specify the most relevant or crucial attributes for making a decision (the largest absolute value of the coefficients for normalized inputs).
- For black box models:
  Methods used to create explanations or support interpretability should be described. Methods should either be model-specific or model-agnostic, as in the sample methods listed below:
  – Partial dependence plots
  – Individual conditional expectation plots

---

**46** According to: Molnar, C. (June 2021). Chapter 4 Interpretable Models | Interpretable Machine Learning: A Guide for Making Black Box Models Explainable. https://christophm.github.io/interpretable-ml-book/simple.html (last accessed: 06/16/2021)

– Accumulated local effects (ALE) plot
– Feature interaction
– Feature importance
– Global surrogate model
– Student-teacher-model with interpretable student
– Local surrogate (LIME)
– Shapley values

The following example-based explanations are also permissible (mostly model-agnostic, but related to individual data points rather than to the complete input-output context):
– Counterfactual explanations
– Prototypes
– Influential instances

**[TR-R-UA-ME-05] Statistical evaluation of explanations**

Requirements: Do | Te

- Tests are performed and documented to illustrate that the explanations of the outputs of the AI application, as presented in **[TR-R-UA-ME-04]**, meet the required characteristics in **[TR-R-UA-CR-01]**. The fidelity of the explanations should be taken into account, especially in the case of error modes. The type of test used for this should have a strong focus on the chosen explanatory method.
  For example, a heat map method could be used to explain image classification. In this case, a check could be carried out regarding whether the highlighting on the heat map is on parts of the image that are relevant for the classification (based on human judgment) and to what extent the heat map is still localized to specific areas in the event of misclassification.
- Parts of this test may coincide with measures to review requirements from the **Risk area: transparency for experts (EX)**, see, for example, **[TR-R-EX-CR-02]**. If tests are identical or overlap, the corresponding measures can be referenced at this point.

### 6.1.3.3 Embedding

**[TR-R-UA-ME-06] Communicating the justifications for decisions**

Requirement: Do

- Documentation should be available that makes it clear in which cases or according to which criteria the AI application communicates the justifications for its decisions to the outside world. The form of communication (e.g., display or visualization of the explanations from **[TR-R-UA-ME-04]**) must also be described. If decisions are not communicated or there is no communication in any cases, justification must be provided as to why this does not happen or why this is not possible in relation to the principle of proportionality and the risk analysis carried out.

**[TR-R-UA-ME-07] Human evaluation of the explanations**

Requirements: Do | Te

- Tests must be carried out and documentation must be available to demonstrate that the requirements regarding the clarity and comprehensibility of explanations for users and affected persons, as defined in **[TR-R-UA-CR-01]**, have been met with regard to both presentation and content. A study involving human testers can be conducted and evaluated for this purpose. The nature and questions of the study must be described in detail and the reasons behind them justified. Documentation must be provided regarding how the study is conducted, including the selection of participants, how many participants there are, their qualifications and the results of the study. The qualification level of the study participants should roughly correspond to that of the expected and/or desired target group. This can be undertaken by referencing the appropriate subsections of the study.

#### 6.1.3.4 Measures for operation

**[TR-R-UA-ME-08] Process for responding to user queries**

Requirements: Pr | Te

- A process that has previously undergone empirical testing is established to ensure that explanations for the outputs of the AI application can be provided upon request to users and affected persons of the AI application. This process must be documented. Alternatively, justification must be provided as to why such a process is not needed to achieve the objective. To do this, reference can also be made to the information normally provided to users and affected persons in accordance with **[TR-R-UA-ME-03]** and **[TR-R-UA-ME-06]**.

### 6.1.4 Overall assessment

**[TR-R-UA-OA] Overall assessment**

Requirement: Do

- Documentation should be available explaining that the criteria defined in **[TR-R-UA-CR-01]** are met on the basis of the measures taken.
- If not all requirements specified in **[TR-R-UA-CR-01]** are met, the deviations must be documented. This also applies to requirements that have only been partially met, e.g., where the criteria have not or not always been met.

## 6.2 Risk area: transparency for experts (EX)

The transparency for experts risk area is closely related to the objective in the **Risk area: transparency in relation to users and affected persons (UA)**. However, this risk area focuses on validation, for example to detect model weaknesses and on the (technical) traceability and reproducibility of outputs of the AI application by experts. The technical level is correspondingly higher.

In many cases, such as the use of neural networks, it is difficult even for AI experts to understand how outputs are generated. In areas with high reliability and/or traceability requirements, this represents a potential risk. The purpose of this risk area is therefore to use introspective methods

- to make decisions transparent,
- to make outputs plausible,
- to test the inherent "logic" of the AI application,
- and to uncover potential causes of errors and, if necessary, systematic model weaknesses.

In doing so, meeting these goals can contribute not only to transparency, but also to the **Dimension: Reliability (RE)** of the AI application.

Transparency and interpretability are usually intrinsic qualities for AI methods with low levels of complexity, e.g., decision trees or clustering. However, in the case of a complex AI application, all four of the objectives specified involve substantial effort. In fact, it is possible that in these cases total transparency cannot be achieved due to trade-offs between effort and scope of introspection on the one hand and also necessity and practicality on the other. This is why the subsequent risk analysis must clarify what level of transparency for experts is targeted in relation to the AI application.

### 6.2.1 Risk analysis and objectives

**[TR-R-EX-RI-01] Risk analysis and objectives**

Requirement: Do

- **Risk analysis:** An analysis is performed to determine what possible damage and hazards can occur due to a lack of transparency of the AI application for experts and, in particular, due to the experts' inability to validate or check the plausibility of the outputs of the AI application. The probability of occurrence as well as the potential amount of damage are assessed.
- **Objectives:** Based on the risk analysis, qualitative targets are set regarding the various transparency aspects of an AI application (interpretability, explainability and traceability) for experts.
  **Example 1**: The pedestrian detection system in an autonomous vehicle should meet high reliability standards. In the event of damage, it may be necessary to closely trace the AI application's decisions.
  **Example 2**: An AI application for flood forecasting models physical processes. To check how reliable they are, it is useful to corroborate and validate the outputs in terms of simple (physical) relationships.

## 6.2.2  Criteria for achieving objectives

The complex set of requirements in the area of transparency poses a challenge when it comes to checking that targets have been achieved. Quantitative specifications should be favored over qualitative ones wherever possible to ensure accurate verifiability. If there is more than one requirement, quantitative and qualitative specifications can be combined. However, the determining factor is that the criteria for all requirements must be achievable separately and at the same time. Exceptions, such as criteria that replicate the same transparency requirement, must be justified separately.

### [TR-R-EX-CR-01] Requirements for the characteristics of transparency/introspection methods

Requirement: Do

- Documentation should be available describing the criteria to be used to evaluate transparency/introspection methods. Firstly, each criterion should specify the circumstance or context to which it refers, which should be explainable/interpretable/comprehensible to experts. Secondly, it should include requirements for the transparency method that is used to explain the corresponding circumstance or context to experts. In particular, target intervals to be achieved (quantitative) or target characteristics to be achieved (qualitative, structural) must be specified. The following points should be addressed and discussed as a minimum when selecting the criteria:
  - The scope, design and level of detail of transparency methods. In particular, it is necessary to investigate whether the level of transparency required in the application context is achieved by the method.
  - The depth and breadth of introspection in relation to model outputs. This involves discussing whether an introspection method has to be used for each model output (breadth) and for which model depth the introspection methods have to be used (e.g., transparency only for final outputs of the model or also for preliminary results within the model).
  - The time frame within which an explanation/introspection must be made available. For this, particular attention must be paid to whether there are real-time requirements for the introspection methods.
  - Complexity of the transparency method, i.e., how costly implementing the method may be.
- It should be justified in detail that the established criteria can be met separately and simultaneously, so there are no conflicting objectives. Requirements that are only collectively meaningful for assessing if an objective has been achieved should be combined into a common criterion.
- For each criterion and its associated target values or qualitative target characteristics, justification must be provided that they are suitable for the application context and conform to the objectives defined in **[TR-R-EX-RI-01]**.

### [TR-R-EX-CR-02] Requirements for the outputs or results of transparency/introspection methods

Requirement: Do

- The requirements specified in **[TR-R-EX-CR-01]** regarding the characteristics of transparency and/or introspection methods must be enhanced based on findings associated with these. The choice of criteria must be documented and justified. The following aspects should be taken into consideration:
  - The stability of the explanation in relation to similar cases
  - The correlation between the confidence of the model and the accuracy of the explanation
  - The comprehensibility of the results of the transparency method for experts (confirmed by human testers)
  - An intrinsic weighting of the statement, i.e., whether one or more statements have a weighting in terms of a confidence level
  - Fidelity, i.e., the smallest possible deviation of the prediction suggested by the explanation from the actual prediction

This list is not exhaustive. Requirements specific to the application context in question should also be included to define the criteria.

**Example**: Possible criteria for evaluating the outputs of a heat map method to be used by experts to corroborate the outputs of an image classification include:

– The heat map method should provide a similar explanation on similar input images (stability).
– The heat map represents the effect of certain image regions and can therefore highlight relevant image features. These features can be interpreted by humans in terms of classification.
– Both the confidence of the ML model and the strength of the heat map should decrease if the input image is distorted, see also **Risk area: uncertainty estimation (UE)** in the **Dimension: Reliability (RE)**.

▪ It should be justified in detail that the specified criteria can be met separately and simultaneously, so there are no conflicting objectives. Requirements that can only be used together to achieve an objective should be combined into a common criterion.

▪ It is demonstrated that sufficient domain knowledge was involved during the selection of transparency criteria. This can be ensured by involving domain experts in the target area of the AI application. If no such knowledge is required when selecting transparency criteria, justification must be provided.
**Example**: On the advice of medical professionals, the decision is made to perform a plausibility check on the outputs of an AI application used to analyze a blood count by determining the relevance of each input parameter.

▪ For each criterion and its associated target values or qualitative target characteristics, justification must be provided that they are suitable for the application context and conform to the objectives defined in **[TR-R-EX-RI-01]**. If qualitative criteria are used, an additional explanation must be provided as to why a quantitative criterion was not chosen.

### 6.2.3 Measures

The objective regarding introspective measures differs depending on the underlying AI application and its application context. This is why possible approaches should ideally be considered at an early stage during the development of the AI application as guiding elements, see **[TR-R-EX-ME-02]**. The transparency of Machine Learning processes is an active field of research. At the time of publication of the assessment catalog, the current state of this field can only be presented here as a snapshot. It is critical to undertake a more in-depth examination of the state of the art applicable at the time of the assessment, especially for applications that require a high level of transparency.

### 6.2.3.1 Data

**[TR-R-EX-ME-01] Suitability of training and test data**

Requirement: Do

▪ Documentation should be available on the training and test data of the ML model that indicates what kind of data is involved and whether it is understandable/interpretable for the intended experts (related to the intrinsic interpretability of the data). If the data is not intrinsically interpretable, then a reason must be provided as to why this particular type of data is being used. Furthermore, there must be a description of the measures taken to give experts the required understanding of the data, for example through documentation or guidance about the data. If necessary, reference can be made to **[TR-R-UA-ME-01]**.

▪ If applicable, the data used for training and for testing the introspective methods is documented and its choice or suitability is justified. Depending on the approach used, the requirements may be higher than would be necessary for purely training the AI component.
**For example,** out-of-domain data may be required to justify sufficient validity of the tests performed. In addition, there may be a requirement for more detailed labeling for correlation analyses regarding the application purpose, or for traceable data sources to better confine the underlying domain. In this instance, there may be overlaps with the requirements from the **Dimension: Reliability (RE)**, and if already covered there, reference can be made to the corresponding section in the reliability dimension for the choice of test data.
– Furthermore, it may be necessary to provide metadata for the test data that increases the interpretability of the test and training data used and/or also of the transparency method used. This is particularly the case for transparency methods in the form of a visual interactive interface (see **[TR-R-EX-ME-06]**).

### 6.2.3.2 AI component

Given that the criteria defined in **[TR-R-EX-CR-01]** and **[TR-R-EX-CR-02]** require transparency/introspection with regard to different circumstances or contexts, measures **[TR-R-EX-ME-03]** to **[TR-R-EX-ME-05]** must be applied separately for each of these circumstances or contexts.

**[TR-R-EX-ME-02] Justified choice of introspection/transparency methods**

Requirement: Do

- Documentation should be available that demonstrates engagement with the current state of the art for implementing traceable Machine Learning in relation to the AI application and the application context in question. In general, transparency can be achieved at different levels of an AI application, e.g., by analyzing outputs but also by defining internal states or auxiliary variables such as gradients. The technical options considered for achieving transparency of the AI application and plausibility of its outputs should be described. The focus of the description must align with the objectives defined in **[TR-R-EX-RI-01]**.
  The following are possible approaches to introspection that can be considered for implementing transparency requirements:
  – Analysis of boundaries, for example in classification
  – Heat maps on the output (e.g., DeconvNet)
  – Analysis of (latent) representations (e.g., TCAV)
  – Relationship between correlation or causality (see brittle features)
  – Use of interpretable local proxy models (e.g., LIME)
  – Sensitivity analyses concerning parameters, both of the model and the input
  – Consideration of the loss function used
  – Evaluation of attention
  Both the purpose and underlying ML models of these methods differ (e.g., random forests, SVMs or DNNs). This list does not claim to be exhaustive.
- Furthermore, the extent to which the transparency methods considered affect the reliability and performance of the AI application must be documented. The characteristics that are likely to be affected should be outlined for each approach/method for achieving transparency considered.
  **Example:** An interpretability requirement of the latent representation of an autoencoder that interferes with the loss (e.g., by a regularization term) usually leads to larger to medium reconstruction errors.
- The documentation covers which approaches or methods for implementing the transparency requirements formulated in **[TR-R-EX-RI-01]** are implemented in the AI application in addition to the explanation methods provided for users and affected persons (see measure **[TR-R-UA-ME-04]**).
  – Detailed justification must be provided regarding the choice of the implemented approaches or transparency methods based on the consideration of the current state of the art and the effect of transparency methods on reliability and performance.
  – If applicable, it must be explained how negative effects on other dimensions (in particular the **Dimension: Reliability (RE)**) were handled during development (by considering transparency requirements in the design, or even by implementing methods to achieve transparency).

- Furthermore, an explanation must be provided regarding the extent to which the selected approaches or methods ("by design" or as transparency methods implemented at a later stage) fulfill the criteria in **[TR-R-EX-CR-01]**.
- It should also be discussed whether an aggregate view of the chosen introspection methods, for example using a visual interactive interface (see **[TR-R-EX-ME-06]**), is meaningful and applicable. If the introspection methods are used individually, i.e., no aggregate view, this decision must be justified.

### [TR-R-EX-ME-03] Sanity check of the approach/transparency method implemented

Requirements: Do | Te

- Documentation should be available covering a test (referred to as a sanity check) that proves the effectiveness or plausibility of the implemented approach. More specifically, the test is designed to check whether the implemented approach actually provides insight into the specific workings of the ML model, rather than, for example, generating an explanation that is de facto separate from the model. The design of the test[47] and the choice of the test data set must be justified with reference to the AI application. How the test is performed and the test results must be documented.
  **Example:** A heat map method can be used to explain an AI application for image classification. In order to check the plausibility of the explanation for a specific classification, the model implemented in the AI application is retrained on a training data set with randomized labels and the heat map method is then applied to the classification of the retrained model. If it appears that the heat maps are very similar in terms of the original and the newly trained model, it is likely that the heat map method is more of an edge detector than something capable of explaining the intrinsic decision processes of the (very different) models. Furthermore, for input data far outside the application domain, such as an untrained class, the heat map should ideally not provide a meaningful explanation for any classification for the known classes.

### [TR-R-EX-ME-04] Quality assurance of the transparency method results

Requirements: Do | Te

- In addition to being suitable for the chosen purpose, the results of the introspective measures taken should meet the requirements of **[TR-R-EX-CR-02]**. This must be verified by suitable tests.
  - The quantitative criteria can be verified by statistical tests, as in **[TR-R-UA-ME-05]**. The test data sets used must be documented.
  - The qualitative criteria can be verified, for example, in tests with test affected persons, as in **[TR-R-UA-ME-07]**.
- How the tests are performed and the test results are documented and the extent to which the test results fulfill the criteria set in **[TR-R-EX-CR-02]** is presented.

### [TR-R-EX-ME-05] Complete fulfillment of a criterion

Requirements: Do | Te

- If several approaches/transparency methods are aligned to jointly fulfill a single criterion, i.e., are applied for a common purpose, the documentation must justify that the criterion is completely fulfilled by the combination of these methods. If this is technically possible, gaps are actively searched for and the outcome is documented. Otherwise, justification can be provided as to why such a gap cannot exist. This measure is only applied if a combination of approaches/transparency methods is used.

---

**47** The following article offers guidance: Adebayo, J. et al. (November 2020). Sanity Checks for Saliency Maps. GitHub. https://github.com/adebayoj/sanity_checks_saliency (last accessed: 06/16/2021)

### 6.2.3.3  Embedding

**[TR-R-EX-ME-06] Visual interactive interface**

Requirements: Do | Pr | Te

- Depending on the complexity of the underlying AI component and the data, it may be helpful and/or necessary to use a visual interactive interface to aggregate different introspection measures and to visually prepare data and metadata. In particular, this type of interface can visually break down the huge amount of data for experts and facilitate an improved understanding of data. Furthermore, experts can investigate semantic hypotheses, identify clusters and thus be supported when searching for weaknesses in the model (e.g., in the sense of closed-loop testing, see **[RE-R-SC-ME-04]**).
  **Example**: When examining an AI application for pedestrian detection, an image is found in which a person wearing a red sweater is not detected. The experts would like to investigate whether this error also occurs with similar input data and thus constitutes a systematic weakness. The corresponding image area with the person in the red sweater can be marked within the visual interface. Based on this example image, the data set is searched for similar images using a similarity metric. This produces a filtered data set that the experts can now use to analyze the performance of the AI application for this specific case.
- A detailed description of a visual interactive interface should be available if one has been created for the AI application. It is important to specify certain details including which interfaces the visual interactive interface has, which data formats are supported and how the embedding of the outputs of the AI application or the ML model itself is implemented. Furthermore, the implemented methods for visually preparing the data within the interface must be described, e.g., which graphs, tables, image data or similar are displayed, which areas are linked and how the data is processed in the background. The interactivity and the visual representation or aggregation of the data should be based on visual analytics methods. In addition, the process of operating the interface should be explained using examples.
- The documentation covers the extent to which the visual interactive interface was used to investigate model weaknesses (for example, in the sense of closed-loop testing, see **[RE-R-SC-ME-04]**) and which weaknesses were identified in the process. If this is related to measures in the **Dimension: Reliability (RE)**, reference can be made to them.
- If a visual interactive interface is available that is also to be used during operation to (systematically) search for weaknesses in the ML model, there must be an established process for this. In particular, it should be specified which aspects the analysis focuses on using the interface, which criteria are used to identify weaknesses and which measures are taken in light of the findings.

**[TR-R-EX-ME-07] Effect of integration**

Requirements: Do | (Te)

- Documentation should be available to show that the transparency properties of the AI application are not impacted by it being embedded in a larger surrounding system. If there is a justified suspicion that this might not be fulfilled, this is tested using the measures **[TR-R-EX-ME-04]** and if required **[TR-R-EX-ME-05]** for the AI component (already embedded and functioning as an AI application). The choice of test data sets, how tests are performed and the test results must be documented.
  **Example:** In an autonomous vehicle, an AI component is embedded in an online monitoring system that monitors all embedded modules based on complex rules. In certain situations, for example in the event of an external error, the embedding can overwrite the outputs of the modules or even disable them to enter a fail-safe mode. If this online monitoring system behaves erratically, there could be a negative impact on the traceability of the AI component's decisions, particularly if the complex rules make it difficult to trace the embedding activities.

- In some circumstances, this test may replace the previous tests (in **[TR-R-EX-ME-04]** and if applicable **[TR-R-EX-ME-05]**) at the AI component level, as long as sufficient justification is provided.
- If adjustments have been made to the embedding in advance due to a breach of transparency requirements, the change history, including tests of previous versions, should be provided if possible.

**[TR-R-EX-ME-08] Contribution of embedding**

Requirement: Do

- Depending on the nature of the AI application, it might be possible to some extent to achieve compliance with transparency requirements through characteristics of the embedding. The extent to which embedding contributes to fulfilling the transparency objectives must be documented.
  **Example:** If transparency is linked to the objective of additional validation and resulting improvement of the AI component, this could perhaps also be achieved by collecting additional training data. The input data that is treated with a high level of uncertainty or as faulty predictions can be examined separately by having the overall function mirror those input data back to a server. In this example, the procedure would also need to be considered in the **Dimension: Data Protection (DP)** and in the **Risk area: uncertainty estimation (UE)** reliability dimension.

### 6.2.3.4 Measures for operation

There are no planned measures for this category.

### 6.2.4 Overall assessment

**[TR-R-EX-OA] Overall assessment**

Requirement: Do

- Referencing the previously described measures and tests, it must be demonstrated that the criteria defined in **[TR-R-EX-CR-01]** and **[TR-R-EX-CR-02]** are met.
- If not all requirements specified in **[TR-R-EX-CR-01]** or **[TR-R-EX-CR-02]** are met, the deviations must be documented. This also applies to requirements that have only been partially met, e.g., where the criteria have not or not always been met.
- There must be a record of the extent to which relevant negative impacts on other dimensions (in particular the **Dimension: Reliability (RE)**) were identified in **[TR-R-EX-ME-02]** that need to be addressed and evaluated in the overall cross-dimensional assessment.

## 6.3  Risk area: auditability (AU)

The auditability of an AI application is an umbrella term for the detailed technical documentation on the structure, development and workings of the AI application and, in particular, the data used for this purpose.

The auditability of an AI application can serve different purposes. Firstly, subsequent audits (internal or external) with potentially different issues can be made easier or actually made possible in the first place. Documentation and logged data can be essential requirements, for example, to trace specific outputs of the AI application or the cause of errors in liability matters. The reproducibility of outputs and of the ML model itself also contributes to this. Secondly, detailed documentation facilitates changes or improvements to the AI application.

Potential auditability requirements may result from the **Risk area: transparency for experts (EX)**, for example. However, conflicts may arise with the **Dimension: Data Protection (DP)**, for example, if the data that to be documented is of a personal nature, and the **Dimension: Safety and Security (S)**, as auditability measures may increase the vulnerability of the AI application. Any such conflicts must be discussed in the cross-dimensional assessment.

### 6.3.1  Risk analysis and objectives

**[TR-R-AU-RI-01] Risk analysis and objectives**

Requirement: Do

- **Risk analysis:** An analysis is performed to determine which hazards and potential damage may arise due to limited auditability, giving consideration to the specific application context of the AI application as well as the legal framework. The significance of each auditability aspect (including documentation as a basis for external audits, traceability and reproducibility) for the application context in question and potential consequences for various degrees of non-compliance should also be examined. In addition, the occurrence probability of the identified potential damage must be estimated.
- **Objectives:** Qualitative objectives are defined regarding the different auditability aspects of the AI application based on the findings of the risk analysis. This involves describing in detail what auditability specifically means in the context of the AI application, e.g., for which parts of the AI application and to what extent documentation is required (training data, model characteristics and hyperparameters, logging of input data, outputs, system architecture), and to what extent the outputs of the AI application should be traceable or even reproducible.

### 6.3.2  Criteria for achieving objectives

**[TR-R-AU-CR-01] Auditability level of the AI application**

Requirement: Do

- Criteria are established and documented to help meaningfully assess the auditability of the AI application. The following aspects should be considered as a minimum when selecting the criteria:
  - Accessibility, source, content, date, revision number and possibility, necessity and duration of storing training and test data
  - Existence of documentation explaining the system architecture
  - Possibility and necessity of reproducing outputs, if necessary within a specified period of time, for example by accessing saved model versions

    − Possibility and necessity of logging decisions and the data required for this (such as input data or additional information, e.g., about the operating status, the operational environment or random seeds, if results are obtained using non-deterministic methods)

The list serves as a starting point for establishing suitable criteria for the application context. Other aspects not listed here can also be used to assess auditability provided they are described and the reasons for choosing them are justified.

- Qualitative target characteristics or, if applicable, quantitative target values are also documented for each specified criterion. These criteria must be met to achieve the required auditability of the AI application in accordance with the target objectives defined in **[TR-R-AU-RI-01]**.
- Justification must be provided that the selected criteria and associated target values conform to the objectives defined in **[TR-R-AU-RI-01]**.

### 6.3.3  Measures

### 6.3.3.1  Data

**[TR-R-AU-ME-01] Availability of training and test data**

Requirements: Do | Pr

- A storage system for archiving training and test data collected before the AI application is put into operation must be developed and documented. Under this system, the training and test data must be available to experts and developers in a form that can be used to perform necessary introspections (see **Risk area: transparency for experts (EX)**). The access rights as well as the storage must be aligned with the **Dimension: Data Protection (DP)** if the data in question also includes personal or other data that needs to be protected.
- If the AI application is capable of relearning, then there must be a process for logging and storing training and test data each time the AI application relearns. In particular, the relevant training and test data should be versioned and it must be possible to trace which data was used to train the version currently in operation. If the AI application learns continually during operation, **[TR-R-AU-ME-03]** must also be considered.

### 6.3.3.2  AI component

There are no planned measures for this category.

### 6.3.3.3  Embedding

**[TR-R-AU-ME-02] Software environment and interfaces**

Requirement: Do

- Documentation of the structure of the AI application must be available which describes the various software components and other system components (e.g., cloud storage) and explains how they interact. If this has already been described in the AI profile or elsewhere, reference can be made to the relevant documentation.
- Documentation should be available on the software libraries used and the exact versions of the packages used.
- Documentation should be available on the embedding of the AI component with descriptions of all interfaces and the respective input and output formats at the interfaces.

### 6.3.3.4 Measures for operation

In the case of transparency requirements at runtime, it may be necessary to log input data, predictions or internal states of the AI application. If such a requirement exists, a process for logging and storing the necessary data must be developed. In some cases, this may also involve the **Dimension: Data Protection (DP)**. The following measures cover the storage of various data during operation.

**[TR-R-AU-ME-03] Availability of data from operation (for training and/or validating the model)**

Requirement: Do

- If the AI application continues to learn during operation, or data on validating or upgrading the model is collected during operation, there is a logging and storage system for the data collected for this purpose in addition to **[TR-R-AU-ME-01]**. In particular, if the model is trained further during operation without revision by means of the collected data, this data should also be made available to the experts and developers afterwards. The system for and scope of logging is documented.
- If the data to be stored is personal information or business data that needs to be protected, then the **Dimension: Data Protection (DP)** must be involved.

**[TR-R-AU-ME-04] Storage of model and training parameters**

Requirement: Do

- The model parameters, e.g., the weights of a neural network, are stored and versioned.
- The hyperparameters that characterize the training procedure when the model relearns or continuously learns during operation are stored and versioned.

**[TR-R-AU-ME-05] Reproducibility and traceability**

Requirement: Do

- A logging and storage system for the outputs of the AI application must be available. The system for and scope of logging is documented. Access to logged outputs (e.g., on the basis of **[RE-R-RO-ME-08]** or to correct outputs as in **[AC-R-TD-ME-02]**) must be aligned with the appropriate measures from the other dimensions.
- A logging and storage system for the inputs of the AI application must be available. The system for and scope of logging is documented. If necessary, reference can be made to **[TR-R-AU-ME-03]**. Access to logged inputs (e.g., on the basis of **[RE-R-RO-ME-07]**) must be aligned with the appropriate measures from the other dimensions.
- If model results are not produced using a deterministic method, and if required for reproducibility or traceability purposes, the intermediate representations computed during ML model inference (e.g., feature maps of a neural network) are stored. In doing so, it is possible to evaluate to what extent logging system states such as random seeds is sufficient to reliably reproduce these intermediate representations. Otherwise, justification must be provided as to why this data is not stored.
- If the data to be stored is personal information or business data that needs to be protected, for example, then the **Dimension: Data Protection (DP)** must be involved.
- It must be demonstrated how the measures taken contribute to the reproducibility and traceability required in **[TR-R-AU-CR-01]**. Justification must be provided if the scope of logging and storage is not sufficient to meet the requirements.

**[TR-R-AU-ME-06] Logging of user queries**

Requirements: Do | Pr

- If a process for responding to user queries has been established under **[TR-R-UA-ME-08]**, it must be documented
  - how user queries and the explanations given in response are logged.
  - for how long this information is stored. It must be ensured that the data of users and affected persons is protected in this process.
  - that information is provided about the storing of the queries.
  - whether a process is established to delete these queries in certain cases. If applicable, the procedure should be aligned with the **Dimension: Data Protection (DP)**.

### 6.3.4  Overall assessment

**[TR-R-AU-OA] Overall assessment**

Requirement: Do

- Documentation should be available explaining that the criteria regarding the auditability of the AI application defined in **[TR-R-AU-CR-01]** are met on the basis of the measures taken.
- There must be a record of the extent to which relevant negative impacts on other dimensions (in particular the **Dimension: Data Protection (DP)**) were identified in this risk area that need to be addressed and evaluated in the cross-dimensional assessment.
- If not all requirements specified in **[TR-R-AU-CR-01]** are met, the deviations must be documented.
  This also applies to requirements that have only been partially met, e.g., where the criteria have not or not always been met.

## 6.4  Risk area: control of dynamics (CD)

The control of dynamics risk area should ensure that the transparency of the AI application is maintained.

First, it deals with the risk that established transparency properties of the AI application may be lost during operation if, for example, the ML model in question changes due to model drift. For example, a change in the ML model may result in a decrease in the robustness of explanations. Secondly, there is the risk that the transparency requirements of the AI application change, for example due to external factors such as user needs or new laws. In addition, advancement in the state of the art may mean that previously unattainable requirements become feasible or others become obsolete. An abstract transparency requirement could involve monitoring and addressing these advancements.

### 6.4.1  Risk analysis and objectives

**[TR-R-CD-RI-01] Risk analysis and objectives**

Requirement: Do

- **Risk analysis:** An analysis is performed to determine which (external) factors or circumstances affect the transparency requirements established in the previous risk areas. This analysis is then used to help estimate the likelihood of foreseeable changes to the transparency requirements of the AI application, and what impact or potential damage may result. In addition, it is necessary to identify possible causes (external or inherent in the AI application) for the loss of existing transparency properties of the AI application during operation and estimate their probability of occurrence. Furthermore, it is necessary to investigate what damage may occur if such a loss takes place.
- **Objectives:** Qualitative objectives for operation are set based on the risk analysis. They outline the extent to which and the approach to be taken for ensuring that the established transparency properties are maintained or, if necessary, adapted during operation so that risks are reduced to an acceptable level.

### 6.4.2  Criteria for achieving objectives

**[TR-R-CD-CR-01] Reviewing and adapting transparency requirements**

Requirement: Do

- Based on the **[TR-R-CD-RI-01]** risk analysis, external factors are defined that are to be observed due to their potential impact on the transparency requirements established in the previous risk areas. The assessment or monitoring interval planned must be defined and documented according to the estimated rate of change of these factors.
- In relation to the change in relevant external factors, the threshold value or the qualitative extent of the deviation should be defined and documented. Anything above this threshold should initiate an adjustment of the previously established transparency requirements (which are regularly reviewed in accordance with **[TR-R-CD-CR-02]**).
- It must be demonstrated that the criteria set conform to the objectives defined in **[TR-R-CD-RI-01]**.

**[TR-R-CD-CR-02] Maintaining transparency properties**

Requirement: Do

- Criteria are defined and documented for a process to regularly review transparency properties against the existing requirements. When selecting criteria for assessing the process, the following points (quantitative or qualitative) should be considered as a minimum:
  - Assessment interval. This shows the extent to which the reviews of the transparency properties are also initiated by other, regular reviews. This approach is useful if a re-evaluation of transparency properties results from direct changes to transparency requirements as well as from other initiated changes to the AI application, such as retraining.
  - Scope and nature of the review or methods used.
  - Threshold value or qualitative extent of deviation from the requirements, above which corrective measures are taken.

  Quantitative target values or qualitative target characteristics must be specified for each criterion.
- It must be demonstrated that the criteria set adequately reflect the objectives defined in **[TR-R-CD-RI-01]**.

### 6.4.3  Measures

#### 6.4.3.1  Data

There are no planned measures for this category.

#### 6.4.3.2  AI component

There are no planned measures for this category.

#### 6.4.3.3  Embedding

There are no planned measures for this category.

#### 6.4.3.4  Measures for operation

**[TR-R-CD-ME-01] Monitoring external factors**

Requirements: Do | Pr

- A monitoring process is established relating to the external factors that may influence the transparency requirements identified in the previous risk areas. The factors specified in **[TR-R-CD-CR-01]** must be considered as a minimum. Depending on the nature of the transparency requirements, the process may involve methods such as monitoring and analysis, but it may also include, for example, gathering user feedback through questionnaires. The nature and scope of the process must be documented and detailed justification must be provided as to how the process contributes to meeting the objectives.
- In addition, it must be documented how, as part of the process, the previously established transparency requirements are adjusted according to **[TR-R-CD-CR-01]**.
- Furthermore, the steps that come after adjustment of the transparency requirements must be described. In particular, it must be outlined how adjustments are initiated or made to the AI application (see **[TR-R-CD-ME-02]**).

**[TR-R-CD-ME-02] Reviewing and maintaining transparency properties**

Requirements: Do | Pr | Te

- A process is established that checks if the existing transparency properties of the AI application are maintained in accordance with the specifications in **[TR-R-CD-CR-02]** and, if necessary, initiates or makes corrections.
  – The nature and scope of the tests performed per assessment interval must be documented. If applicable, the methodology in **[TR-R-UA-ME-05]**, **[TR-R-UA-ME-07]** and **[TR-R-EX-ME-04]** can be used. In particular, it is important to specify which test data sets are used and the reasons for using them. If test data sets are gathered during operation, these data sets should be favored in the assessment process over the test data used before the application was put into operation (appropriate handling of concept drift). Otherwise, the reason why these test data sets were not used must be provided.
  – How to proceed with the documentation from the assessment intervals and in particular the test results should be described. In doing so, a description must be provided regarding which steps are taken to restore the transparency properties if the checks yield insufficient test results.

### 6.4.4 Overall assessment

**[TR-R-CD-OA] Overall assessment**

Requirement: Do

- It must be demonstrated that processes have been established to monitor external factors and regularly review the AI application. These processes must meet the requirements in **[TR-R-CD-CR-01]** and **[TR-R-CD-CR-02]**.
- If not all requirements specified in **[TR-R-CD-CR-01]** and **[TR-R-CD-CR-02]** are met, the deviations must be documented. This also applies to requirements that have only been partially met, e.g., where the criteria have not or not always been met.

## Summary

**[TR-S] Summary of the dimension**

Requirement: Do

- If there is a medium or high protection requirement for this dimension, documentation must be prepared for the remaining residual risks. First of all, the residual risks from the various risk areas in this dimension are summarized. Subsequently, and taking into account the protection requirement, the identified residual risks are collectively assessed as negligible, non-negligible (but acceptable) or unacceptable. The result of the analysis must be explained.
- If risks or measures under this dimension have been identified as having potentially negative effects on other dimensions, such as reliability or security, they must be documented.
- A conclusion must be made about the dimension that includes the assessment of residual risks.

# 7. Dimension: Reliability (RE)

## Description and objectives

From a technical perspective, the reliability of an AI application is an umbrella term that involves different aspects of the quality of its AI component: the correctness of the outputs, the assessment of the ML model uncertainty, robustness to corrupted or manipulated inputs, and unexpected situations, and of course, intercepting errors.[48, 49]

Profound application knowledge is necessary to evaluate these different reliability aspects and to determine under which conditions the AI application should be classified as reliable. The description of reliability requirements with quantitative measures and target values needs mathematical expertise as well as domain knowledge and is naturally never complete. The same applies to the description of the application domain which is defined as the input data to be expected during operation that the AI application is meant to process correctly.

In the **Risk area: reliability in standard cases (SC)**, this application domain is specified and formalized as precisely as possible to ensure that the training and test data used sufficiently covers the application domain. In most cases, corruption in the input data that may occur during regular operation and are part of the application domain are also intrinsically represented in the database. At the same time, for some application contexts, it may make sense to single out such corruptions and target them to strengthen the coverage of the data as well as the performance of the AI application. This is done in the **Risk area: robustness (RO)**, which is designed to ensure that the AI component performs as consistently as possible, even at the boundary of the application domain. To this end, it addresses corrupted or manipulated input data such as sensor noise or adversarial examples.

The reliability of an AI application also includes appropriate handling of deviations from standard cases. Especially in direct human-machine interaction, new error modes that are unexpected or unknown to humans can lead to potentially critical situations that have not been dealt with previously. This is why input data that is well outside the defined application domain and therefore cannot be expected to be processed correctly by the AI component should be intercepted. The **Risk area: intercepting errors at model level (IM)** examines the reliability of the AI application by determining to which extent potential errors are already prevented by detection strategies at the AI component level. This risk area supplements the traditional methods for fault tolerance and fail-safe operation at the embedding level as addressed in the **Risk area: functional safety (FS)** in the safety and security dimension.

Various aspects of AI application reliability can also be improved by performing a realistic uncertainty estimation. On the one hand, an uncertainty estimation provides an opportunity to identify and respond to weaknesses in the ML model. On the other hand, an uncertainty estimation can provide an indication of whether inputs are on the boundary or outside the application domain. In doing so, the assessment can intercept potential errors in the AI application. Performing a proper uncertainty estimation is a necessary prerequisite for making the most of these benefits. The **Risk area: uncertainty estimation (UE)** examines the requirements associated with this.

---

The risk areas under the reliability dimension are:

1. **Reliability in standard cases:** This risk area addresses the risk of incorrect predictions by the AI component on regular input data.[50]

2. **Robustness:** This risk area addresses risks that arise when input data is corrupted or manipulated, but for which accurate processing by the AI component is intended. Both qualitative and quantitative input data pertubations are considered, such as noise or adversarial examples.

3. **Intercepting errors at model level:** This risk area addresses risks from input data that is not in the application domain and that the AI component is not expected to process correctly. This data should be intercepted by a detection strategy.

4. **Uncertainty estimation:** This risk area examines risks that arise due to an unrealistic, unusable or absent uncertainty estimation.

5. **Control of dynamics:** This risk area addresses the risk that the ML model implemented in the AI component will experience a decline in performance or losses in relation to other requirements due to unintended model drifts or changes in application context (concept drift).

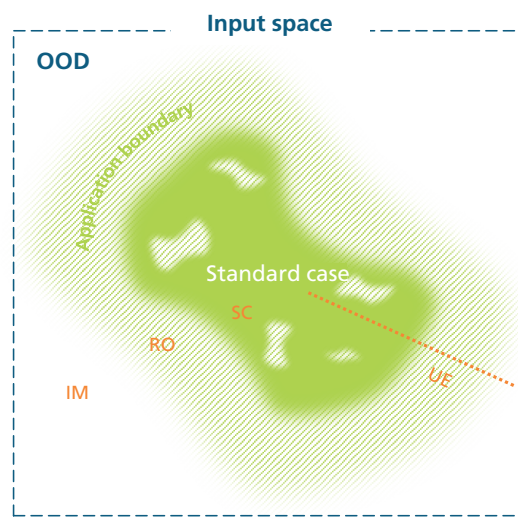

Figure 7: Representation of the different areas of the input space and classification into the risk areas of the reliability dimension.

## Protection requirements analysis

The protection requirement of the reliability dimension is determined by the maximum amount of damage that (incorrect) outputs of the AI application can cause without deliberate subsequent human revision. For example, there is generally a high protection requirement for AI applications embedded in autonomous robotics systems or in decision-making systems that determine aspects such as the distribution of goods or the medical treatment of individuals.

---

**50** The following defines regular input data as input data that originates from the application domain, i.e., data that the AI application is intended to process correctly.

In contrast to the other dimensions, the reliability dimension cannot have a *low* protection requirement. This is due to the fact that a low protection requirement in terms of reliability would mean that the outputs of the AI application could produce little or no damage in the event of a malfunction. In particular, it would mean that no bodily injury would occur in the event of incorrect outputs, and potential financial damage, such as loss of earnings or reputational damage, would either not be a threat or would only be a threat to a minor extent. In other words, this would mean that the AI application is completely non-critical and could equally be replaced by a non-AI based system that generates random outputs. This is why assessing an AI application with a *low* protection requirement in terms of reliability does not seem necessary, therefore this case is excluded.

The protection requirement is categorized as follows:

| | |
|---|---|
| High | An incorrect prediction of the AI application can lead to bodily injury or major financial damage. **Examples:** Pedestrian detection in autonomous vehicles, automated credit lending decisions, medical treatment recommendations |
| Medium | An incorrect prediction of the AI application can at worst lead to medium financial damage. **Examples:** A poor-quality route planner causes increased energy consumption and a longer time of travel, a faulty forecast of machine utilization in a manufacturing facility causes delays, a defective obstacle detection of a robot vacuum cleaner causes property damage |

**Note:** The amount of damage in all of the stated examples depends on the particular circumstances and does not necessarily come under the category chosen here.

**[RE-P] Protection requirements analysis documentation**

Requirement: Do

- The protection requirement of the AI application for the reliability dimension is defined as *medium* or *high*. The choice of category (*medium or high*) is justified in detail with reference to the table above.

In both cases (*medium* or *high* protection requirement), each risk area must be addressed in detail below.

## 7.1 Risk area: reliability in standard cases (SC)

The risk area reliability in standard cases considers all incorrect predictions of an AI application that can potentially occur on input data from the application domain and lead to damage. The criteria and measures in this risk area address the overall risk of incorrect predictions of the AI component, with the objective of limiting it to an acceptable level in the context of the application. In addition to the universal analysis in this risk area, specific aspects for avoiding incorrect predictions or errors in the application domain are singled out in the **Risk area: robustness (RO)** and in the **Risk area: uncertainty estimation (UE)** and covered in more depth.

An important first step in addressing the standard cases is to specify the application domain. To ensure adequate quality of the AI application on input data to be used in operation, it must be ensured that the training data sufficiently covers the application domain. The correct implementation of training routines and the final trained model are another prerequisite for the quality of the AI application.[51]

The requirements for reliability in standard cases are largely determined by the context in which the AI application is applied. Existing requirements for the AI application must first be quantified so they can serve as an objectively verifiable criterion. Particularly in the case of qualitative requirements such as "customer satisfaction" or "simplifying the work of employees", the creation of quantitative metrics and target intervals to be used to perform tests is always affected by losses. To avoid overfitting to a specific metric, i.e., one-sided optimization of the AI component with respect to a single target aspect, as many different performance metrics as possible should be considered to assess reliability. To assess whether the target intervals have been met, tests should be established that are tailored to the AI application in question and, in particular, to its application context.

### 7.1.1 Risk analysis and objectives

To perform the risk analysis for the reliability in standard cases risk area, the (permissible) application domain must first be specified. Based on this defined domain, a comprehensive assessment is made of the type and amount of possible damage that can be caused directly or indirectly by using the AI application for its intended purpose.

**[RE-R-SC-RI-01] Determining the application area and risk assessment**

Requirement: Do

Documentation should be available that addresses the following points in a qualitative and conceptual manner:

- **Application domain:** The application domain is defined as the input data to be expected during operation that the AI application is meant to process correctly. This must be described in detail. In addition, the application domain should be roughly defined using examples (faulty input data or limiting cases of the application domain, including non-semantic corruptions such as noise, are specified in more detail in the **Risk area: robustness (RO)**).
  If possible, the application domain defined here should be distinctly separate from related or supposedly similar application domains for which the AI application was not developed and the reliability for which is thus not assessed.
- **Risk analysis**: An assessment is made of the risks created by insufficient reliability of the AI application when using regular input data. This also describes the potential damage resulting from this insufficiency and, if possible, how frequently they can occur. For each identified risk, an assessment must be made as to whether the level of associated potential harm is acceptable or not.

---

**51** The description in this section draws heavily on section "3.4 Reliability" in the white paper: Poretschkin, M.; Rostalski, F.; Voosholz, J. et al. (2019). Trustworthy Use of Artificial Intelligence. Sankt Augustin: Fraunhofer Institute for Intelligent Analysis and Information Systems IAIS. https://www.ki.nrw/wp-content/uploads/2020/03/Whitepaper_Thrustworthy_AI.pdf (last accessed: 06/18/2021)

- **Objectives**: It is documented which residual risks are acceptable for the application context in question. Furthermore, objectives are set for developing and operating the application so that the tolerable residual risk is not exceeded when these objectives are met.

### 7.1.2 Criteria for achieving objectives

The following covers methods and metrics for risk assessment and for defining the application area. This involves determining value ranges (target intervals) for all metrics used, which quantify the desired level of acceptable risk in the objectives. The reliability of the AI application can be assessed in the overall assessment based on these target intervals.

**[RE-R-SC-CR-01] Quantification of reliability**

The reliability of an AI application is illustrated by mathematical-statistical measures or metrics. It is important to draw a distinction between general performance metrics and the loss function.

The latter is a crucial part of the training of many ML models. In the case of supervised Machine Learning in particular, learning algorithms such as stochastic gradient descent are generally used to adjust the model at each training step in a way that the loss function of the training data set is gradually optimized. The value of the loss function (also referred to as loss) with respect to the training data, compared to the loss with respect to the test data, provides information about the quality of the training and can indicate aspects such as possible overfitting.

However, a performance metric is the general term for a measure or metric that assesses how good/bad a model or the AI application based on it is at solving a given task. The performance metric used to assess the quality of a model can differ from its loss function (if there is one). In particular, it is not part of the training process, but is actually determined on the basis of validation or test data to measure the performance of the AI component in relation to a given task.

Common loss functions are listed below for regression and classification, which represent two major problems of supervised Machine Learning. In practice, a regularization term for the model parameters (e.g., the weights of a neural network) is also often used in addition to these loss functions to counteract overfitting.

Regression:[52]
- Squared error loss
- Absolute error loss

Classification:[53]
- Cross-entropy loss
- Hinge loss
- Kullback-Leibler divergence
- Brier score[54]

---

The following list provides an overview of key performance metrics for different (also unsupervised) Machine Learning tasks. Various other metrics also exist.

Regression:[55]
- (Mean) squared error
- (Mean) absolute error

Classification:[56]
- (Mean) accuracy
- F1 score
- Precision and recall
- Sensitivity and specificity
- AUC value[57]

Ranking:
- Mean reciprocal rank[58]
- Discounted cumulative gain[59]

Clustering:
- Silhouette value[60]
- Adjusted mutual information score[61]
- Completeness score[62]

Computer vision:[63]
- Peak signal-to-noise ratio (SNR)
- Structural similarity index[64]
- (Mean) intersection over union (mIoU)

Natural language processing:
- Perplexity score[65]
- BLEU score[66]

---

Requirement: Do

- The documentation covers which performance metric(s) should be used to assess the reliability of the AI application in the application domain. Justification must be provided as to why the chosen metrics are appropriate for evaluating the fulfillment of a task by the AI application and adequately reflect the specific quality requirements of the model that for example stem from the business context of the AI application. If the model is trained with a loss function, it is also necessary to specify and justify the choice of the loss function. If none of the evaluation metrics specified in the list above are used, the reason for this must be provided. If this is the case, documentation must be provided as to why the loss function or metric chosen instead is reasonable and appropriate for the AI application.
- Furthermore, target intervals for the chosen performance metric(s) and, if applicable, the loss function should be defined in the application area. The choice of values to be achieved should be comprehensively justified in view of the application context in question and the criticality of the task to be solved by the AI application.

**[RE-R-SC-CR-02] Quantification of the application domain coverage**

Requirement: Do

- Coverage of the application area by the training, test, and validation data set is formalized and captured using a quantitative metric, if possible. For example, if the input space can be formalized as a low-dimensional vector space and the application domain as a subset of it, then the simplest metric to check is whether each cell of a grid in that area contains data points.
  – In general, it may be useful to rely on methods that generate additional input data, for example, to achieve better coverage through the enriched training data (see **[RE-R-RO-ME-01]** and **[RE-R-RO-ME-02]**, as well as **[RE-R-RO-ME-04]** for augmentation techniques). It is often possible to make use of additional criteria to evaluate the "newness" of created data points[67].
- If possible, application-specific target intervals should be defined and justified for the coverage metric chosen previously for the application domain.
- If adequate coverage of the application domain justifiably cannot be demonstrated using quantitative means, qualitative reasoning may be used instead. Particularly in contexts such as open-world applications in which this type of review can never be conclusive, there must be a strategy to consider (e.g., logging) new scenarios and input data during operation.

**[RE-R-SC-CR-03] Quality of training and test data**

Adequate (quantitative) coverage of the application domain by training and test data is a key factor for ensuring that the ML model considers all possible application scenarios and therefore learns optimal decision rules. However, in order to ensure reliable performance of the AI component in the application domain, qualitative requirements should also be set for the data, as these can have an equal effect on the quality of the ML model. For example, the truthfulness of the data or the accuracy of annotations/labels is essential to ensure that the AI component draws the right conclusions from the learned context. Data processed and stored must be accurate, particularly with regard to personal data. In addition, depending on the application context, there may be additional qualitative data requirements that are related to technical constraints or arise from operational processes and requirements, such as the need to reproduce outputs (see **[TR-R-AU-CR-01]**) and associated inputs, from analyzing causes of errors/accidents (see e.g., **[S-R-CD-ME-03]** and **[S-R-FS-ME-14]**) or from the ability to provide information about data and models (see also **[TR-R-UA-ME-01]**, **[TR-R-EX-ME-01]** and **[DP-R-PD-ME-11]**).

---

**67** See for example. Odena, A. and I. Goodfellow (2018) TensorFuzz: Debugging Neural Networks with Coverage-Guided Fuzzing. Cornell University. https://arxiv.org/pdf/1807.10875.pdf (last accessed: 06/23/2021).

Requirement: Do

- Criteria are defined for assessing data quality. The following aspects should be considered as a minimum when selecting the criteria:
  – Technical requirements (format, file size)
  – Completeness of the data (e.g., all attributes present)
  – Truthfulness of the data
  – Correctness of annotations/labels
  – Relevance of the data to the application domain
  – Availability/access to data and metadata
- A qualitative objective/target characteristic is also formulated for each of the defined criteria, which, if met, ensure an acceptable risk level with respect to data quality.
- The choice of criteria and associated targeted properties must be justified. It should also be demonstrated that this choice is consistent with the objectives defined in **[RE-R-SC-RI-01]**.

## 7.1.3  Measures

### 7.1.3.1  Data

**[RE-R-SC-ME-01] Origin and quality of the database**

Requirement: Do

- The origin of the training and test data must be documented and the integrity of the data source(s) assessed.
- For annotated data, there must be documentation on how the annotations or labels were created. It should also describe how the correctness of data annotations is ensured (e.g., through checks by two people or special software).
- It must be demonstrated that the data is qualitatively suitable for training. In particular, the extent to which the criteria in **[RE-R-SC-CR-03]** are met must be explained.
- There should also be a description of the "compatibility" of training and test data taking the application domain of the AI component into account.
  – There should be a particular emphasis on whether the data is structurally identical and belongs to the same distribution or whether there are significant deviations.
  – Measures taken to prevent training and test data overlaps (data leakage) must be described (e.g., by local-sensitive hashing of the data or statistical methods).

**[RE-R-SC-ME-02] Choice of database**

Requirement: Do

- The choice of training and test data must be justified in detail in terms of the regular input data to be expected when the AI application is in operation.
- It must be documented that the training and test data sufficiently cover the application domain. Specifically, the following points should be addressed:
  – Coverage of the application domain should be documented in a, where possible, quantitative fashion using the metric and target interval specified in **[RE-R-SC-CR-02]**. For the coverage of the application boundary, reference can be made to **[RE-R-RO-ME-01]** to **[RE-R-RO-ME-03]** if necessary.
  – Comprehensible, meaningful documentation should be available describing any measures taken to improve coverage of the application domain, such as data augmentations. The choice of these measures must be justified. If applicable, reference can be made to **[RE-R-RO-ME-01]**, **[RE-R-RO-ME-02]** or **[RE-R-RO-ME-04]**.
  – If it is not possible to quantify the coverage of the application domain, detailed justification as defined in **[RE-R-SC-CR-02]** must be provided.

- If the training data does not correspond to the "real" data in operation, e.g., because it was generated using a different process, the evaluation of the AI application should be adjusted accordingly. The evaluation should be performed on related data sets that data sets that, while within the application domain, differ qualitatively from the training and test data (domain adaptation test set, e.g., simulation data, different data acquisition method, different data pre-processing). If applicable, the suitability of the selected evaluation data sets must be justified.

### 7.1.3.2 AI component

**[RE-R-SC-ME-03] Component design choice**

Requirement: Do

- Documentation should be available that links the choice of model components (training algorithm, loss function, etc.) to the selected reliability requirements. In addition, justification must be provided as to why the design and architecture of the AI component are appropriate for the application area in question. If different ML models were considered for the AI application, the documentation should describe the associated trade-offs and the rationale for the choice made. It should also be indicated here if frameworks were used to configure the ML model. In addition, there should be an explanation of how the features of the input data for the ML model were selected.
- If the training data does not match the "real" data in operation, a description must be provided of the measures taken to meet the challenges of concept, covariate and/or prior shift (e.g., transfer learning methods) and those taken to ensure the generalizability of the model.

**[RE-R-SC-ME-04] Systematic search for weaknesses**

Requirements: Do | Te

- It must be documented how a systematic search for AI component weaknesses by iterative data adjustment (closed-loop testing, see also **[TR-R-EX-ME-06]**) or introspective methods (see **[TR-R-EX-ME-02]**) is performed. If weaknesses were identified in the process, these must be recorded along with the measures taken in response.
  **Example:** An image classification application which separates ships from other objects was analyzed using a heat map-based method. For the "ship" category, the analysis showed that the wave pattern on the water was the decisive factor and not the ship as such. In response, the ML model was retrained on an augmented data set with a wider variety of wave patterns in the images.

**[RE-R-SC-ME-05] AI component reliability tests**

Requirements: Do | Pr | Te

- Tests of the AI component should be performed on data not seen during training (test data) that sufficiently covers the application domain as described in **[RE-R-SC-ME-02]**. (The performance of the AI component, with a particular focus on the application boundary, is evaluated in the **robustness** risk area tests, see **[RE-R-RO-ME-05]**.) How the tests are performed as well as the metrics considered according to **[RE-R-SC-CR-01]** and the achieved metric values should be documented. In addition, it must be demonstrated how the tests specifically look for model weaknesses.
  - In the event that training data from a different domain or distribution was used (such as transfer learning on synthetic data), the test data must correspond to the actual application domain (see **[RE-R-SC-ME-01]** and **[RE-R-SC-ME-02]**).
- If model weaknesses were uncovered during the development of the AI application by missing target intervals of relevant metrics according to **[RE-R-SC-CR-01]**, these cases must be documented as well as the corrective measures taken and the lessons learned.

### 7.1.3.3 Embedding

**[RE-R-SC-ME-06] AI application real-world tests**

Requirements: Do | Pr | Te

- Extensive real-world tests of the AI component (already embedded and functioning as an AI application) must be performed and documented. This involves determining the relevant performance metrics according to **[RE-R-SC-CR-01]** and also checking embedding-specific requirements such as runtime metrics and input distributions changed by the embedding. (The performance of the AI application, with a particular focus on the application boundary, is tested in the **Risk area: robustness (RO)**, see **[RE-R-RO-ME-06]**.)
  - For real-world tests, it is particularly important to ensure that all regular application situations have been covered. If possible, this can be achieved by testing all possible parameter combinations under real-world conditions. Taking the example of a system for sorting apples, this would involve testing all (categorical) combinations of color, variety and size. If it can be demonstrated that complete coverage is not possible, a suitable option must be chosen instead that illustratively covers the most important cases. This selection must be justified in the context of the application.
- Model weaknesses uncovered during real tests must be documented along with the corrective measures taken and lessons learned.

### 7.1.3.4 Measures for operation

**[RE-R-SC-ME-07] Supplement to open-world coverage**

Requirements: Do | Pr

- If the AI application is used in an open-world context or complete coverage of the application area according to **[RE-R-SC-CR-02]** cannot be guaranteed, a documented process must exist for the quality control of new input data during regular operation. This supplements the continuous testing procedures from **[RE-R-RO-ME-07]**. Furthermore, it must be documented how the findings from this process are used to enable continuous adaptation and improvement of the AI application (see, for example, **[RE-R-CD-ME-02]** as well as federated learning approaches in **[DP-R-PD-ME-08]**). If data is logged in this process, the **Dimension: Data Protection (DP)** must be taken into account.

## 7.1.4 Overall assessment

**[RE-R-SC-OA] Overall assessment**

Requirement: Do

- With reference to the measures taken, it must be demonstrated that the performance metrics defined in **[RE-R-SC-CR-01]** and the coverage metrics defined in **[RE-R-SC-CR-02]** for the application domain are within each of the target intervals defined there. Furthermore, justification must be provided that the data quality requirements defined in **[RE-R-SC-CR-03]** are met.
- If not all requirements specified in **[RE-R-SC-CR-01]** to **[RE-R-SC-CR-03]** are met, the deviations must be documented. This also applies to requirements that have only been partially met, e.g., where the criteria have not or not always been met.

## 7.2  Risk area: robustness (RO)

The robustness risk area accounts for risks that arise from a minor change or disturbance to a use case that the AI application is expected to be able to handle without error under normal circumstances. Different types of such deviations exist, such as image distortion, sensor noise or failure, or imprecise data collection such as measurement or typing errors. A special class of errors that also falls into this risk area are adversarial examples. These are characterized by a slight deviation from input data that can be processed correctly, but cause a major deviation from the expected result. Adversarial examples may be deliberately designed as an attack (referred to as an "adversarial attack" in this case), but are generally an expression of model weaknesses that also carry significance beyond attack scenarios.

The deviations in the input data examined here all stay within the application boundary and thus the AI component is still meant to correctly process the input despite the deviation. This application boundary must be specified. To do this, possible disturbances, including common data perturbations must first be recorded and potential damages estimated. Based on this, a check can be performed regarding whether the overall risk of each disturbance is acceptable in view of the application context.

The purpose of the measures designated for this risk area is to improve the robustness of the AI component against the deviations or perturbations that can occur within the application domain and to demonstrate that error-free processing of these deviating inputs has been achieved as far as possible. This clearly separates robustness from the **Risk area: intercepting errors at model level (IM)**, which deals with disturbances and inputs that, if they occur, would not be expected to lead to any meaningful processing by the AI application, or for which processing could only be performed with an unacceptable level of risk. In the **Risk area: intercepting errors at model level (IM)**, the measures are aimed at detecting these inputs instead of ensuring that they are processed properly by the AI component. In both risk areas mentioned, the measures listed are exclusively related to the AI component. In addition, there is the option of intercepting perturbations in the input space through measures that do not involve the AI component, but concern its embedding instead (e.g., redundant sensor design or hardware monitoring). These measures can be found in the **Dimension: Safety and Security (S)**.

### 7.2.1  Risk analysis and objectives

**[RE-R-RO-RI-01] Risk assessment and definition of the application boundary**

Requirement: Do

Documentation should be available that addresses the following points in a qualitative and conceptual manner:

- **Risk assessment:** The types of disturbances that can be expected given the application domain specified in **[RE-R-SC-RI-01]** are examined.
  Examples of possible disturbances are:
  - Expected sensor noise, e.g., from an optical sensor
  - Delayed data transfer or lower data quality (e.g., for real-time applications)
    - For example, audio transmission in an AI application for live translation
  - Distortion or other manipulation of input data
    - Typically changed object position during object detection, for example if a target object moves out of place or is not optimally positioned
  - Expected change in environmental conditions
    - If the AI application is used outside controlled environments, external parameters such as weather may change.

–  Adversarial examples (e.g., in image or audio data)
·  This is especially relevant for AI applications with a publicly accessible interface, including both public online use (e.g., free service offering) and an application positioned in public space (e.g., surveillance camera)

The list is a selection of examples and does not claim to be exhaustive. It should be explored which other types of disturbances can occur in the application domain. Where possible, these disturbances should be formalized and purposefully used in later tests. The **Risk area: functional safety (FS)** is referenced again here. This considers possible classical, AI-independent error sources, which can be handled using conventional algorithms or measures, such as defective pixels as sensor errors. If these error sources are also dealt with using AI-specific methods, these methods and possible overlaps with the **Dimension: Safety and Security (S)** must also be documented here.

For each type of disturbance identified, it is necessary to estimate the frequency and intensity of its potential occurrence. Based on this, it is necessary to analyze the possibility of a erroneous prediction or failure of the AI component for each identified disturbance depending on the intensity, and the potential resulting damage must be estimated.

- **Defining the application boundary:** The application boundary is defined based on the risk assessment. The application boundary describes the crossover of the application domain into input areas for which no meaningful processing is expected. This periphery is defined in terms of expected disturbances and describes the disturbance level up to which the AI component should correctly process input. In particular, input data (possibly containing perturbations) should only be processed by the AI component if this input data allows the AI application to be operated safely. This is why the risk assessments from the **Dimension: Safety and Security (S)** should also be considered during the process of defining the application boundary. Using this application boundary, it should be possible to associate which of the previously specified disturbances are covered to which extent in the **Risk area: robustness (RO)** (within the application boundary) and which ones are beyond this in the **Risk area: intercepting errors at model level (IM)** (outside the application boundary).

- **Objectives:** For disturbances that are within the application boundary, objectives are set in relation to the AI component. If these are achieved, the remaining residual risk is deemed acceptable. The objectives may differ according to the nature and level of disturbance. In particular, they should align with the requirements and measures in the **Dimension: Safety and Security (S)**, so that the AI application is sufficiently safeguarded within the application boundary through combined robustness, safety and security measures. This is discussed in the overall cross-dimensional assessment.

**Example:** A facial recognition model is deceived by a photo it is shown. The AI application could intercept this by taking dynamics such as blinking into account. A measure from the **Dimension: Safety and Security (S)**, however, would be to also use an infrared camera or a lidar system.

### 7.2.2 Criteria for achieving objectives

The following lists methods for quantifying the robustness and the application boundary, which make it objectively verifiable whether the set objectives are achieved and the identified risks are acceptable.

**[RE-R-RO-CR-01] Quantification of the application boundary**

Requirement: Do

- The application boundary described in **[RE-R-RO-RI-01]** is – as far as possible – formalized, for example as a degree of occlusion for pedestrian detection or as a signal-to-noise ratio for conversational AI. The application boundary may be defined, as described in **[RE-R-RO-RI-01]**, in terms of different expected disturbances and can encompass a quantitative grading of different disturbance levels. For example, if different target intervals related to performance of the AI application are provided for different disturbance levels.
- As a minimum, qualitative or semantic requirements should be specified for the input data to further characterize the application boundary.

**[RE-R-RO-CR-02] Quantification of robustness**

Requirement: Do

- The robustness of the AI component in relation to the boundary of the application area must be determined using mathematical-statistical metrics (compare **[RE-R-SC-CR-01]**). Documentation and justification must be provided regarding which of the metrics are subsequently used to evaluate the robustness.
- Target intervals are defined for the specified mathematical-statistical metrics for each type of disturbance within the application boundary. The robustness of the AI component to these disturbances can then be objectively verified by meeting the target interval in a test adapted to the specific disturbance. The target intervals may differ depending on the disturbance risk and level, but justification must be provided as to why the choice of target intervals is appropriate.

**[RE-R-RO-CR-03] Coverage of the application boundary**

Requirement: Do

- If possible, the coverage of the application boundary is formalized and quantified. The criteria in **[RE-R-SC-CR-02]** should be used as a starting point for this, especially if data point pertubations are favored or described by factors that are not fully covered in **Risk area: reliability in standard cases (SC)**. In addition, the application boundary can be graded differently depending on the type and level of disturbance (see **[RE-R-RO-CR-01]**). For example, if the input space can be formalized as a low-dimensional vector space, the simplest coverage metric to check is whether each cell of a grid in the application boundary contains data points.
- Application-specific target intervals for the coverage of the application boundary must be defined.

### 7.2.3 Measures

### 7.2.3.1 Data

**[RE-R-RO-ME-01] Data for testing robustness**

Requirement: Do

Documentation should be available describing test data properties and selection according to the following structure:

- The choice of data sets for evaluating the AI component in relation to possible disturbances must be justified. It must be demonstrated how the chosen data sets relate to the specific requirements of the application domain and that the data is quantitatively or otherwise qualitatively or semantically appropriate for the disturbances described in **[RE-R-RO-CR-01]**.

– The disturbances included in the data set should be semantically close to an adversarial example or technical fault relevant to the AI application.

– If the data set is created using augmentation or other methods, such as applying adversarial attacks, the suitability of the chosen method (regarding quality, proximity to the disturbance, etc.) and additionally of the underlying data set the method uses must be demonstrated. For this purpose, reference can be made to **[RE-R-SC-ME-02]**, for example.

– If the data sources or the method used to collect the data were also used in **[RE-R-RO-ME-02]**, it must be discussed whether the test data has sufficient variability in terms of the disturbances they represent compared to the associated training data so that the test can still detect possible overfitting to relevant disturbance patterns.
  **Example:** Using only slightly variable augmentations (e.g., 90° image rotations in the case of image recognition) can lead to overfitting to the chosen augmentation (and therefore no robustness to 45° rotations, for example).

- It must be documented how the integrity of the data source(s) is assessed. If already described there, reference can be made to **[RE-R-SC-ME-01]**.

#### [RE-R-RO-ME-02] Data for robust training

Requirement: Do

- Documentation should be available to demonstrate whether specific training data is used to achieve increased robustness ("robust training data"). To do this, the data sources mentioned in **[RE-R-RO-ME-01]** can be referred to, provided that their scope allows an appropriate split into training and test data. Alternatively, augmentation techniques[68] can be used to create a more diverse data set from existing data. These techniques can already be dynamically integrated into the training process, see **[RE-R-RO-ME-04]**.
- It must be examined to what extent the choice of training data is statistically representative for the application domain or why it is not critical or even appropriate to use data with possibly limited representativeness. For example, a model for collision detection could be trained using an over-representative number of (near-) collision examples in order to improve its performance in rare but critical situations (see also **[RE-R-RO-ME-03]**).

#### [RE-R-RO-ME-03] Examining corner cases

Requirement: Do

- Documentation should be available that demonstrates effort has been made to search for challenging input data (referred to as corner cases). These cases involve "difficult" data, such as the type of data from class boundaries, or input data that occurs so rarely that it is unlikely to be found in a randomly selected data set, even though correct processing of the data is meant to be possible.
  **Example:** For image recognition in autonomous driving, a ball rolling onto the road behind a parked vehicle represents a corner case.

### 7.2.3.2  AI component

#### [RE-R-RO-ME-04] Development and training procedure

Requirement: Do

Documentation should be available that uses the following structure to describe the extent to which development and training improve the robustness of the AI component:

---

**68** For example, see: Hendrycks, D. et. al. (December 2019). AugMix: A Simple Data Processing Method to Improve Robustness and Uncertainty. Cornell University https://arxiv.org/pdf/1912.02781.pdf (last accessed: 06/22/2021).

- The possible disturbances specified in **[RE-R-RO-RI-01]** must be linked to the development or training of the ML model. Specifically, the measures taken that contribute to the desired level of robustness must be described, such as
  – Augmented training (e.g., AugMix, adversarial training[69]),
  – Regularization (e.g., label smoothing, self-distillation, dropout, batch normalization),
  – Transfer approaches (e.g., pre-trained backbones, multi-task learning),
- It must be demonstrated how the measures taken, as well as the choice of loss function(s) and training algorithm,
  – promote reliability or robustness with respect to the disturbances specified,
  – (if applicable) achieve the desired level of generalizability, e.g., through multi-task or transfer learning approaches. If already described there, reference can be made to **[RE-SC-ME-03]**.

**[RE-R-RO-ME-05] Testing of AI component robustness**

Requirements: Do | Pr | Te

- Using the data sets specified in **[RE-R-RO-ME-01]**, tests of the AI component, which reflect identified relevant disturbances, are performed and documented. The target values achieved and the lessons learned during the tests are recorded.

### 7.2.3.3 Embedding

**[RE-R-RO-ME-06] Real-world generalization/exploration testing of the AI application**

Requirements: Do | Pr | Te

- In addition to the real-world tests performed in the **Risk area: reliability in standard cases (SC)**, the robustness of the AI application is tested using embedding-specific requirements, such as changed input distributions and other conceivable pertubations/errors/deviations. These generalization/exploration tests focusing on disturbances should be performed and documented while the AI component is already embedded and functioning as an AI application. In particular, a description must be provided regarding which of the disturbances identified as relevant in **[RE-R-RO-RI-01]** are taken into account by these tests.
- Justification must be provided regarding which of the previously established target intervals (from **[RE-R-SC-CR-01]** or **[RE-R-RO-CR-02]**) is used to evaluate the generalization/exploration tests. The target values achieved and lessons learned must be documented.

### 7.2.3.4 Measures for operation

**[RE-R-RO-ME-07] Monitoring input data in operation**

Requirements: Do | Pr | Te

- Where possible, tests must be carried out to determine whether the input data meets minimum requirements (technical quality, correct format) and is admissible (e.g., data cleansing through outlier detection). Depending on the data complexity, it is also necessary to check the semantic proximity to the use case (for complex inputs, a possible breach of the application boundary is reviewed in depth in **[RE-R-IM-ME-08]**). The test for monitoring the input data should be performed continuously during live operation. The methods on which this test is based, as well as possible follow-up responses, must be documented and justified if not already described in the **Risk area: functional safety (FS)** of the safety and security dimension.

---

**69** See also Carlini, N. et. al. (February 2019). On Evaluating Adversarial Robustness. Cornell University https://arxiv.org/pdf/1902.06705.pdf (last accessed: 06/22/2021); Kolter, Z. and Madry, A. (2021) Adversarial Robustness – Theory and Practice. https://adversarial-ml-tutorial.org/ (last accessed: 06/22/2021) and Zheng, S. et. al. (April 2016). Improving the Robustness of Deep Neural Networks via Stability Training. Cornell University. https://arxiv.org/pdf/1604.04326.pdf (last accessed: 06/22/2021).

**[RE-R-RO-ME-08] Monitoring outputs in operation**

Requirements: Do | Pr | Te

- If possible, a monitoring process should be established with regard to the outputs of the AI component in live operation (sanity check). If it is detected that outputs are moving away from the application domain, these outputs must be intercepted. The technical implementation of the monitoring process must be documented and checked.
  **Example:** The speed of a pedestrian is estimated to be 35 km/h. This dubious prediction is detected and results in verification by a redundant unit.

### 7.2.4  Overall assessment

**[RE-R-RO-OA] Overall assessment**

Requirement: Do

- Considering the tests performed in **[RE-R-RO-ME-05]** and **[RE-R-RO-ME-06]**, it must be demonstrated that the AI component is robust according to the criteria in **[RE-R-RO-CR-02]**. It should also be documented that the **[RE-R-RO-CR-01]** and **[RE-R-RO-CR-03]** criteria are met.
- If not all requirements specified in **[RE-R-RO-CR-01]** to **[RE-R-RO-CR-03]** are met, the deviations from these requirements must be documented. This also applies to requirements that have only been partially met, e.g., where the criteria have not or not always been met.

## 7.3 Risk area: intercepting errors at model level (IM)

While the **Risk area: robustness (RO)** targets error-free outputs of the AI component in relation to the application boundary, the intercepting errors at model level risk area specifically addresses cases where failure of the AI component is foreseeable or unavoidable, or where an excessive number of erroneous predictions results in unacceptable risk. An example of foreseeable failure is an optical detection system faced with unfavorable conditions for capturing shots. For example, unacceptable risk can be caused by an unanticipated domain shift, such as when an autonomous drone with an AI control system for open area operation is deployed in a street canyon.

The transition from robustness to failure can be fluent. A certain level of noise is unavoidable even in normal operation and therefore only falls into this category if it is above a threshold value. This application boundary has already been examined in the **Risk area: robustness (RO)** and described as far as possible.

The intercepting errors at model level risk area sits alongside the **Risk area: robustness (RO)** and also the **Risk area: functional safety (FS)**, as both chapters deal with monitoring the functional status. While the **Risk area: functional safety (FS)** examines risks from the embedding side, this section focuses on the AI component and, in particular, the ML model. Additional reference can be made to approaches from the **Dimension: Safety and Security (S)**, where possible. A number of constraints, such as the noise mentioned above, can be detected using conventional algorithms (i.e., measures from the **Risk area: functional safety (FS)**). However, the application domain of an AI application is often too broad to be fully covered using conventional monitoring. For example, it is reasonable to assume that conventional approaches are usually unable to detect whether a home user is trying to operate a voice assistance system in a language other than the one the system is expecting. A detection method is required at model level to prevent the AI application from misinterpreting the unknown input as an instruction to order a product. Therefore, a key focus in this risk area is to reliably test problematic input data for which meaningful processing cannot be expected and unobstructed processing by the AI component may lead to unacceptable risk in terms of bodily injury, property damage or financial damage. A model-level detection mechanism is designed to intercept these inputs and can also initiate follow-up responses.

It is important not to assume that detection strategies to be implemented at ML model level can be generalized to other (or even similar) problems. This is why the detection scope must first be defined, which presents a particular challenge with respect to disturbances on the semantic level. Here, a distinction can be made between a closed scope and an open-world context. An example of the former would be a static machine on a conveyor belt. An example of an open-world application, however, would be a smart appliance used by untrained users in day-to-day life, such as the speech recognition system mentioned previously. In the second case, the data is inherently more widely distributed, but even a closed scope does not provide protection against unknown input data. For example, there could be a person on a conveyor belt.

### 7.3.1 Risk analysis and objectives

**[RE-R-IM-RI-01] Scope, risk analysis and objectives**

Requirement: Do

- **Scope:** It is outlined what input data outside the application domain should be considered in order to intercept errors at the model level. To do this, it is first necessary to analyze the extent to which the unobstructed processing of out-of-distribution data (OOD data)[70] poses an unacceptable safety risk or would result in major financial damage. Based on the analysis, it is then necessary to consult the application boundary described in **[RE-R-RO-RI-01]** and determine which aspects that overstep this boundary should be considered in this risk area, i.e., by detection mechanisms at model level, and which aspects (instead or additionally) should be handled in the **Risk area: functional safety (FS)**. This part of the documentation is completed in line with the risk analysis in **[S-R-FS-RI-01]**. However, input data ranges that are well beyond the application boundary, but which could still occur, should also be considered. In addition, justification must be provided if detection is deemed unnecessary for identified OOD input areas.

---

[70] Out-of-distribution data is input data that is outside the application domain and not part of the original problem.

**Note:** The considerations at model level are to be distinguished from the **Dimension: Safety and Security (S)** (see **[S-R-FS-RI-01]**), insofar as technical input variability, e.g., due to sensor failure, can be covered there. The focus of the intercepting errors at model level risk area is on inherent properties of the AI component, such as the ability to process complex input data. Problems such as complete sensor failure that can be handled using conventional methods should ideally be addressed in the **Dimension: Safety and Security (S)**.

- **Risk analysis:** The documentation supplements and extends the risk assessment of probabilities of occurrence and potential damage in **[RE-R-RO-RI-01]** to include the input areas discussed above for which detection at model level is targeted. The damage in this instance is based on the assumption of faulty processing by the model, unless the input is intercepted. For this purpose, both a worst-case scenario, i.e., the most unfavorable output from the AI application, and a random output must be examined. Using image segmentation as an example, the first case would equate to "overlooking" critical elements, such as missing tumor segmentation on a CT image in computed tomography, and the second case would equate to a randomly distorted output in which all segmentation regions deviate from their usual shape and only loosely correspond to the actual image content.

- **Objectives:** The risk analysis is used to set qualitative requirements or objectives for the detection measures, which can classify the risk as controllable if achieved (assuming suitable follow-up responses if successfully detected). These requirements are described and examined in more detail below.
  The choice of follow-up responses to detections at model level is addressed in the risk analysis in the **Risk area: functional safety (FS)**.

### 7.3.2 Criteria for achieving objectives

The wide variety of problems in this risk area presents a challenge when it comes to formulating criteria that enable any AI application to be quantitatively assessed. The criteria must therefore be adapted to each specific case according to the objectives defined in **[RE-R-IM-RI-01]**. As a minimum, the criteria defined below should be used for this purpose.

**[RE-R-IM-CR-01] Out-of-distribution coverage**

Requirement: Do

- In line with **[RE-R-RO-CR-03]**, the coverage of the input space OOD areas to be intercepted as defined in **[RE-R-IM-RI-01]** should be formalized and quantified. If justification can be provided as to why quantification is not possible, a qualitative categorical argument can be used instead, as in **[RE-R-RO-CR-03]**, for example based on Zwicky boxes, provided that this is sufficiently justified.
- In particular, if it is possible to justify that the coverage cannot also be formalized in full using a qualitative categorical argument or if generalizability is required in **[RE-R-IM-CR-03]**, the coverage of the OOD data should also be assessed using other available OOD data sets. These data sets may contain "noise", but should preferably contain data for another related application purpose. The choice of the data range must be justified.

**[RE-R-IM-CR-02] Existence of mitigation strategies**

Requirement: Do

- Similarly to **[S-R-FS-CR-03]**, each input area from **[RE-R-IM-RI-01]** to be used for detection is paired with a mitigation strategy that will be used if the detection procedures from that risk area take effect. For example, this may involve passing control over to the user. A strategy does not have to be applicable to all of the areas mentioned, but can also apply only to specific areas (or even sub-areas). However, each (sub-)area must be covered by at least one measure. This classification is recorded in a table.

**[RE-R-IM-CR-03] Requirements for the detection methods**

Requirement: Do

- Requirements (quantitative if possible) for the detection methods should be recorded for each of the input areas mentioned in **[RE-R-IM-RI-01]** on the basis of the risk analysis and the assigned mitigation strategies defined in **[RE-R-IM-CR-02]**. At least the following points should be addressed:
  - Reliability. (Definition of a metric to measure the performance of the detection methods and a target interval or upper limit up to which detection failure would be acceptable. One of the metrics from **[RE-R-SC-CR-01]** can be selected.)
  - Response time. (The maximum detection time allowed is determined depending on the subsequent mitigation strategy.)
  - Application threshold. (If the shift from "robust" behavior and failure of the AI component in the input space is continuous, specifying a detection threshold is justified.)
  - Generalizability. (If the coverage in **[RE-R-IM-CR-01]** is qualitative rather than quantitative, the requirement for the detection method to be able to generalize or extrapolate is discussed. In this context, generalizability refers to the detection method's ability to work, even in the case of input data outside the application boundary, which is not necessarily to be expected in operation and thus was not explicitly included when designing the detection method.)
  - (If applicable) specific additional requirements resulting from each of the subsequent mitigation strategies defined in **[RE-R-IM-CR-02]**.

### 7.3.3 Measures

#### 7.3.3.1 Data

**[RE-R-IM-ME-01] Out-of-distribution data set**

Requirement: Do

- Out-of-distribution (OOD) data sets are prepared for testing the detection measures. If necessary, input data from relevant OOD input areas can be generated from the existing data in **[RE-R-RO-ME-01]**. This data must be enriched with further data in such a way that coverage is achieved for relevant input data outside the application domain according to **[RE-R-IM-CR-01]**. If the required quantitative or qualitative categorical coverage cannot be achieved, the suitability of the test data sets must be justified. This also applies to the scope of the test data.

**[RE-R-IM-ME-02] Data set splits for extrapolation**

Requirements: Do | Pr

- If **[RE-R-IM-CR-03]** requires generalizability of the detection method or the coverage as defined in **[RE-R-IM-CR-01]** cannot be achieved, "artificial" OOD data should be generated based on the overall data set. For this purpose, the overall data set is split so that the resulting sub-data sets are structurally different from each other. To create this split, the following measurable criteria can be used, for example:
  – semantic properties, e.g., on the basis of existing label information
  – statistical properties, e.g., distribution properties of subsets
  – latent representations of Deep Learning approaches, such as variational auto encoders (in which use of DNN features should also consider the lessons learned from the **Risk area: transparency for experts (EX)**)
- The choice of the criterion for splitting data sets must be justified. This choice should always be based on the application context and each method used, taking into account the purpose of obtaining additional "artificial" OOD data sets.
- The data set splits created must be documented and can be used for extrapolation tests in **[RE-R-IM-ME-05]**.

## 7.3.3.2  AI component

**[RE-R-IM-ME-03] Design for intercepting errors in outputs with correlation-based methods**

Requirement: Do

- There are numerous design approaches that can be used for complex tasks at the AI component level that lead to more reliable detection of inadmissible situations. One option concerns certain multi-sensor approaches that create an ensemble of ML models, each combining information from different "perspectives". Another design approach is multi-label learning, which involves training an ML model to solve different but related tasks simultaneously and thus output multiple labels, for example. While these approaches improve the robustness of the AI component (see **[RE-R-RO-ME-04]**), conflicts between the outputs can also serve as an indicator for the unreliability of the overall output. It must be demonstrated to what extent the design of the AI component helps intercept errors at model level, or it must be justified if intercept errors was not explicitly considered in the design.

**[RE-R-IM-ME-04] OOD tests**

Requirements: Do | Pr | Te

- The OOD test data set from **[RE-R-IM-ME-01]** is used to statistically analyze the proportion of the cases in which implemented detection mechanisms are effective. This analysis falls under binary classification (success/no success) and should at least be statistically evaluated with the associated criteria from **[RE-R-IM-CR-03]**. The qualitative requirements for the detection mechanism specified there should also be checked. The findings must be documented.
- It should be documented if, during the tests, the existing test data has been (iteratively) adjusted, new data has been created, or the detection mechanism has been adjusted, e.g., in reaction to inadequate performance.

**[RE-R-IM-ME-05] Extrapolation test**

Requirements: Do | Te

- If methods defined in **[RE-R-IM-CR-03]** are also to be applied to unknown data/data that is not sufficiently specifiable, it should be documented that the tests from **[RE-R-IM-ME-04]** were performed on the data sets created through **[RE-R-IM-ME-02]**. In each case, part of the data, being separated according to the criteria for each split, should be considered as outside the application domain and the same AI component should be retrained on the rest of the data. Furthermore, it is necessary to document how the effectiveness of the detection methods for the newly trained AI components were tested in the resulting OOD input area, and the test result must be recorded.
- If not all splits were used because a data split is not deemed relevant for one of the detection methods, for example because the risk to be avoided cannot occur in principle, this must be documented and justified in summary for the splits that were ignored.
- It should be documented if (iterative) adjustment of the existing data set has taken place during the tests in the event of potentially inadequate performance of the detection method or has led to new data being created or requested or to the adjustment of the AI component or its detection method.

**[RE-R-IM-ME-06] Uncertainty estimation**

Requirement: Do

- In principle, an intrinsic uncertainty estimation of the AI component can be used as a form of self-assessment to detect failures due to input data outside the application domain or due to model uncertainty. If this is intended, the **Risk area: uncertainty estimation (UE)** must include corresponding measures. In this case, explicit reference must be made in the measures to the intercepting errors at model level risk area.
  - In particular, the OOD data sets specified in **[RE-R-IM-ME-01]** and data set splits in **[RE-R-IM-ME-02]** should be considered when performing tests related to the uncertainty estimation in the **Risk area: uncertainty estimation (UE)**. The **[RE-R-IM-CR-03]** criteria must be addressed when evaluating the corresponding findings and in the final assessment in **[RE-R-UE-OA]**.

### 7.3.3.3 Embedding

**[RE-R-IM-ME-07] Real-world tests**

Requirements: Do | Te

- Where possible, the detection of error modes that may result from the OOD input areas to be intercepted according to **[RE-R-IM-RI-01]** should be tested under real conditions. This supplements and extends the existing tests under **[RE-R-RO-ME-06]** and should take place under the same conditions and requirements. The real-world test results as well as how the test was performed should be documented. If there is no separate test for this measure, it should be demonstrated, where necessary, that the measure is already covered by **[S-R-FS-ME-09]**, **[S-R-FS-ME-11]** or **[S-R-FS-ME-13]**.

### 7.3.3.4 Measures for operation

**[RE-R-IM-ME-08] Monitoring of input and output data**

Requirements: Do | Te

- The existing measures for monitoring input and output data (see **[RE-R-RO-ME-07]** and **[RE-R-RO-ME-08]**) should be evaluated with respect to their suitability for detecting potential error sources on the basis of the OOD data set from **[RE-R-IM-ME-01]**. These measures can be supplemented at model level by additional pre-processing, post-processing and monitoring procedures[71]. The findings and adjustments must be documented.

## 7.3.4 Overall assessment

**[RE-R-IM-OA] Overall assessment**

Requirement: Do

- Considering the testing performed and documented in this risk area, it must be demonstrated that the **[RE-R-IM-CR-01]** and **[RE-R-IM-CR-03]** criteria are met. Where measures complement each other, it must be demonstrated that the risk of correlated failure of these measures can be considered manageable.
- Furthermore, a table overview should be available that assigns corresponding mitigation strategies to the detection mechanisms for intercepting errors from this risk area. If necessary, reference can also be made to **[S-R-FS-ME-07]**. It must be demonstrated that this assignment is consistent with the risk analysis and objectives defined in **[S-R-FS-RI-01]** and meets the requirements in **[RE-R-IM-CR-02]**.
- It must be documented if the planned detection measures are not feasible or are not sufficient to meet the criteria in this risk area. The problems that cannot be addressed here can be re-examined in the **Risk area: functional safety (FS)** and the residual risk can be considered in the overall cross-dimensional assessment.

---

**71** For example, MetaSeg could be used for segmentation tasks, see: Rottmann, M. et. al. (November 2018). Prediction Error Meta Classification in Semantic Segmentation: Detection via Aggregated Dispersion Measures of Softmax Probabilities. Cornell University. https://arxiv.org/pdf/1811.00648.pdf (last accessed: 06/22/2021).

## 7.4 Risk area: uncertainty estimation (UE)

The uncertainty estimation risk area is designed to ensure that the AI application makes an accurate statement about the confidence of its outputs when required. Performing a realistic uncertainty estimation through the AI component enables to predict the risk of further processing its output and to implement an application-specific response. Similarly to the actual outputs of an AI component, the confidences are subject to misjudgments, so the resulting risk must be assessed.

Self-estimation of uncertainty through the AI component is not required for all application areas. However, it may be helpful or even necessary for more complex applications, particularly in an open-world context. Even uncertainty estimations containing errors can, in some cases, be subsequently corrected and mapped to realistic variables by means of calibration. A (correctly) calibrated uncertainty estimation can help detect unusual situations early on or even signal a departure from the application area. There are a number of methods that already intrinsically include uncertainty estimation (e.g., the probabilistic output of a DNN-based image classification), but these are usually poorly calibrated. Especially in the case of neural networks, it is common for intrinsic uncertainty estimation to significantly underestimate risks and specify high confidences even in cases of blatant misclassification. For this risk area, it therefore important to first assess to what extent an uncertainty estimation is needed for the AI application in question and the quality level required.

**Example:** The quality of the end product of an automated production line can be determined either by laborious quality control by experts or by an AI application using suitable sensor data. While the quality control method is costly, the AI-based method provides less reliable assessments. An uncertainty estimation can increase the benefit of the AI application for quality control by identifying products where further manual inspection could be worthwhile. The level of uncertainty estimation should be good enough to ensure that there are not too many rejected products and that not too much testing is required.

### 7.4.1 Risk analysis and objectives

**[RE-R-UE-RI-01] Defining and illustrating an estimation of uncertainty**

Requirement: Do

- **Risk analysis:** It must be examined how the residual risk resulting from the preceding risk areas can be further mitigated in terms of reliability by performing a realistic uncertainty estimation. In doing so, the specific use case must be addressed. If uncertainty estimation is used to intercept model-level errors, as defined in **[RE-R-IM-ME-06]**, this analysis should also take into account the follow-up responses that are supposed to start when high uncertainty is indicated. If an uncertainty estimation is not considered necessary or feasible in the application context in question, detailed justification must be provided.
- **Objectives:** Based on the risk analysis, it must be determined whether an uncertainty estimation should be implemented in the AI application. If this is the case, formal and qualitative objectives must be set to ensure that the planned uncertainty estimation contributes to mitigating existing residual risks in relation to reliability.
  - In particular, the formal objectives should specify the output of the uncertainty estimation, e.g., whether it is categorical or probabilistic. The former refers to categorization as discrete levels (certain/uncertain or low/medium/high), while the probabilistic approach is supposed to make probability statements that are as accurate as possible about the failure of the AI component for given input.

### 7.4.2 Criteria for achieving objectives

The risk analysis produces one or more primary requirements that the uncertainty estimation must fulfill. These relate to its quality, and in a sense the reliability of the uncertainty estimation itself, that can be measured using a suitable (calibration) metric. In this context, it is important to consider that uncertainty estimations may be subject to bias, causing incorrect uncertainty estimations to accumulate in certain use cases. For example, a system for detecting people might overestimate the confidence in detecting people wearing bright clothes, but generally still meet the quality requirements for the uncertainty estimation. It is important to consider ways to measure these irregularities and minimize the resulting risks.

**[RE-R-UE-CR-01] Documentation of uncertainty metrics and uncertainty estimation quality**

Requirement: Do

- At least one metric for assessing the quality of the uncertainty estimation and target intervals must be specified. The choice should correspond to the targets described in **[RE-R-UE-RI-01]** and must be justified. One of the options listed below should be chosen as the metric. If a different metric is used, it must be described in detail and this alternative choice must be justified.
- In addition, semantic dimensions of the use case can be specified along which the quality of the uncertainty estimation should be constant. For example, if it is known that the model performance is low for a certain subset of the data, the uncertainty estimation should also reflect this.
  **Example:** If facial recognition on images of people of color has an increased error rate, then the uncertainty estimation should be increased in line with this. These semantic dimensions must be specifically examined in subsequent tests to avoid problems such as incorrect uncertainty estimations accumulating for certain cases.

Types of evaluation for uncertainty estimation discussed in the literature include:

- The negative log-likelihood evaluates a probabilistic output of the AI component and accumulates – for each example of the test data set – the "probability" of drawing the associated correct label under the predicted probability distribution.
- For classifications, a popular choice is the Brier score, which calculates the squared Euclidean distance of the confidence output on the simplex (the interval [0,1] in the case of binary classification) to the label considered correct as the corner of the simplex (in binary increments of {0,1}) and accumulates it for the test data set.
- The expected calibration error (ECE) represents whether statements with e.g., 90 percent confidence are also correct in 90 percent of the cases. This method could also be adapted to categorical uncertainty scores, provided that these correspond to probability intervals.
- For categorical uncertainty statements, usually a heuristic evaluation scheme is needed that should take into account the specific risks and be documented in detail.

### 7.4.3 Measures

#### 7.4.3.1 Data

**[RE-R-UE-ME-01] Choice of a data set annotated with uncertainties**

Requirement: Do

- In some cases, uncertainty can already be a property of the ground truth data set. For example, several experts may provide (slightly) different labels for segmenting a medical image data set, or there may be intrinsic (referred to as aleatoric) uncertainty due to the aggregation of complex or large data sets, e.g., regarding averaged data points and their associated standard deviation. If a data set is available that contains information about the uncertainty of labels, it must be documented whether and in what way or to what extent the AI component was trained with this data set, and in what form the uncertainty annotation was used. Furthermore, the process for gathering or estimating the uncertainty annotations must be documented and an estimate of the quality of these annotations must be provided.

**Note:** Data sets of this nature are only available in rare cases. This measure therefore has limited applicability, but can provide significant added value.

### 7.4.3.2  AI component

#### [RE-R-UE-ME-02] Selecting an appropriate uncertainty estimation method

Requirement: Do

- Depending on the ML model used, it is possible to choose between different approaches with varying levels of quality and expenditure when implementing an uncertainty estimation. Based on the objectives and criteria, it is necessary to justify the relevant choice, highlighting both positive and negative characteristics of the method. Established methods for performing uncertainty estimation include (combinations of):
  - Bayesian network (e.g., *Monte Carlo dropout*[72]),
  - Parametric uncertainty,[73]
  - Deep ensembles.[74]

#### [RE-R-UE-ME-03] Post-processing for calibration

Requirement: Do

- To improve uncertainty estimations, procedures can be implemented which post-process the confidence output for the purpose of better calibration. The choice of post-processing method must be justified and documented on the basis of the quality criteria. Established methods for post-processing confidence outputs include temperature scaling[75] or isotonic regression[76].

**[RE-R-UE-ME-04] Testing the uncertainty estimation**

Requirements: Do | Pr | Te

- The uncertainty estimation must be tested on data not used in training or for calibration. If specific semantic dimensions have been specified in **[RE-R-UE-CR-01]**, test data should be chosen that allows the uncertainty estimation to be examined along these dimensions. Furthermore, the requirements in **[RE-R-IM-ME-06]** should be taken into account if uncertainty estimation is used as a detection method for intercepting errors. The metrics and target intervals defined in **[RE-R-UE-CR-01]** are used to evaluate quality. The test results must be documented.
- It must be documented if (iterative) adjustment of the uncertainty estimation has taken place (e.g., by means of additional post-processing measures) during the tests in the event of potentially inadequate quality.

### 7.4.3.3  Embedding

**[RE-R-UE-ME-05] Assessing follow-up responses**

Requirements: Do | Te

- If the uncertainty estimation can or should initiate follow-up responses, these responses and how they are implemented must be documented, if necessary in line with the **Risk area: intercepting errors at model level (IM)** (see **[RE-R-IM-ME-06]** and **[RE-R-IM-CR-02]**). Real-world tests are also performed in which the initiation of the follow-up responses by the uncertainty estimation is purposely triggered and tested. If the tests required by this measure are already documented elsewhere, for example in **[RE-R-IM-ME-07]** or **[S-R-FS-ME-13]**, reference can be made to this instead.

### 7.4.3.4  Measures for operation

There are no planned measures for this category.

### 7.4.4  Overall assessment

**[RE-R-UE-OA] Overall assessment**

Requirement: Do

- Documentation should be available demonstrating that the criteria specified in **[RE-R-UE-CR-01]** have been achieved.
- If not all requirements specified in **[RE-R-UE-CR-01]** are met, the deviations must be documented. This also applies to requirements that have only been partially met, e.g., where the criteria have not or not always been met.

## 7.5  Risk area: control of dynamics (CD)

The control of dynamics risk area deals with risks linked to model and concept drift. This ensures that AI application reliability is maintained during operation.

The reliability of the AI application can be weakened by changing external circumstances. If, for example, the statistical properties of the variable to be predicted change, the ML model may no longer be suitable to optimally describe this variable and its performance would suffer in line with this. This could be the case, for example, with an AI application that was trained to recognize faces before the onset of the COVID-19 pandemic and now has to deal with faces partially covered by masks. Furthermore, changes in the external conditions, such as changes in legislation, may require measures to be taken after the AI application has been put into operation.

To ensure AI application reliability at all times during live operation, the application should be checked on a regular basis and at appropriate intervals to make sure that it is functioning correctly. Furthermore, suitable measures can be established, such as logging challenging scenarios in live operation, in order to continually increase the reliability of the AI application.

In summary, this risk area deals with the following two risk categories:

1. **Model drift:** The AI application has reduced reliability after retraining on new training data collected during operation.
2. **Concept drift:** Changed external conditions impose new demands on reliability.

### 7.5.1  Risk analysis and objectives

**[RE-R-CD-RI-01] Risk analysis and objectives**

Requirement: Do

- **Risk analysis:** It must be documented whether and to what extent the AI application continues to learn during operation. If this is the case, it is also necessary to explain what the requirements are for the new incoming training data and what processes or mechanisms are in place to check the new incoming data. Based on this, it is necessary to estimate the probability of incorrect continual learning during operation and the damage that may arise as a result. It is also necessary to analyze the types of concept drift that can conceivably or potentially occur, and to document the potential consequences or damage that could result if external (possibly changed) requirements are not met.
- **Objectives:** Objectives should be set to detect, record and handle new error cases that emerge during operation. An acceptable risk level is achieved if these objectives are met.

### 7.5.2  Criteria for achieving objectives

**[RE-R-CD-CR-01] Intervals and quality requirements for assessing during operation**

Requirement: Do

- Appropriate assessment intervals for the AI application to assess reliability must be defined and documented according to the metrics and target intervals chosen in the previous risk areas in the reliability dimension. Prioritization of the risk areas in the planned regular testing must be application-specific and justified in detail. The size of the assessment intervals should specifically relate to the expected speed of relevant concept drifts (e.g., in an application that detects road users, the decision may be made to carry out a check once a year regarding whether new road users, such e-scooter riders, are detected with sufficient accuracy).

- It is necessary to define and document (quantitative) criteria (e.g., thresholds regarding performance) that describe in which scenarios re-evaluation of the AI application is required, possibly combined with retraining.
- Furthermore, it is necessary to develop and document criteria for assessing the quality of procedures used to record new application and error cases as well as how they are handled.
- It must be demonstrated that the chosen criteria are consistent with the **[RE-R-CD-RI-01]** objectives.

### 7.5.3 Measures

### 7.5.3.1 Data

There are no planned measures for this category.

### 7.5.3.2 AI component

There are no planned measures for this category.

### 7.5.3.3 Embedding

There are no planned measures for this category.

### 7.5.3.4 Measures for operation

**[RE-R-CD-ME-01] Avoiding catastrophic forgetting on new training data**

Requirement: Pr

- A process must be established which – when new training data is used, e.g., when incremental training of an AI component takes place over multiple cycles – checks that the incremental training steps do not lead to performance losses on the previous database, provided that it is still relevant to the application.
- The new training data and model versions should be saved after each incremental training step. Furthermore, the distribution of the new training data should be analyzed and the AI application should be tested on both the new and old data. It is possible to use additional measures such as a linear combination of outputs from the different model versions to counteract previously learned patterns/models being forgotten.
- The exact procedures of the established process are documented.

**[RE-R-CD-ME-02] Relearning with newly available training data**

Requirement: Pr

- A process must be established that initiates systematic retraining/relearning of the model based on **[RE-R-CD-CR-01]** when new training data is available (e.g., due to the need to keep the data constantly up to date or due to new data related to concept drift), while complying with all training and testing requirements of this catalog.
- The need for new categories or data collection procedures must be reviewed, documented and implemented each time new data is collected.
- The exact procedures of the established process are documented.

**[RE-R-CD-ME-03] Regular review of the AI application**

Requirements: Do | Pr

- There must be a procedure for ensuring periodic model monitoring with respect to reliability according to the assessment intervals defined in **[RE-R-CD-CR-01]**. The process can be a combination of review by humans at regular intervals, e.g., by the user or IT staff, and continuous automatic monitoring.
  - As part of this review, which can be supplemented by **[RE-R-IM-ME-08]** if required, it should also be examined whether the distribution of the input data changes during operation. To detect changes, either related to the input data or to a performance metric for instance, an (online) drift detection algorithm can be implemented depending on the complexity of the variable, such as ADWIN (adaptive windowing)[77]. The choice of method must be justified.
  - Critical and new input data that represents concept drifts, for example, is stored permanently with the aim of improving reliability in the future, for example through further training or improving detection measures (see **Risk area: intercepting errors at model level (IM)**). This data must be stored in accordance with the **Dimension: Data Protection (DP)**.
  - If the reliability targets are no longer met or if there are significant changes in the data basis, this must be communicated to the user or affected person and, if necessary, a process must be initiated to update or shut down the AI application in a controlled manner. It must be ensured that all relevant scenarios are covered in the **Risk area: intercepting errors at model level (IM)** and in the **Risk area: functional safety (FS)**. Furthermore, the measures in the **Risk area: control of dynamics (CD)** in the safety and security dimension must be taken into account.
  The process and the nature and scope of the reviews must be documented.

### 7.5.4 Overall assessment

**[RE-R-CD-OA] Overall assessment**

Requirement: Do

- It is demonstrated that a process has been established to regularly review the AI application that meets the criteria in **[RE-R-CD-CR-01]**.
- If not all requirements specified in **[RE-R-CD-CR-01]** are met, the deviations to the requirements must be documented. This also applies to requirements that have only been partially met, e.g., where the criteria have not or not always been met.

## Summary

**[RE-S] Summary of the dimension**

Requirement: Do

- If there is a medium or high protection requirement for this dimension, documentation must be prepared for the remaining residual risks. First of all, the residual risks from the various risk areas in this dimension are summarized. Subsequently, and taking into account the protection requirement, the identified residual risks are collectively assessed as negligible, non-negligible (but acceptable) or unacceptable. This analysis should specifically assess the impact of measures from the safety and security dimension in terms of whether they help mitigate or prevent AI component errors. The result of the analysis must be explained. In addition, if trade-offs exist between the **Risk area: reliability in standard cases (SC)** and **Risk area: uncertainty estimation (UE)**, justification of the selected prioritization must be provided.

---

- If risks or measures under this dimension have been identified as having potentially negative effects on other dimensions, such as **Dimension: Fairness (FN)**, **Dimension: Transparency (TR)** or **Dimension: Safety and Security (S)**, these effects must be documented.
- A conclusion must be made about the dimension that includes the assessment of residual risks.

# 8. Dimension: Safety and Security (S)

## Description and objectives

The safety and security dimension addresses risks relating to the areas of functional safety and IT security, the latter in the sub-areas of integrity and availability. Functional safety refers to protection against threats to the outside world in the event of a functional failure of the AI application. For example, a typical measure of this type is the installation of airbags in a car to protect the passengers in the event of an accident. By contrast, IT security is concerned with protecting the AI application from its environment, such as external attacks that can lead to a change in or impairment of functionality (integrity). Closely related to this is damage or loss of availability in which the system fails to respond or react to the required extent. The **Dimension: Data Protection (DP)** separately deals with information security breaches that lead to information being disclosed to unauthorized persons, for example.

The High-Level Expert Group on AI (HLEG) has defined abstract safety and security objectives for AI applications. However, these abstract objectives (and further-reaching ones) are far from being operationalized through a standard, for example. Conversely, there are already a number of operationally verifiable specifications and standards in the safety and security domain, but they do not specifically reference the features of AI applications. The purpose of the safety and security dimension is to merge the requirements from existing standards, which are essential to ensure protection against attacks on and threats from AI applications, and to expand them with further AI-specific requirements[78]. Thus, the risks and measures specified in this document refer to or are related to AI applications. Safety and security risks of the conventional software modules surrounding the AI component must be managed in accordance with the existing standards and are not covered in this catalog.

Threats in the safety and security domain can manifest themselves as a functional failure or a major functional change to the AI component. The **Dimension: Reliability (RE)** also examines these types of scenarios, for example in the form of adversarial examples and functional change due to concept drift. However, the criteria and measures in the **Dimension: Reliability (RE)** focus on the AI component, i.e., only the causes of threats that stem from the AI component are addressed there. By contrast, the measures in the safety and security dimension are mainly related to the embedding and are not directly applicable to the AI component itself. They are particularly applicable if the measures in the **Dimension: Reliability (RE)** cannot guarantee total mitigation of risks. For example, the **Risk area: intercepting errors at model level (IM)** of the reliability dimension, which covers detection mechanisms for significantly deviating inputs, explicitly refers to the **Risk area: functional safety (FS)** of the security dimension for the mitigation strategies following detection.

---

**78** The description in this section draws heavily on section "3.5 Security" in the white paper: Poretschkin, M.; Rostalski, F.; Voosholz, J. et al. (2019). Trustworthy Use of Artificial Intelligence. Sankt Augustin: Fraunhofer Institute for Intelligent Analysis and Information Systems IAIS. https://www.ki.nrw/wp-content/uploads/2020/03/Whitepaper_Thrustworthy_AI.pdf (last accessed: 06/18/2021)

The risk areas under the safety and security dimension are:

1. **Functional safety:** This risk area addresses the risk of accidental bodily injury or property damage that is facilitated or even caused by the malfunction or failure of the AI application as a result of flawed embedding design.

2. **Integrity and availability:** This risk area addresses risks that arise when data relevant to operating the AI application is falsified, resulting in the AI application being manipulated and possibly no longer available in some cases.

3. **Control of dynamics:** This risk area addresses risks that arise as a result of new threats of the above risk areas occurring or established safeguarding methods becoming less effective.

## Protection requirements analysis

The safety and security dimension considers two potential damage scenarios that form the basis for determining the protection requirement: bodily injury and damage to property, on the one side, and financial damage, on the other.

From a functional safety perspective, autonomous robotic systems, such as autonomous vehicles, which can cause bodily injury or property damage in the event of an accident, have an increased protection requirement. However, from an IT security perspective, these types of AI applications are particularly critical in scenarios where manipulated or incorrect functioning could result in considerable financial damage. Ensuring low latency can also be a concern for applications with real-time requirements, such as the automated processing of financial transactions.

The potential amount of damage results from whether a malfunction of the AI application (e.g., incorrect functioning or a failure) can lead to bodily injury or property damage. In addition, the extent of financial damage that a malfunction of the AI application may cause is taken into account.

The protection requirement is categorized as follows:

| High | At least one of the following applies: |
|------|----------------------------------------|
|      | ▪ The AI application interacts with people in such a way that they can be injured if it malfunctions. |
|      | ▪ A malfunction of the AI application (caused by errors, failures, manipulation or attacks) can result in very high financial damage (e.g., due to property damage). |
|      | **Example**: An AI application used to detect people and objects in an autonomous vehicle. Incorrect functioning can result in injury to people and high financial costs due to damage to property. |
|      | **Example**: An AI diagnostic application that makes decisions about the type of medical treatment given to individuals. Manipulation of the AI application can result in incorrect treatment of patients and have a serious impact on their health as a result. |
| **Medium** | The AI application cannot cause direct physical injury to people. However, a malfunction of the AI application (caused by errors, failures, manipulation or attacks) can result in high financial damage.<br>**Example:** An AI-supported application for transporting goods in storage can damage goods, for example, if goods are unintentionally unloaded in areas or at heights where this is not allowed. |
| **Low** | A malfunction of the AI application can at worst lead to medium financial damage.<br>**Example:** An AI application that is used to compose pieces of music. In case of failures or incorrect functioning, no financial damage is expected. |

**[S-P] Protection requirements analysis documentation**

Requirement: Do

- The protection requirement of the AI application for the safety and security dimension is defined as *low, medium* or *high*. The decision regarding the *low/medium/high category* is justified in detail with reference to the table above.

If the protection requirement for the safety and security dimension is *low*, the individual risk areas do not need to be examined more closely. The exception to this is that if measures are taken in the **Risk area: intercepting errors at model level (IM)** of the reliability dimension that are related to the **Risk area: functional safety (FS)** of the safety and security dimension, then the **Risk area: functional safety (FS)** should always be examined appropriately. If a *medium* or *high* protection requirement has been identified, each of the following risk areas must be examined in more detail.

## 8.1 Risk area: functional safety (FS)

Functional safety is part of the safety domain that is concerned with protecting the outside world from threats caused by the AI application. In this respect, the assessment catalog focuses on AI-specific accident risks[79] in the sense of unintended and harmful effects of the AI application that are facilitated or even caused by defective design. In particular, the functional safety risk area is designed to reduce to an acceptable level the risk of bodily injury or material damage in the event of a malfunction or even failure of the AI application.

As this risk area covers the consequences of malfunction or failure of the AI application, it addresses a risk class that is also dealt with in the **Dimension: Reliability (RE)**. The measures in the reliability dimension have an impact on various aspects including the avoidance of malfunctions (**Risk area: robustness (RO)**) and detection of errors (**Risk area: intercepting errors at model level (IM)**, **Risk area: uncertainty estimation (UE)**) through methods at ML model level. Functional safety continues in the vein of these measures in that it helps to detect errors or threats by using methods at the embedding level and deploys (partly conventional) mitigation strategies at the embedding level to ensure that safety is maintained in the case of detected threats.

Typical threats in the context of functional safety are those that lead to minor to severe injury, death or property damage and result from incorrect behavior or failure of the AI application. One possible cause is incorrect or harmful inputs outside the application domain. Problematic inputs can be detected at the embedding level in many cases. For example, an error message can be implemented to indicate when the camera of an image-processing AI application has malfunctioned. This would prevent the AI application from processing interfering signals, which could possibly lead to errors.

However, for AI applications that process complex input data, conventional methods are often not sufficient to detect harmful inputs. This is especially true in the case of open-world applications such as speech recognition where conventional methods would not be able to detect if a user is operating the AI application in the wrong language. For these scenarios, the **Risk area: intercepting errors at model level (IM)** of the reliability dimension covers model-level detection mechanisms. This allows the AI component to intercept problematic input data that would otherwise lead to unacceptable risk in terms of bodily injury, property damage or financial damage if (further) processing was not obstructed. The procedure in the **Risk area: intercepting errors at model level (IM)** of the reliability dimension must be aligned with the implementation of possible (additional) conventional function monitoring methods in this risk area. In particular, follow-up responses (mitigation strategies) must be established if errors or threats are detected at model level.

Another approach for intercepting errors or detecting threats is to perform a sanity check of the AI component's outputs before they are processed further by the surrounding software modules. In the case of an AI application used to detect objects in video footage, for example, the current segmentation mask could be compared with those of previous frames and checked for significant deviations. Furthermore, it is important to ensure that errors are not caused by the embedding itself, e.g., when interpreting the output of the AI component. Appropriate follow-up responses (mitigation strategies) must also be established in the event that errors or threats are detected at the system level.

---

**79** The concept of safety risks in the context of Artificial Intelligence in this risk area is based on the definition of accident risk in the paper: Amodei, D. et al. (2016). Concrete Problems in AI Safety. arXiv: 1606.06565 https://arxiv.org/abs/1606.06565 (last accessed: 06/30/2021)

If an error occurs or a threat is detected that only results in an isolated or short-term failure with manageable risk, the subsequent mitigation strategy should work toward fault tolerance. For this purpose, a safety function is usually triggered to ensure that the AI application is protected during the isolated failure while maintaining rudimentary functionality. Once the error is fixed, the AI application can return to operating as usual without much effort. For example, if an AI component for continuous environment recognition fails, the data no longer being supplied could be approximated for a short period using recently created environment descriptions to briefly bypass the failure.

However, the AI application must be put into a fail-safe state if an error occurs or a threat is detected that would lead to a sequential failure or even result in the loss of the overall functionality of the AI application under uncontrollable risk. Fail-safe means transitioning to a safe state where the top priority is to minimize the damage to the lowest possible level. The fail-safe state may deactivate the intended function of the AI application and, in extreme cases, even willfully destroy the AI application or the larger surrounding system. In the above example of environment recognition, if the AI application cannot be restored in a timely manner due to a sensor failure, for instance, the AI application could be transferred to a fail-safe state by giving control to the user (in this case, the AI application would be dysfunctional). In human-out-of-the-loop systems, which have no direct user, the transition to the fail-safe state must be designed based on the type and operating state of the AI application. In the case of AI-based environment recognition in a robot vacuum cleaner, for example, a fail-safe state can be achieved simply by switching off the application. However, if the environment recognition is linked to the control of a moving driverless vehicle, a fail-safe state might need to include the gradual reduction of speed and the sending of warning signals.

Aside from the scenario of a faulty function or even a failure of the AI application, safety risks can arise even when the AI application is operating correctly, for example due to external factors. Taking the AI-based control of an autonomous vehicle as an example, a human could suddenly run into the road. If the AI application correctly detects this person, there may still be a safety threat if, for example, the braking distance between the vehicle and detected human is too long. However, the purpose of functional safety is only to mitigate the types of risks and threats that arise from a malfunction of the AI application. For example, the risk of collision with people or objects correctly detected by the AI application is not a functional threat and is therefore not the focus of the functional safety risk area.

### 8.1.1 Risk analysis and objectives

**[S-R-FS-RI-01] Risk analysis and objectives**

Requirement: Do

- **Risk analysis:** Threats or potential damage due to low reliability for regular inputs (see **[RE-R-SC-RI-01]**) or a malfunction in case of emerging disturbances (see **[RE-R-RO-RI-01]**) have already been analyzed and their probability of occurrence estimated. The risk analyses specified are relevant to the objective of this risk area, but do not need to be reiterated again here.
  In addition, the risk of embedding malfunctions, which were not considered in the **Dimension: Reliability (RE)**, contribute to the safety of the AI application. It is necessary to analyze which AI-specific factors may lead to a malfunction of the embedding (among other aspects, it must be clarified whether certain outputs of the AI component may cause errors during further processing by the embedding), what damage may arise as a result and what the probability of this occurring is. User behavior (that is foreseeable) should also be taken into account when investigating possible causes of a malfunction or failure in the AI application, at least including:
  – foreseeable misuse of the application
  – unexpected start of the application
    (based on ISO 10218-1 4)

- **Scope for intercepting errors through embedding:** Justification should be provided for the scope of disturbances or errors and the input/output scenarios of the AI component to be intercepted by embedding methods, i.e., at the system level.
  - On the one hand, errors or failures of the AI application can be counteracted by intercepting input data that sits outside the application boundary defined in **[RE-R-RO-RI-01]**.
    **Example:** The optical environment recognition of an autonomous device does not recognize glass as an obstacle. To avoid damage at the embedding level, pressure sensors could stop the movement of the device in case of resistance.
    The documentation of which input areas should be intercepted using methods from this risk area should be drawn up in accordance with the risk analysis in **[RE-R-IM-RI-01]** so that all relevant harmful input areas are covered. In particular, input data areas that are not related to the AI application or its application boundary, but that could still occur, should also be considered.
  - On the other hand, errors or faults can be intercepted that only arise as a result of embedding. However, this assessment catalog addresses the AI-specific risks of an AI application and does not have the intention or pretense to fully map out classic functional safety or information security. This is why, in relation to embedding, only the errors or malfunctions of the embedding that are directly related to the AI component or the interpretation of its output are considered in this documentation. This does not include, for example, intercepting hardware faults or avoiding the operational limits of the embedding from being exceeded, unless this is related to the function of the AI component. However, intercepting outputs of the AI component that would compromise functional safety if processed further by the embedding is explicitly included in the scope of this risk area. An example of this type of threat is that an AI component for object detection outputs so many detections that the components of the embedding cannot handle them in a reasonable amount of time. It must be determined which output scenarios should be intercepted by the embedding on the basis of the risk analysis.
  It must be explained for which areas fault detection is not considered necessary. Furthermore, justification should be provided that sufficient risk mitigation is possible for the different kinds of disturbances, errors or input or output scenarios that are to be detected.
- **Objectives:** Qualitative objectives are set for the embedding to safeguard the AI application with regard to functional safety based on the risk analysis of this risk area and also considering **[RE-R-SC-RI-01], [RE-R-RO-RI-01]** and **[RE-R-IM-RI-01]**. Firstly, it is necessary to roughly describe with which approach the scopes defined in the previous section (of errors, disturbances, input/output scenarios) are to be intercepted. Secondly, it should be outlined to what extent and with which fundamental approach the embedding threats identified as relevant as well as the errors intercepted are to be mitigated or eliminated. Potential follow-up responses to the detection mechanisms from the **Risk area: intercepting errors at model level (IM)** must also be addressed. In particular, based on the identified risks, a broad classification should be provided detailing in which threat scenarios a fault-tolerant approach (maintenance of basic functionality) is desirable, and at what point the transition to a fail-safe state (safe but dysfunctional state) should take place.

### 8.1.2 Criteria for achieving objectives

Based on the identified threats for this risk area, appropriate safeguard measures must be taken that aim to reduce the risks covered here to an acceptable level. In order to be able to objectively check during the final assessment of the measures if these risks have been successfully mitigated, the objectives described in **[S-R-FS-RI-01]** must be translated into quantitative criteria. For this purpose, the desired risk level that is considered acceptable for the application context in question must be specified. In addition, requirements for the test data as well as the mitigation strategies must be specified.

**[S-R-FS-CR-01] Quantification of acceptable risk**

Requirement: Do

- Appropriate criteria for assessing the risk of potential harm caused by malfunction of the AI application must be specified for the particular application context. These criteria should at least include:

  Amount of damage:
  – Type and severity of possible injuries
  – Number of people affected
  – Costs (loss of work, compensation and additional effort to rectify the incident and for support)
  – Expenditure to repair property damage

  Probability of occurrence:
  – Specifying possible causes

- Target values must be defined for the established criteria (possibly for each damage scenario). A transparent argument must be provided of why the risk reaches an acceptable level when these target values are met and is in line with the ethical-legal framework. In addition, it must be demonstrated that the criteria defined here reflect the objectives.

**[S-R-FS-CR-02] Requirements for the test data**

Requirement: Do

- In view of the aspects considered in this risk area that go beyond the **Dimension: Reliability (RE)**, such as
  – the risk of an embedding malfunction,
  – intercepting harmful inputs or outputs of the AI component through the embedding,
  – embedding mitigation strategies,
  it may be necessary to supplement the criteria formulated in **[RE-R-RO-CR-03]** and **[RE-R-IM-CR-01]** for formalizing and quantifying the coverage of the application boundary as well as relevant OOD input areas for test data to be used in this risk area. If quantification is justifiably not possible, a qualitative categorical argument can be used instead, for example based on Zwicky boxes. If supplementing the criteria is not considered necessary and the test data sets are to be taken from the **Dimension: Reliability (RE)**, justification must be provided that the previously mentioned aspects are already sufficiently covered by this.

**[S-R-FS-CR-03] Existence of mitigation strategies**

Requirement: Do

- Each (fault) area to be intercepted by embedding methods defined in **[S-R-FS-RI-01]** is paired, similarly to **[RE-R-IM-CR-02]**, with a mitigation strategy (from **[S-R-FS-ME-08]** or **[S-R-FS-ME-10]**) to be taken if the detection method fails. For example, this can may involve passing control over to the user. A strategy does not have to be applicable to all of the areas mentioned, but can also apply only to specific designated areas (or even just sub-areas). However, each (sub-)area must be covered by at least one mitigation strategy. This classification is recorded in a table.

**[S-R-FS-CR-04] Requirements for error detection**

Requirement: Do

- Requirements (quantitative if possible) for detection must be recorded for each of the input areas to be intercepted mentioned in **[S-R-FS-RI-01]** on the basis of the risk analysis and the mitigation strategies to be assigned according to **[S-R-FS-CR-03]**. At least the following points must be addressed:
    – Reliability (Definition of a metric to measure the performance of the detection strategy (see **[RE-R-SC-CR-01]** for details) and a target interval or upper limit up to which detection failure would be acceptable)
    – Response time (maximum time offset allowed, depending on the subsequent mitigation strategy)
    – Application threshold (If the shift from "robust" behavior to failure of the AI component in the input space is continuous, specifying a threshold for detection to occur is justified.)
    – Generalizability (If the coverage in **[S-R-FS-CR-02]** is qualitative rather than quantitative, requirement for the method of generalization or extrapolation is discussed.)
    – (If applicable) specific additional requirements resulting from each of the subsequent mitigation strategies to be defined according to **[S-R-FS-CR-03]**

**[S-R-FS-CR-05] Requirements for mitigation strategies targeting fault tolerance**

Requirement: Do

- A mitigation strategy targeting fault tolerance must have the following properties for the security-related parts of the AI application (characteristic features):
    – A single error or an isolated failure of the AI component does not lead to a loss of safety.
    – If proportionate and reasonably feasible, each individual error is reported to the user and, ideally, to the developer or operator. It is important to ensure that error messages containing sensitive data are protected accordingly (e.g., personal information, see **[S-R-IA-ME-07]** and **[S-R-IA-ME-11]**).
    – If an individual error occurs, basic functionality should be maintained in the safe state until the detected error is corrected. A transition to a fail-safe state must be initiated (see **[S-R-FS-ME-10]**) if the detected error cannot be corrected and safe operation can no longer be maintained.
    (based on: ISO 10218-1 5.4)
- For each mitigation strategy targeting fault tolerance, the three characteristic features specified must be formulated into (quantitative if possible) application-specific criteria based on the risks determined in **[S-R-FS-RI-01]**. For this purpose, at least the following points must be addressed, and further requirements should be added depending on the application context:
    – Error scenarios or (if possible) threshold values or qualitative criteria whereby the mitigation strategy should take effect
    – Type and scope of functions to be maintained by deploying the mitigation strategy
    – Reliability (definition of a metric and target value that can be used to assess the performance of the mitigation strategy, see **[RE-R-SC-CR-01]**)
    – If required, the time frame for which reliability must be kept at least at this level
    – Response time (maximum time offset allowed for deploying the mitigation strategy)
    – Failure scenarios or (if possible) threshold values or qualitative criteria from which the mitigation strategy must transition to a fail-safe state (taking into account the operational limits of the embedding, among other factors).

**[S-R-FS-CR-06] Requirements for mitigation strategies targeting a fail-safe state**

Requirement: Do

- Based on the particular application context and the associated risks identified in **[S-R-FS-RI-01]**, (quantitative if possible) requirements must be recorded for each mitigation strategy to be assigned according to **[S-R-FS-CR-03]** with the purpose of achieving a fail-safe state. At least the following points must be addressed:
  – Error scenarios or (if possible) appropriate threshold values or qualitative criteria from which the mitigation strategy targeting a fail-safe state should take effect. Operational limits of the embedding, e.g., maximum rotation speed for motor control, must be taken into account. In addition, this also includes the scenarios in which a fail-safe strategy is triggered following a mitigation strategy targeting fault tolerance (see **[S-R-FS-CR-05]**).
  – Response time (maximum time offset allowed before the mitigation strategy is deployed)
  – Description of the fail-safe state to be achieved in the best-case scenario
  – Reliability (if applicable) of the fail-safe strategy, i.e., the reliability with which the fail-safe state is achieved
  – Justifiable maximum damage that may be accepted during transition to the fail-safe state

### 8.1.3 Measures

The measures in this risk area include embedding methods and tests to intercept errors or a failure of the AI application, to bypass them (in terms of fault tolerance) as well as to enter a fail-safe state if necessary.

**[S-R-FS-ME-01] Safety guidelines and instructions for use**

Requirements: Do | Pr

- Safety objectives for functional safety must be derived from business objectives, business processes, relevant laws, regulations and potential threats, and these must be documented. This safety guideline also contains strategic guidance on how to achieve these objectives.
- Based on the safety guideline, instructions are provided in a standard format on
  – using the application safely and
  – developing the application
  Measures should be described to ensure that all users take note of them.
- A process must be established and documented that makes users and developers of the application aware of risks relating to functional safety and indicates how to act correctly with respect to the safety of data, model and embedding.
  (based on BSI C5 SA-01)

### 8.1.3.1 Data

**[S-R-FS-ME-02] Scenario coverage**

Requirement: Do

- Documentation should be available describing which test data is used to verify measures in this risk area. It should be explained in a transparent way that the test data contains sufficient potential accident scenarios and critical situations and thus fulfills the criteria defined in **[S-R-FS-CR-02]**. If necessary, reference can be made to the documentation from the **Dimension: Reliability (RE)**.

### 8.1.3.2 AI component

**[S-R-FS-ME-03] Role of the AI component**

Requirement: Do

- It must be demonstrated to what extent the architecture and design of the AI component contribute to the functional safety of the AI application, in particular to the prevention of accidents. If applicable, it is necessary to describe whether the learning function (or the creation of the learning function) takes into account accidents and injuries with negative feedback. If this has already been described in the **Dimension: Reliability (RE)**, reference can be made to the relevant documentation.

### 8.1.3.3 Embedding

**[S-R-FS-ME-04] Embedding design**

Requirement: Do

- Documentation should be available detailing the extent to which the design and architecture of the embedding (e.g., through redundant design or the integration of classic assistance systems) contribute to the prevention or even bypassing of a malfunction and thus to the strengthening of functional safety.
- As this assessment catalog primarily addresses the AI-specific risks of the AI application and does not have the intention or pretense to fully map out existing conventional standards (on functional safety, product safety, information security, etc.), the documentation should focus on aspects of the embedding that are related to the further processing or interpretation of the output of the AI component. Reference should be made to the current classic norms and standards when dealing with the risks that are not specific to AI.

**[S-R-FS-ME-05] Intercepting harmful input data**

Requirements: Do | Te

- It should be explained in a transparent way which methods at the embedding level are used to detect harmful inputs outside the application boundary that would represent an unacceptable safety risk if processed. For example, the malfunction of a sensor is a typical error mode that can be detected by conventional methods. In the case of image data, harmful inputs could be detected and intercepted by measuring defective pixels, for example.
  - If methods are implemented at embedding level to intercept harmful inputs in addition to detection mechanisms at model level, these must be aligned with the approach in **Risk area: intercepting errors at model level (IM)** of the reliability dimension. Furthermore, it must be demonstrated that all input areas for which embedding level detection is envisaged in **[RE-R-IM-RI-01]** are covered effectively by the measures presented here.
- The effectiveness of the methods for intercepting errors or detecting threats must be demonstrated in suitable tests. The tests must be documented and their choice justified. If there is no separate test for this measure, it must be demonstrated that the methods have already been adequately examined through **[S-R-FS-ME-09]**, **[S-R-FS-ME-11]** or the **[S-R-FS-ME-13]** final real test.
- An explanation must be provided regarding the extent to which the method(s) described contribute to fulfilling the criterion **[S-R-FS-CR-04]**.

**[S-R-FS-ME-06] Intercepting errors during interpretation of the AI component output**

Requirements: Do | Te

- It should be explained in a transparent way which methods at the embedding level are used to detect outputs of the AI component that would represent an unacceptable safety risk in the event of further processing/interpretation by the embedding. For example, an AI-based collision avoidance system could provide an additional safeguard for the interpretation of the AI component by comparing the results of distance and speed sensors separately from the output of the AI component or by checking the output of the AI component for consistency (e.g., in terms of time).

- The effectiveness of these measures at embedding level for intercepting errors or detecting threats must be demonstrated in suitable tests. The tests must be documented and their choice justified. If there is no separate test for the measure, it must be demonstrated that the measure has already been adequately examined through **[S-R-FS-ME-09]**, **[S-R-FS-ME-11]** or the **[S-R-FS-ME-13]** final real-world test.
- An explanation must be provided regarding the extent to which the method(s) described contribute to fulfilling the criterion **[S-R-FS-CR-04]**.

**[S-R-FS-ME-07] Choice of mitigation strategy**

Once an error is intercepted or a threat scenario is detected by embedding methods or the model, a mitigation strategy must be deployed to ensure safety in this exceptional situation. The type of error, the operational and environmental state of the AI application and the associated risk situation determine whether the mitigation strategy contributes to the fault tolerance of the AI application in terms of bypassing the exceptional situation for a brief period while maintaining basic functionality, or whether the mitigation strategy transitions to a fail-safe state.
**Example:** An object detection system in an autonomous car can temporarily override detections with an expected uncertainty so high that the position can no longer be reliably determined, if the last known (safe) distance to the object is big enough in relation to the car's speed. Otherwise, it should brake immediately, for example.

Requirement: Do

- The procedure (e.g., "by design" or depending on situation) and criteria used to determine which subsequent mitigation strategy is used to deal with an intercepted error or detected threat (by the embedding or the AI component itself) must be documented.
  – It must be demonstrated that the risks and consequences of a fault-tolerant approach have been adequately considered, taking into account the risk analyses **[S-R-FS-RI-01]** and **[RE-R-IM-RI-01]**.
  – It is also necessary to demonstrate that the approach to selecting the subsequent mitigation strategy is consistent with the objectives defined in **[S-R-FS-RI-01]**.
- Justification must be provided that the described approach for choosing the mitigation strategy is appropriate for the particular application context and contributes to the functional safety of the AI application.
- In accordance with **[S-R-FS-CR-03]**, a table overview should be available that illustrates the subsequent mitigation strategy including its purpose (i.e., fault tolerance or fail-safe state) for each detection mechanism (at model and embedding level).

**[S-R-FS-ME-08] Mitigation strategies targeting fault tolerance**

Fault tolerance relates to the ability of the AI application to maintain its functionality (at least basic functionality) even if a malfunction or failure of the AI component occurs. To achieve fault tolerance, the AI application must have mitigation strategies that, for example, deal with harmful inputs or briefly bypass a failure until the fault is corrected.

Requirements: Do | Te

- Documentation should be available that clearly describes the fault tolerance of the AI application. For this purpose, the table overview from **[S-R-FS-ME-07]** is used as a starting point to explain in detail which mitigation strategies targeting fault tolerance the AI application has.
  – If applicable (and not already addressed as a model-level detection mechanism), follow-up responses related to an uncertainty estimation (see **Risk area: uncertainty estimation (UE)** under the reliability dimension) should also be established and documented.

- The effectiveness of the mitigation strategies targeting fault tolerance in the intended application scenarios defined in **[S-R-FS-CR-05]** must be demonstrated through suitable tests. The tests must be documented and their choice justified. If there is no separate test for the mitigation strategies, it must be demonstrated that this has already been adequately examined through **[S-R-FS-ME-09]** or the **[S-R-FS-ME-13]** comprehensive real-world test.
- It must be demonstrated that the existing mitigation strategies targeting fault tolerance meet the criteria specified in **[S-R-FS-CR-05]**.
- It must also be justified in detail that the fault tolerance achieved is sufficient and appropriate for the application context in question.

**[S-R-FS-ME-09] Testing fault tolerance and its detection mechanisms**

Requirements: Do | Te

- Tests of the AI application in unexpected situations and in various threat environments in which the AI application is expected to be fault-tolerant must be performed and documented.
  - Detection mechanisms are specifically triggered at model and embedding level for harmful inputs in order to test subsequent mitigation strategies targeting fault tolerance (see **[S-R-FS-ME-08]**). The test data used here must at least contain the scenarios from the data set from **[S-R-FS-ME-02]** and **[RE-R-IM-ME-01]** (and if applicable also **[RE-R-IM-ME-02]**) for which the AI application is meant to be fault-tolerant.
  - The test data must be expanded to also trigger AI component outputs that should be intercepted at the embedding level for the purpose of fault tolerance.
- The choice of test data sets must be justified. In addition, it must be demonstrated that they meet the **[S-R-FS-CR-02]** criteria.
- Documentation should be available that justifies that the tests performed and their results are sufficient to confirm that the AI application has suitable fault-tolerant behavior in unexpected situations and in threat environments.

**[S-R-FS-ME-10] Mitigation strategies targeting a fail-safe state**

Requirements: Do | Te

- Documentation should be available that explains in detail the mitigation strategies of the AI application targeting a fail-safe state based on the table overview from **[S-R-FS-ME-07]**. In addition, the fail-safe state to be established by each mitigation strategy must be described. Among others, the following aspects are addressed:
  - If a fail-safe strategy is potentially capable of causing damage in an operational scenario, it must be demonstrated that different strategies have been considered and that the chosen mitigation strategy minimizes the damage.
  - In the case of mitigation strategies targeting a fail-safe state that involve the user (e.g., by passing control to the user), it is important to indicate that the user is informed about this option and has been instructed on how to act if required. If this has already been addressed in the **Dimension: Autonomy and Control (AC)** or in **[S-R-FS-ME-12]**, or fulfilled by **[S-R-FS-ME-01]**, reference must be made to the relevant section.
- The effectiveness of the mitigation strategies targeting a fail-safe state in the intended application scenarios defined in **[S-R-FS-CR-06]** must be demonstrated through suitable tests. The tests must be documented and their choice justified. If there is no separate test for the mitigation strategy, it must be demonstrated that this has already been adequately examined through **[S-R-FS-ME-11]** or the **[S-R-FS-ME-13]** comprehensive real-world test.
- Detailed justification must be provided for each fail-safe strategy to demonstrate that it is proportionate and appropriate for the application scenarios in which it is to be used in accordance with **[S-R-FS-CR-06]** (and **[S-R-FS-CR-05]**).

**[S-R-FS-ME-11] Testing fail-safe strategies and their detection mechanisms**

Requirements: Do | Te

- In tests, the AI application is subjected to various threats that should initiate a fail-safe response. How the tests are performed must be documented.
  - In the tests, these types of detection mechanisms are specifically triggered at the model and embedding level and the subsequent mitigation strategy should trigger a transition to a fail-safe state.
    - · The test data used here must at least contain the scenarios from the data set from **[S-R-FS-ME-02]**, **[RE-R-IM-ME-01]** (and if applicable also **[RE-R-IM-ME-02]**), which should lead to a **fail-safe state**.
    - · The test data must be expanded to also cover the detection mechanisms at the embedding level that intercept harmful outputs of the AI component and lead to a fail-safe state.
  - In addition, if they exist, the scenarios in which the AI application must transition from fault-tolerant behavior to a fail-safe state must be triggered.
- The choice of test data sets must be justified. In addition, it must be demonstrated that they meet the **[S-R-FS-CR-02]** criteria.
- Documentation should be available that justifies that the tests performed and their results are sufficient to confirm that the AI application adequately transitions to a fail-safe state in all required situations.

**[S-R-FS-ME-12] Option for human intervention**

Requirement: Do

- Documentation should be available on the extent to which human intervention in the workings or operation of the AI application is necessary and possible in order to avoid an accident. If already addressed there, reference can be made to relevant mitigation strategies (see **[S-R-FS-ME-08]** and **[S-R-FS-ME-10]**) or to the **Dimension: Autonomy and Control (AC)** (see **[AC-R-TD-ME-02]**).
- If the AI application has been identified as having a high protection requirement in the safety and security dimension, it must be demonstrated that it has a designated stop function that
  - has priority over all other functions of the application,
  - causes the shutdown of all threats arising from controlled parts,
  - provides the option to control threats arising from the application,
  - remains active until reset and
  - can only be reset by explicit confirmation.
    (based on: ISO 10218-1 5.5)
  If, in the case of a high protection requirement, use of the stop function is restricted to a certain group of people (or if certain groups are excluded from using it), this must be justified in detail in terms of the risks.
- As spontaneous shutdowns and other forms of human intervention can potentially involve risks (e.g., a threat may still exist due to inactivity), it is also necessary to document that potential users have been instructed in how to handle the intervention options appropriately and the consequences that may arise as a result. If already discussed there, reference can be made to **[S-R-FS-ME-01]** or to the **Dimension: Autonomy and Control (AC)**.

**[S-R-FS-ME-13] Comprehensive real-world test**

Requirement: Do | Te

- Once the AI application has been installed in the larger surrounding system, if applicable, and is functional, it must undergo a real-world test before it is put into operation. This test must cover all the safety functions examined in this risk area and how they are linked. The AI application must be presented with all malfunction triggers that are identified as relevant; these triggers must first be detected by the application and then trigger a mitigation strategy targeting fault tolerance or the transition to a fail-safe state. The performance of the tests must be documented.

- The choice of test scenarios must be documented and justified. In addition, it must be demonstrated that they meet the **[S-R-FS-CR-02]** criteria. If necessary, reference can be made to the corresponding sections in the **Dimension: Reliability (RE)**. It must be demonstrated that the test and the test results obtained are sufficient to confirm that the AI application behaves appropriately in threat environments and hazardous situations as defined in **[S-R-FS-CR-01]**.
- Testing of functional safety measures that are not specific to AI and are thus not addressed in this risk area should be performed separately and in accordance with existing norms and standards.

### 8.1.3.4  Measures for operation

**[S-R-FS-ME-14] Dealing with accidents**

Requirements: Do | Pr

- The nature of accidents occurring in connection with the AI application and the way in which they develop must be logged and it must be specified how the AI application deals with each accident situation. The accidents that have occurred and their cause must be analyzed.
- Continuous checks must be performed to determine whether the way the AI application works in an accident situation meets the requirements set out in **[S-R-FS-RI-01]** and whether the measures in this risk area contribute sufficiently to meeting the criteria specified in **[S-R-FS-CR-01]**. If deviations are identified, adjustments must be made to the safety measures or the AI component itself (see **Dimension: Reliability (RE)**). These adjustments must be documented.

### 8.1.4  Overall assessment

**[S-R-FS-OA] Overall assessment**

Requirement: Do

- With reference to the tests performed and documented, it must be demonstrated in a transparent manner that, in accordance with **[S-R-FS-CR-01]**, operation of the AI application is ensured under acceptable accident risk, in particular also with regard to unknown input data, if this is required.
- It must be documented if the measures in this risk area are not feasible or are not sufficient to meet criteria **[S-R-FS-CR-01]** to **[S-R-FS-CR-06]**. The problems that cannot be addressed here can be considered in the overall cross-dimensional assessment.
- If safety measures from this risk area and detection measures from the **Risk area: intercepting errors at model level (IM)** under the reliability dimension complement each other, it must be documented that the tests only correlate weakly with each other, so that the risk of simultaneous or interrelated failure can be considered manageable.
- Furthermore, functional safety standards or norms must be documented that are used in addition to this assessment catalog to test the AI application.

## 8.2 Risk area: integrity and availability (IA)

Integrity and availability are protection objectives of classic IT security. They are revisited in this AI assessment catalog because – firstly – existing risks with regard to these protection objectives are increased by the use of Machine Learning and – secondly – new types of risks are emerging, for example because data represents a more sensitive attack vector for AI technologies than is the case with classic IT systems. In particular, the fact that AI applications are data-driven IT systems results in a larger overlap between AI-specific risks related to integrity and availability, which is why these protection objectives are combined into one risk area.

In the context of information security, integrity is synonymous with intactness in the sense that no unauthorized or unintended changes are made. This risk area considers the integrity of the AI application. The nature of Machine Learning means that data has an essential impact on the quality and functionality of the AI application and thus creates an attack surface for its integrity. Integrity breaches can take on different dimensions.

Firstly, the integrity of an AI application can be undermined in specific circumstances. In particular, adversarial examples breach the integrity of the AI application if they are deliberately created by attackers, for example to force a specific output. These are known as adversarial attacks. Adversarial attacks exploit weaknesses of the ML model and – in terms of their cause – are to be assigned to the **Dimension: Reliability (RE)**. However, the vulnerability of ML Models represents a security risk that, in addition to the measures in the **Risk area: robustness (RO)**, which mainly concern the design and development of the AI component, should also be mitigated through classic IT security measures as appropriate in this dimension. Usually, the more precise the attacker's knowledge of the ML model, the more successful an attack on the AI application will be. This means the integrity of the AI application is indirectly related to data confidentiality (weights, training data, etc.), and data protection measures thus help to mitigate security risks. Relevant conventional measures relating to confidentiality, such as encryption, are covered in this risk area to avoid duplication of measures in the **Dimension: Safety and Security (S)** and in the **Dimension: Data Protection (DP)**. However, the **Dimension: Data Protection (DP)** focuses on AI-specific confidentiality measures that relate to the design and modeling of the learning algorithm implemented in the AI application.

In addition, the integrity of an AI application can be breached to the extent that its functionality is (permanently) changed. This can happen, for example, if attackers make unauthorized changes to the code or the weights. Unlike classic IT systems, it is sometimes possible to make functional changes to AI applications by deliberately manipulating the database, which is referred to as data poisoning. In the case of AI applications that (continue to) learn their decision rules/models online based on user inputs, for example, this can even be achieved by sending targeted requests or queries to the application.

Availability in the context of information security means that the AI application is executed or retrieved in a timely manner and as intended. In classic IT systems, unavailability is usually experienced either when the hardware is overloaded due to a high number of requests, or when the system is in error mode and thus does not process requests at all. The first scenario (in terms of denial-of-service attacks) is relevant for AI applications, as they are usually CPU-intensive and the number of user requests can often change significantly, e.g., in the case of AI applications with a public interface such as an online translation service. This is why scalability should be considered in the architecture of an AI application, but hardware is not the focus of this assessment catalog. The second scenario, i.e., unavailability due to an error mode as well as other types of failures, is addressed in the **Dimension: Reliability (RE)** and in the **Risk area: functional safety (FS)** if such scenarios form part of the assessment object of the assessment catalog. Restrictions on the availability caused by embedding errors or failures when the AI component is functioning correctly are not addressed in this assessment catalog.

In addition to the scenarios related to unavailability in classic IT systems, there is another form of unavailability that is specific to AI applications. It stems from the fact that the AI component may (continue to) learn its decision rules/model during operation and thus change how it functions based on this. If this continual learning causes the AI application to lose its original form and quality during operation, this constitutes a form of unavailability. This can be illustrated by the example of a chatbot[80] that (continues to) learn online through user interaction. If a chatbot faces large quantities of unwanted inputs, such as expletives, meaning that it mainly outputs inappropriate expressions, then – even if the chatbot continues to answer messages in a timely manner and is not noticeably in error mode – this should be interpreted as a negative impact on its availability, as its AI-based functionality is no longer provided in its original form and quality. This example of data poisoning also illustrates the connection between risks relating to the integrity and availability of an AI application.

Similarly to adversarial examples, the cause of the change of the AI application during operation (in the sense of model drift) falls under the **Dimension: Reliability (RE)**. Since a major functional change also poses a security risk, particularly if it is the result of targeted attacks such as data poisoning, this risk area also addresses AI application drift. However, the availability measures in this risk area do not relate to the design and development of the AI component (as in the **Dimension: Reliability (RE)**), but are based on classic IT security measures to limit potential unavailability and to quickly restore the availability of the AI application if required. Specifically, the focus is on (potentially financial) damage that the operator of the AI application may incur due to unavailability. Measures with the purpose of preventing or mitigating damage to users or the environment during or due to unavailability in terms of failure of the AI application, however, are addressed in the **Risk area: intercepting errors at model level (IM)** of the reliability dimension and in the **Risk area: functional safety (FS)**. The two risk areas mentioned include measures to prevent damage in the event of failure of system-relevant components, e.g., due to harmful inputs.

## 8.2.1 Risk analysis and objectives

**[S-R-IA-RI-01] Risk analysis and objectives**

Requirement: Do

- **Integrity risk analysis:** Integrity threats must be specified for the AI application being examined. Typical threats to integrity are those that lead to undesired modification of the AI application and specifically its outputs. Risks related to manipulation of training data or the training environment, as well as manipulation of the ML model, must be examined considering the application context in question. Furthermore, plausible possibilities of attack on the AI application must be analyzed and their probability of occurrence must be estimated. In this respect, it is also necessary to consider the risk of data leakage regarding training data or model parameters, if this creates an increased risk of manipulation of the AI application. Lastly, it is necessary to examine the effects of an integrity breach of the AI application and specifically the damage that may result from this.
- **Availability risk analysis:** Availability threats must be specified for the AI application being examined. Considering the particular application context, the analysis should examine risks related to attacks that can cause a change in function, with regard to delays or interruptions in communication between the AI component and the embedding (for example, when online services are integrated), and with regard to the scalability of requests. In particular, it is necessary to take into account the frequency and duration of use of the AI application. Furthermore, the (non-physical) damage that could potentially occur if the AI application is unavailable must be estimated. (Physical damage is covered in the **Risk area: functional safety (FS)**.)
- **Objectives:** Based on the risk analysis, objectives regarding safeguarding and testing with respect to integrity and availability must be documented. In particular, the objectives should describe the circumstances in which control of the threats or risks identified as relevant in the risk analysis is gained.

---

## 8.2.2 Criteria for achieving objectives

Appropriate safeguard measures must be implemented based on the threats identified in the integrity and availability risk area. In order to be able to objectively check during the final assessment of the measures if identified risks have been successfully mitigated, the objectives described in **[S-R-IA-RI-01]** must be translated into quantitative criteria. This should involve considering a more specific definition of the potential level of damage (see **[S-R-IA-CR-01]**) to be used for evaluating the residual risk.

**[S-R-IA-CR-01] Quantification of acceptable risk**

Requirement: Do

- Criteria must be defined for assessing integrity and availability risks based on the level of potential damage and its probability of occurrence. The following aspects should be considered as a minimum when defining the criteria:

  Amount of damage:
  – Relevance of the unintentionally disclosed or stolen data/information to the integrity of the AI application (in terms of vulnerability)
  – Extent of the availability breach (i.e., is the entire AI application or a subfunction affected?)
  – Duration of downtime (of a part) of the application
  – Number of persons and dependent systems affected in case of unavailability
  – Maximum permissible latency (latency requirements are usually dependent on the application context)
  – Costs due to downtime

  Probability of occurrence, taking into account possible causes such as:
  – Attacks
  – Unauthorized access to model/data
  – Limited scalability
  – High number of inappropriate uses (prank requests, trolling), e.g., through automated requests or malware

- The expected damage is determined over the lifetime of the AI application (or an annual average in the case of open-ended use) using the estimated probabilities of occurrence. It is important to work on the basis that these situations will arise frequently, particularly in the case of automated breaches, such as those caused by scripts for attacking the AI application. In addition to the expected average damage, possible deviations should also be taken into account, for example on the basis of worst-case scenario considerations.
- With regard to the expected damage, it is important to specify threshold values for the above criteria, which – provided they are observed – would ensure an acceptable level of risk or expected damage. A transparent argument must be provided demonstrating that the specified criteria and threshold values are appropriate and sufficient for the application context in question and that they reflect the objectives defined in **[S-R-IA-RI-01]**.

**[S-R-IA-CR-02] Data access options**

Requirement: Do

Based on the risks identified in **[S-R-IA-RI-01]** and for all data related to the AI application which, if accessed or modified, could reduce the quality of the AI application, it is necessary to record

- for which group of people,
- how often,
- for what reason,
- and under which other conditions

insight into or modification of the data point should be possible or, conversely, when this must not be permitted.

**[S-R-IA-CR-03] Number of requests/queries of the AI application**

Requirement: Do

- Under certain circumstances, risks related to targeted attacks, functional changes, for example through data poisoning, and denial-of-service attacks, may mean the number of user requests have to be limited. If this is the case, the type and scope of the AI application's request/query options (per user or overall) must be designed in a way that guarantees integrity and availability. Otherwise, justification must be provided to explain why it is not considered necessary to restrict user requests to the AI application.

## 8.2.3 Measures

**[S-R-IA-ME-01] Security guidelines and instructions for use**

Requirements: Do | Pr

- Documentation should be available in which security objectives relating to integrity and availability are compiled from the business objectives, business processes, relevant laws, regulations and possible threats based on the **[S-R-IA-RI-01]** risk analysis. This security guideline also contains strategic guidance on how to achieve these objectives.
- A process must be established and documented that makes users and developers aware of risks relating to integrity and availability and indicates how to act correctly with respect to the integrity of data, model and embedding as well as in relation to the availability of the AI application.
- Measures should be described that ensure that all users take note of them. The extent to which these measures contribute to mitigating or controlling the identified risks in accordance with **[S-R-IA-CR-01]** must also be explained.
  (based on BSI C5 SA-01)

### 8.2.3.1 Data

**[S-R-IA-ME-02] Data integrity**

Requirement: Do

- Documentation should be available on what measures, such as signatures or checksums, are taken to ensure the integrity of the training data, the trained model (i.e., hyperparameters and weights) and other stored data or new incoming data during operation in the training and production environments. Data integrity measures can prevent attackers from manipulating weights or data sets, for example, to make the AI application susceptible to certain types of attacks (data poisoning).

**[S-R-IA-ME-03] Data confidentiality**

Requirement: Do

- Documentation should be available on what measures, such as encryption, are taken to ensure the confidentiality of the training data, the trained model (i.e., hyperparameters and weights) and other stored data or new incoming data during operation in the training and production environments.
  Measures to ensure the confidentiality of the model make it difficult, for example, for attackers to design specific attacks (known as white-box attacks) that could, for example, force a specific output.

**[S-R-IA-ME-04] Data backup and restoration**

Requirement: Do

- Documentation should be available on technical and organizational measures to prevent data loss with regard to training data, the trained model or other settings and data, so that these are available again within a reasonable time after a failure resulting from loss. With respect to ML models developed using incremental learning in particular, it should be possible to roll back to the last version of the model at any time. The measures established to regularly back up and restore data, as well as the scope, duration and frequency of them, must be documented. If this is already documented elsewhere, e.g., in the **Dimension: Transparency (TR)**, reference can be made to the corresponding section.

### 8.2.3.2  AI component

There are no planned mitigation measures related to development and modeling of the AI component for the integrity and availability risk area. However, as described in the introduction, the integrity of the AI application is indirectly related to the confidentiality of data. The **Dimension: Data Protection (DP)** describes measures for protecting data related to the AI component. In addition, there is the risk of model drift for AI applications that continue learning. Measures to combat this at the level of the AI component are provided in the **Risk area: control of dynamics (CD)** of the reliability dimension.

### 8.2.3.3  Embedding

**[S-R-IA-ME-05] Physical protection of the storage location**

Requirement: Do

- Documentation should be available on which locations are used to store data related to the AI application. In particular, it is necessary to describe the storage locations of training data, weights and hyperparameters of the trained model, as well as other data such as logging data.
- In addition, documentation must be provided of the measures taken to adequately protect the data and the training and productive environment from unauthorized physical access and the associated theft or damage. In the case of external providers, such as cloud services, it must be demonstrated that they ensure an adequate security concept.
  (based on BSI C5 PS)

**[S-R-IA-ME-06] Protection against malware**

Requirements: Do | Pr

- Documentation should be available that describes what protection against malware is available in the training and production environment. To this end, it is also important to consider AI-specific malware, such as software for executing adversarial attacks (see **Dimension: Reliability (RE)**).
- Furthermore, there must be an established process for monitoring and identifying possible new types of malware, as well as for checking the state of the art of the protection software, including regular updates.

**[S-R-IA-ME-07] Communication security**

Requirement: Do

- Documentation should be available on the measures taken to ensure secure, confidential communication. This should cover both the user's communication with the AI application and the communication within the application e.g., between the AI component and the embedding. Existing norms/standards relating to information security must be consulted to identify appropriate measures depending on the protection requirement. (based on BSI C5 COS-01)

**[S-R-IA-ME-08] AI application timeout**

Requirement: Do

- Documentation should be available that demonstrates a timeout within the AI component or in the interaction of the AI component with the embedding does not represent an unacceptable risk. Firstly, it is necessary to document to what extent it can be guaranteed that the AI component always responds in the required time. If this is not the case, e.g., because resources cannot be made available permanently, the response time is not deterministic, or there are delays in the communication of the AI component to the embedding, it is necessary to explain how a timeout situation is managed without risks becoming unacceptable.
- Secondly, it is necessary to examine to what extent feedback loops with other components of the embedding are exited/terminated in the required time for AI applications with direct feedback loops between classic software components of the embedding and the AI component.

**[S-R-IA-ME-09] Scalability testing**

Requirements: Do | Te

- Documentation should be available that records how the AI application deals with an influx of requests where there is a high risk of unavailability. A typical example where there is an unavailability risk is outsourced AI components, such as smart home systems, which rely on server-side computing capacity to process user requests. However, chatbot systems, e.g., for automated acquisitions or customer support, on websites can also be affected by a substantial change in user behavior, for example, if demand or interest in a product increases unexpectedly. Compared to concepts such as software-as-a-service, AI applications thus often have higher requirements for "real-time availability" (see previous example) with a computing load that can cope with them, as information must be processed before it is made available.
- Tests must be performed and documented that specifically trigger a failure of the AI application. The test scenarios examined and any data sets used for this purpose must be documented. It must be justified that the test results confirm appropriate behavior with respect to scalability according to the defined objectives in **[S-R-IA-CR-01]**.

### 8.2.3.4 Measures for operation

**[S-R-IA-ME-10] Identity and access rights management**

Requirements: Do | Pr

- Documentation should be available that defines a rights and roles concept, as well as a method for managing access and access rights for the training and production environment. The following areas must be addressed:
  – Assigning and modifying access rights to the data, training environment, trained model and embedding based on the principle of least privilege and as necessary for performing tasks
  – Registering users and measures to ensure unique user identification
  – As well as possibly (see **[S-R-IA-CR-03]**), restricting the number of queries/request that users can make to the AI application. The type and extent of the restriction must be described. In particular, it must be demonstrated that the restriction is imposed based on quantitative measures on the risk of loss of integrity, i.e., on the number of queries that are required to reconstruct sensitive information or to manipulate the database in a way that reduces the quality of the AI application
  (based on BSI C5 IDM-01)
- How the measures taken here help meet the **[S-R-IA-CR-02]** and **[S-R-IA-CR-03]** requirements must be explained.

**[S-R-IA-ME-11] Logging and monitoring**

Requirements: Do | Pr

- Documentation should be available detailing which technical and organizational measures are used to log and monitor defined events in the training and production environment that may affect the integrity or availability of the AI application. The following points should be considered as a minimum:
  – Activating, stopping and pausing of logging
  – Creating, modifying or deleting users or user rights with regard to the areas defined in **[S-R-IA-ME-10]**
  – Creating or inserting data deemed relevant under **[S-R-IA-CR-01]**
  – Training the model and creating an up-to-date version of the model
  – If this does not contradict the **Dimension: Data Protection (DP)**, the user queries and the associated output of the AI application, if necessary without referring to the user (it should be checked whether conclusions can be made about the user by the type of query and if this would be critical). For example, this can be helpful or relevant in the context of processing security incidents, adapting the AI application for general use and the traceability of decisions made by the AI application.
  – Metadata of users, separated from their requests. This can be helpful or relevant in scenarios such as troubleshooting and dealing with security incidents. The extent to which metadata is used commercially and when it is deleted must be described.
  If the monitoring or logging of the listed events is already documented elsewhere (e.g., in the **Dimension: Transparency (TR)** or in the **Dimension: Reliability (RE)**), the corresponding reference can be inserted instead.
- In addition, integrity or availability breaches must be recorded. The type, scope and, if possible, cause of the incident should be documented as well as how the AI application dealt with the incident.

- The logs are reviewed as required by authorized personnel in the event of unexpected or conspicuous events to enable malfunctions and security incidents to be investigated promptly and appropriate measures to be initiated. Personal data must be stored and protected in accordance with the applicable data protection requirements (see also **Dimension: Data Protection (DP)**).
  (based on BSI C5 RB-10 and 11)

**[S-R-IA-ME-12] Organization of information security**

Requirement: Pr

- There must be a process in place for organizing information security. The operator should initiate, control and monitor an information security management system that takes into account the availability of the AI application, based on the ISO 27001 standard for planning, implementing, maintaining and continually improving an information security framework for the application.

**[S-R-IA-ME-13] Procedure for loss of integrity or availability**

Requirements: Do | Pr

- The measures taken to avoid or minimize potential further damage in the event of an integrity or availability breach must be documented.
- The procedures that are triggered when a loss is detected must be described. In the case of a lack of integrity, depending on the severity, this may include a diagnostic check of the AI application, shutdown or transition to a fail-safe state. For example, if the weights of the ML model have been revealed, the risk of a white-box attack can be mitigated by retraining or rolling back to a previous version.

**[S-R-IA-ME-14] Restoring the AI component**

Requirement: Pr

- There should be a procedure in place for restoring the AI component in the event of non-functionality or undesired changes within the embedding, if necessary using the data backups (in particular the stored weights and the architecture) from **[S-R-IA-ME-04]**. Reference can be made to classic IT security measures here.

**[S-R-IA-ME-15] Detecting loss of integrity or availability**

Requirements: Do | Pr

- Documentation should be available on measures for detecting loss of integrity or availability. Reference can be made to measures including those for detecting model drift (see **Risk area: control of dynamics (CD)** of the reliability dimension). If required, a description must also be provided of reporting mechanisms for users and processes for handling user requests in a reliable, appropriate and timely manner.

## 8.2.4 Overall assessment

**[S-R-IA-OA] Overall assessment**

Requirement: Do

- Detailed justification must be provided that the **[S-R-IA-CR-02]** and **[S-R-IA-CR-03]** criteria are met. It must also be demonstrated that the residual risk is acceptable according to the **[S-R-IA-CR-01]** criteria.
- If not all requirements specified in **[S-R-IA-CR-01]** to **[S-R-IA-CR-03]** are met, the deviations must be documented. This also applies to requirements that have only been partially met, e.g., where the criteria have not or not always been met.

## 8.3  Risk area: control of dynamics (CD)

The control of dynamics risk area should ensure that the safety and security of the AI application are guaranteed, including during operation.

Concept drift poses a significant risk in this context and can lead to implemented safety and security measures no longer being effective or no longer being sufficient due to changing requirements or conditions during operation. For example, research in the field of Artificial Intelligence could produce new findings that help attackers develop new malware and attack methods. In addition, changes in external conditions can create new safety and security threats. Last but not least, legal or regulatory changes may result in new safety or security requirements.

An example of a risk emerging due to concept drift is a service for automated translation of digital communication that experiences a significant increase in demand due to changes in user behavior and can no longer guarantee availability (see **[S-R-IA-ME-09]**). Another example would be if the actuators of a cyber-physical system were replaced with actuators that expect parameters with different tolerances, meaning that threshold values for mitigation (see **[S-R-FS-ME-08]**) would have to be adapted in this instance.

Model drift represents another aspect of the dynamics of AI applications. However, this plays a marginal role for this risk area, as challenges related to the continued learning of the model based on newly recorded data are already covered in full in the **Risk area: control of dynamics (CD)** of the reliability dimension. In addition, the **Risk area: intercepting errors at model level (IM)** and the **Risk area: functional safety (FS)** reference measures that help intercept AI safety-related errors or faults, which specifically include performance decline during operation. Thus, the present risk area only deals with measures for model drift that refer to safety mechanisms dependent on the performance of the AI application. In particular, these safety mechanisms must be refined as required if the performance of the AI component weakens.

### 8.3.1  Risk analysis and objectives

The risk analysis examines threats in the control of dynamics risk area in terms of safety and security. Typical threats in this area include safety and security threats that may occur because of

- concept drift due to a change in external conditions,
- changes in user behavior,
- changed requirements related to changes in underlying frameworks or hardware.

These threats can affect any of the previously considered risk areas, including the **Risk area: functional safety (FS)** and **Risk area: integrity and availability (IA)**. The exact process in the risk analysis for the control of dynamics risk area is described in the following section.

**[S-R-CD-RI-01] Risk analysis and objectives**

Requirement: Do

- **Risk analysis:** The **[RE-R-CD-RI-01]** risk analysis of the reliability dimension has already examined the risk of model and, in part, concept drift and can be referenced here. However, it must be analyzed to what extent the implemented safety or security measures depend on the performance of the AI application and how the measures must be adjusted in case of drift. The risk analysis must also include regulatory changes. Furthermore, it is necessary to investigate which external conditions in the application context in question can have an influence on the safety and security requirements of the AI application. This involves analyzing the possible attacks on the AI application by new or conceivable malware. It must also examine the extent to which changes in user behavior can affect safety and security. It may be necessary to add further context-specific threat scenarios for controlling the dynamics.

Lastly, an assessment must be made of potential damage that may result from vulnerabilities due to continued learning in operation as well as changed safety or security requirements due to external conditions and their probability of occurrence considering the application context in question.
- **Objectives:** Based on the risk analysis, objectives are set for avoiding, detecting, recording and handling potential safety or security incidents due to model or concept drift.

### 8.3.2 Criteria for achieving objectives

Appropriate safeguard measures must be implemented based on the threats in the control of dynamics risk area. In order to be able to objectively check during the final assessment of the measures if existing risks have been successfully mitigated, the objectives described in **[S-R-CD-RI-01]** must be translated into quantitative criteria.

**[S-R-CD-CR-01] Framework for dealing with changing safety and security risks**

Requirement: Do

- Criteria used to assess how changing safety and security risks are dealt with must be documented. The criteria can be quantitative or qualitative in nature and should at least include:
  – Assessment interval for checking how up to date the implemented safety and security measures are
  – Scope of the review
  – User behavior requirements
  – Threshold value or qualitative criterion at which (depending on the metrics from the **Dimension: Reliability (RE)** as well as criteria from the safety and security dimension) adjustment of the safety and security measures is required
- Target values (or qualitative target characteristics) must be specified for each criterion established. It must be demonstrated in a transparent way that these objectives are consistent with the target values and are appropriate for the particular application context.

### 8.3.3 Measures

### 8.3.3.1 Data

There are no planned measures for this category.

### 8.3.3.2 AI component

There are no planned measures for this category.

### 8.3.3.3 Embedding

There are no planned measures for this category.

### 8.3.3.4 Measures for operation

**[S-R-CD-ME-01] Training and raising awareness of employees**

Requirement: Do

- Documentation should be available on how to ensure that employees and service providers are aware of their safety and security responsibilities. The following measures are considered for this purpose:

- Reliability assessment of employees (e.g., verification of the person by identity card, verification of CVs or also the request for a police clearance certificate in the case of sensitive roles)
- Employment agreement with commitment to comply with laws, rules and regulations
- Safety and security education and awareness-raising program, including regular briefings and training on secure development and maintenance of the application on appropriate handling of training and user data, regular briefing on potential – particularly AI-specific – attacks and regular training on the course of action to take when safety- or security-related events occur
- Disciplinary action for breaches of policies and instructions
  (based on BSI C5 HR-01 to HR-04)

**[S-R-CD-ME-02] Monitoring external conditions**

Requirement: Pr

- A process must exist for monitoring external conditions as well as potential new threats to safety and security that may require adjustments to requirements and measures. As a minimum, the process should address the
  - advances in the field of research, especially the possibility of attacks and malware for the AI application,
  - correctness of the basic assumptions made about the application context in question and according to which the safety and security measures are aligned,
  - changes to the software framework used so that, for example, the current version of the code cannot be read, or security vulnerabilities emerge,
  - changes in legal and regulatory conditions, including changes that may also affect the **Dimension: Data Protection (DP)** and **Dimension: Reliability (RE)**.
  The type and scope of the process are documented.
- Responses or next steps that will be taken if a relevant change in external conditions is identified must also be defined.

**[S-R-CD-ME-03] Emergency management**

Requirement: Pr

- A process must exist for managing emergencies. An emergency can be an accident, failure or safety/security incident. In this process, emergency concepts (see also **[S-R-FS-ME-14]**) must be planned, implemented, tested, monitored and regularly reviewed and improved. This includes recognizing the emergency, dealing with it and using the incident to improve safety and security. The following are performed following an emergency:
  - Emergency analysis (incident management): The history of emergencies that have occurred is logged and the resulting data is used to improve the safety and security of the application.
  - A review of all safety and security measures to determine how effective they are.
- If there are overlaps in terms of content, conformity with the guidelines **[S-R-FS-ME-01]** and **[S-R-IA-ME-01]** must be ensured.
  (based on BSI C5 BCM, SIM)

### 8.3.4 Overall assessment

**[S-R-CD-OA] Overall assessment**

Requirement: Do

- Considering the measures taken, it must be demonstrated that procedures or external conditions have been created that meet the **[S-R-CD-CR-01]** criteria and thus achieve an acceptable level of risk with regard to the dynamics of the AI application in terms of safety and security.
- If not all requirements specified in **[S-R-CD-CR-01]** are met, the deviations from the requirements must be documented. This also applies to requirements that have only been partially met, e.g., where the criteria have not or not always been met.

## Summary

**[S-S] Summary of the dimension**

Requirement: Do

- If there is a medium or high protection requirement for this dimension, documentation must be provided for the remaining residual risks. First of all, the residual risks from the various risk areas in this dimension are summarized. Subsequently, and taking into account the protection requirement, the identified residual risks are collectively assessed as negligible, non-negligible (but acceptable) or unacceptable. This analysis should specifically assess the impact of measures from the reliability and data protection dimensions in terms of whether they help to mitigate or control safety and security risks. The result of the analysis must be explained.
- If risks or measures under this dimension have been identified as having potentially negative effects on other dimensions, such as autonomy and control, transparency or data protection, they must be documented.
- A conclusion must be made about the dimension that includes the assessment of residual risks.

# 9. Dimension: Data Protection (DP)

## Description and objectives

According to their properties, it is possible that AI applications affect a variety of legal positions. Particularly often, this concerns infringements on privacy or the right to informational self-determination. For example, AI applications often process sensitive information, such as personal or private data including voice recordings, photos or videos. Therefore, it is necessary to ensure compliance with the relevant data protection regulations, such as the General Data Protection Regulation (GDPR) and the German Federal Data Protection Act (BDSG). However, AI applications can pose a risk to more than the privacy of individuals: (Business) secrets or data under license may also be affected, which do not constitute personal data in the sense of the GDPR, but can still require legal protection. For example, machine data – entirely separate to the issue of which person was active as a machine operator – can contain information about process utilization or error rates and thus represent sensitive business-related data.[81]

The challenges related to data protection are potentially much greater for AI applications than for classic IT systems. In particular, this is due to the fact that AI applications often combine previously unlinked data and just create new methods of linking data through Machine Learning. The more data that is linked (data linkage), the greater the risk of being able to identify people or, for example, specific operating sites even without directly specifying corresponding attributes. For example, it is possible to re-identify people with 95 percent certainty through their typing behavior on a computer keyboard[82, 83]. If there were now a public (or purchasable) database that assigned keyboard stroke patterns to people, the keyboard stroke pattern would become what is referred to as a quasi-identifier, which would make it possible to make a personal reference.

AI methods can potentially create personal or business references when processing text, voice and image data, as well as logged usage data. However, in addition to the stored or processed data, the actual ML model implemented in the AI application can also be exposed. The targeted systematic querying of the AI application to reconstruct model parameters or other model characteristics is called model extraction. If an attacker gains access to the model parameters and also knows the learning algorithm and the structure of the model, the person could try to reconstruct the model. For example, this could significantly affect the competitive position of a targeted company. In addition, targeted attacks could be developed to extract personal (training) data from a model. For instance, it is possible to reconstruct a training image of a person based on softmax values of an AI application for facial recognition[84]. Methods for deducing raw data from a model, which usually should not be disclosed, are referred to as model inversion and are currently being researched.

---

Federated learning is a method for protecting large data sets and models. With federated learning, only Machine Learning parameters are exchanged, meaning potential attackers have to expend a lot of energy to deduce the training data. First, the attacker would have to access the model parameters. These are held by the coordinator and the local agents and are exchanged between them. The transmission can be protected by encryption. Virtual data spaces ensure distributed data is securely exchanged and guarantee fine granular control over its use.

In the traditional sense, data confidentiality is a protection goal of classic IT security. While the assessment catalog classifies the other protection goals of integrity and availability as risk areas under the **Dimension: Safety and Security (S)**, it gives data protection its own dimension. This is because the properties of ML processes create new types of risks with regard to confidentiality that go beyond access to stored data – as is the case in classic IT security. As described in the previous sections, AI technologies can intelligently link data and thus establish or reconstruct personal references. In addition, there is the risk of model extraction and the risk of extracting training data from the model. Overall, data protection in the context of AI applications opens up its very own risk landscape, which requires new measures specifically for AI and must therefore be seen as a dimension in its own right.

Nevertheless, data protection is still directly related to risks regarding the integrity of an AI application. Thus, the traditional IT security measures described in the **Risk area: integrity and availability (IA)**, such as encryption or identity and access rights management, should also be taken into account where necessary to protect data. However, the measures in the data protection dimension focus on established AI-specific methods. Due to the abundance of possible methods, only a selection of them can be presented. In particular, new measures for mitigating risks that are not listed in the assessment catalog are also allowed.

The purpose of the data protection dimension is to ensure that data protection risks and measures of the AI application, taking into account the particular challenges that AI poses, are sufficiently analyzed and documented to meaningfully support data protection officers in carrying out the investigation and ultimately making decisions about data protection declassification.

The risk areas under the data protection dimension are:

1. **Protection of personal data:** This risk area covers risks associated with the AI application using personal data that is not GDPR-compliant, as well as the risk of re-identification of individuals in a data set.

2. **Protection of business-relevant information:** This risk area addresses risks that arise from the unwanted disclosure of business-relevant information by the AI application.

3. **Control of dynamics:** This risk area addresses the risks that new background information will emerge, such as the creation of a personal reference, or that the requirements for processing data with an AI application will change.

Lastly, with regard to the split into risk areas, it should be noted that the **Risk area: protection of personal data (PD)** and the **Risk area: protection of business-relevant information (BI)** only differ slightly in terms of the potential measures to take listed there. Nevertheless, these topics are split into two risk areas because the protection requirements for personal and business-related data may differ significantly. If there are repetitions, these can be avoided by referring to existing documentation from another risk area.

## Protection requirements analysis

Although personal and business-related data typically presents different damage scenarios, the amount of potential damage still depends in both cases on the type or category of data processed or stored by the AI application.

The handling of personal data is governed by the European General Data Protection Regulation as well as the German Federal Data Protection Act. It is important to note that unauthorized access by third parties constitutes a violation of legal requirements. The same applies, for example, to the mere occurrence of unauthorized access, to unsuitably long storage periods or the inability to provide information about the stored data. Both the non-material damage caused by violation of the personal rights of the data subject(s) and the amount of potential financial damage, e.g., through fines or damage to reputation, depend on the significance/category of the personal data stored.

The same is true with respect to the financial damage to an organization or company in the event of unjustified access to licensed data or the disclosure of business secrets depending on their nature and significance. The term "licensed data" is used in the following for ease to refer to all types of data to which third-party rights exist. More generally, business-related data is data that contains information about the operator, in particular business secrets.

The protection requirement is categorized as follows:

| | |
|---|---|
| **High** | The protection requirement is classified as high if one of the following three scenarios applies: Personal data is processed that contains particularly sensitive personal information, or disclosing it would have economic or security-critical consequences for the person in question. **Example:** Patient file, certificate of good conduct, account information, application documents Licensed data is processed for which the disclosure/access by third parties would violate contractual agreements. **Example:** Data from other companies was purchased to train the model. Access to this data by third parties would violate the contractual agreements. Organization/business-related data is processed which, if it became known/was accessed by third parties, would severely damage the integrity or competitiveness of the organization. **Example:** Model extraction would mean that the corresponding AI application could be copied or deliberately manipulated by other organizations. |
| **Medium** | The protection requirement is evaluated as medium if one of the following three scenarios applies and there is no high potential for damage for any of the named data categories (personal/business-related or licensed) according to the top row of this table. The AI application only processes/stores data that does not contain sensitive personal information or that would not cause a major economic disadvantage or threaten the security of the subject if accessed by a third party. The AI application processes/stores licensed data, which when accessed by third parties could result in negligible consequences. The AI application processes/stores business-related data, the disclosure of which could result in medium economic damage that does not threaten the existence of the company. **Example:** Leisure interests of a person, played tracks, videos viewed in anonymized form **Example:** An AI application that performs trend analysis based on publicly available social media data. |

| Low | The AI application does not request, process or store personal data. |
|---|---|
| | In addition, the AI application does not store/process any licensed data. |
| | Disclosure of the processed data and model characteristics (e.g., model parameters) would have no or negligible impact on the integrity or competitiveness of the organization. |
| | **Example:** A company uses a standard AI solution to predict market development. Other companies in the industry have similar solutions and it is assumed that there is no incentive on the part of the competition to expose or copy this system. For example, data from the DAX or other economic indicators that are freely available are used. |

**[DP-P] Protection requirements analysis documentation**

Requirement: Do

- The protection requirement of the AI application for the data protection dimension is defined as *low, medium* or *high*. The choice of the *low/medium/high* category is justified in detail with reference to the table above.

If the protection requirement for the data protection dimension is *low*, the individual risk areas do not need to be examined more closely. However, if a *medium* or *high* protection requirement has been identified, each risk area must be examined in more detail below.

## 9.1 Risk area: protection of personal data (PD)

If an AI application processes personal data, there is a risk that specific AI procedures or the addition of background knowledge (e.g., other data sets) will allow individuals to be re-identified from the data set. Depending on the sensitivity of the information about a person disclosed in this way, this constitutes a serious violation of their personal rights. This creates the requirement that the data queried, processed or stored by the AI application must be effectively protected both during training and in operation.

Under the General Data Protection Regulation (GDPR), AI applications can only access personal data with the consent of the data subjects. Further processing and disclosure to third parties – subject to further restrictions – can only take place with the consent of the data subject. It must be ensured that there are no protection gaps that would enable unauthorized access. Individuals have the right to have personal data erased.[85]

Regarding the protection of personal data, the GDPR also ensures data subjects have an extensive right to object to the processing of their data at any time. In particular, the legal implementation of this objection to processing can pose specific technical and organizational challenges in connection with AI applications. The obligations arising from the GDPR that would apply specifically to operators in relation to the deletion of personal data have not been fully clarified from a legal perspective. However, it is possible to avoid a potential obligation to provide evidence that personal data has been completely deleted as well as the potentially complex technical and organizational consequences associated with this by taking appropriate measures. For example, anonymization of training data can largely prevent personal references from being made. This would avoid the need to provide evidence that there is no longer any personal reference affected by an objection to processing after the data has been deleted, and in particular the extreme case of having to completely retrain the model in the event of an objection.

The required measures also include informing data subjects about the purpose and use of their personal data or data derived from it. In addition to the consent, information, objection and revocation mechanisms to be provided for the use of personal data, the principles of data minimization and use for a specific purpose must be observed.

The purpose of the risk analysis described below is to determine which specific threats to the protection of personal data are possible for the AI application being examined. This should specifically examine the type and significance of the data queried or stored in connection with the AI application and where potential protection gaps exist. In particular, taking into account any measures used to anonymize or aggregate data, the objective is to achieve low risk of re-identification or of the possibility of establishing a personal reference by linking it to background knowledge. While it is usually not possible to completely rule out the possibility of re-establishing a personal reference, the effort required to re-identify persons in a data set should be disproportionately high.

---

85 The description in this section and also partly in the following sections draws heavily on section "3.6 Data protection" of the white paper: Poretschkin, M.; Rostalski, F.; Voosholz, J. et al. (2019). Trustworthy Use of Artificial Intelligence. Sankt Augustin: Fraunhofer Institute for Intelligent Analysis and Information Systems IAIS. https://www.ki.nrw/wp-content/uploads/2020/03/Whitepaper_Thrustworthy_AI.pdf (last accessed: 06/18/2021)

### 9.1.1  Risk analysis and objectives

**[DP-R-PD-RI-01] Risk analysis of training data**

Requirement: Do

- The properties of the training data used must be described and their choice or suitability must be documented and justified. Furthermore, it must be explained whether the training data of the AI component contains information that enables a personal reference to be made. In particular, the attributes, the volume, as well as the possibility of linking the data with other (personal) background information must be documented.
- Sample training data must be available which can be used to understand the characteristics of the training data described in the documentation with regard to the protection of personal data.

**[DP-R-PD-RI-02] Risk analysis of input and usage data**

Requirement: Do

- It is necessary to document and explain which of the input or usage data that is collected and stored during the operation of the AI application enables a potential personal reference to be made. In particular, the documentation must describe which potentially personal attributes are available and what possibilities there are for linking them to other data sets.
- The volume of potentially personal inputs and outputs as well as other usage data (e.g., through logging) that is collected and stored must also be described.
- Lastly, it is necessary to specify which of the collected data is to be used as training data (see **[DP-R-PD-RI-01]**), or is stored and used only for information and verification purposes, or for other purposes (e.g., load analysis).
- Example data must be available that can be used to understand the characteristics of the input and usage data described in the documentation with regard to the protection of personal data.

**[DP-R-PD-RI-03] Biometric features**

Requirement: Do

- It must be documented how biometric data (e.g., images, handwriting, health data, fingerprints, key and mouse operation) is collected and used by the AI application. In particular, it is necessary to explain whether and with which background knowledge and AI procedure a personal reference could be created from this data.

**[DP-R-PD-RI-04] Model results and side channels**

Requirement: Do

- It must be analyzed and documented to what extent the results of the AI application are susceptible to an unintentional person reference being created. In addition to simply analyzing the output of the model, this also includes analyzing the possibility of linking the output with background information. Furthermore, the AI application should be examined with regard to possible side channels. For example, the processing time of input data could allow us to make an inference to person-related information.

**[DP-R-PD-RI-05] Risk analysis of the overall AI application**

Requirement: Do

- **Risk analysis:** The overall risks and potential damage associated with the potentially personal data described in **[DP-R-PD-RI-01]** to **[DP-R-PD-RI-04]** must be documented, both during development and operation of the AI application. This involves analyzing the risks with regard to unintended or unauthorized processing of and access to personal data that has been collected and used in accordance with data protection requirements. In addition, it is necessary to examine the risk that unwanted or unauthorized personal references could be created by using background information and/or AI techniques. An assessment must also be made of the damage that could result if the potentially personal data was not viewed and used as intended.
- **Objectives:** Based on the risk analysis, formal protection objectives are set in connection with personal data, which must at least comply with the GDPR and are implemented and examined in more detail below.

### 9.1.2  Criteria for achieving objectives

**[DP-R-PD-CR-01] Quantification of data protection risk**

Requirement: Do

- The criteria to be used to assess the risk of creating a personal reference must be documented. Criteria should preferably be quantitative, such as

    – Group sizes when using anonymization methods (e.g., k-anonymity, l-diversity or t-closeness),
    – Group sizes in statistical aggregation,
    – Privacy budgets (differential privacy) or limits on query options,
    – Effort to consult useful background information,
    – Effort to create a personal reference through calculations (e.g., when using encryption methods).
    In addition, qualitative criteria may be considered, such as evidence of legitimate interest within the meaning of the GDPR as well as GDPR-compliant handling of explicit consent for the use of personal data. Other criteria not listed here may also be specified. Specifying different criteria for different data sets requires detailed justification for each data set.

- Target intervals must also be defined for the selected quantitative criteria, which, if met, produce an acceptable level of risk.
- The choice of criteria and, if applicable, target intervals must be justified in a transparent way. In particular, it must be made clear to what extent these are consistent with the objectives defined in **[DP-R-PD-RI-05]**.

### 9.1.3  Measures

### 9.1.3.1  Data

**[DP-R-PD-ME-01] Anonymization**

Requirement: Do

- It is necessary to document which procedures are used to anonymize data. From the many existing procedures, the following are examples of established anonymization procedures:
    – K-anonymity
    – Differential privacy

The choice of anonymization methods used must be justified. In addition, the effectiveness of the procedures must be evaluated, including in relation to potentially available background information.

**[DP-R-PD-ME-02] Pseudonymization**

Requirement: Do

- It is necessary to document which procedures are used to pseudonymize personal data (such as hashing) in order to make it more difficult to re-identify individuals in a data set.
- Pseudonymization is not the same as anonymization and does not usually provide sufficient protection with regard to creating a personal reference. Therefore, it must be demonstrated to what extent the procedures used for pseudonymization are effective in combination with other measures taken (e.g., from the **Risk area: integrity and availability (IA)** of the safety and security dimension) and, if applicable, what gaps exist.

**[DP-R-PD-ME-03] Perturbation of modeling data**

Requirement: Do

- It is necessary to document the ways in which data is changed (perturbed), if necessary, by adding intentional random distortions to the modeling to prevent or impede the extraction of personal data. Possible perturbation methods include:
  – Adding additive or multiplicative random noise. The type (e.g., white or uniform noise) and volume (e.g., amplitude and standard deviation) of the noise must be specified.
  – Random shuffling of attribute values.
  If other methods are used, justification must be provided as to why they are suitable as a way of making the extraction of personal data more difficult.

**[DP-R-PD-ME-04] Aggregation and generalization of data for modeling**

Requirement: Do

- The extent to which data is linked or aggregated and generalized during modeling to prevent or impede the extraction of personal data must be documented. Furthermore, it is necessary to evaluate the effectiveness of the aggregations or generalizations in terms of data protection.

### 9.1.3.2 AI component

**[DP-R-PD-ME-05] Data minimization for modeling**

Requirement: Do

- Documentation and justification must be provided that the modeling performed is not also possible using other data that is less sensitive (in terms of a personal reference).

**[DP-R-PD-ME-06] AI application purpose**

Requirement: Do

- If the AI application processes personal data on the basis of consent for a specific purpose, it must be documented that the implemented ML model only uses this data in accordance with the authorized purpose. If necessary, reference can be made to the documentation from the **Dimension: Reliability (RE)**.

**[DP-R-PD-ME-07] Novelty of outputs**

Requirement: Do

- In principle, there is the risk that the outputs of the ML model reflect parts of the training data, particularly for generative models, but also for predictive models e.g., for expanding inputs. If the training data according to **[DP-R-PD-RI-01]** potentially allows a personal reference to be made, it should be ensured, depending on the model type, that the outputs of the AI component deviate sufficiently from the training data and do not unintentionally reveal them to a disproportionate extent. In this case, it is necessary to document what measures have been taken to prevent the immediate disclosure of potentially personal training data through AI application outputs. For this purpose, reference can be made to data pre-processing measures taken, such as **[DP-R-PD-ME-01]** to **[DP-R-PD-ME-04]**, as well as to documentation from the **Risk area: integrity and availability (IA)** of the safety security dimension, for example on the restriction of query options in **[S-R-IA-ME-10]** in accordance with **[S-R-IA-CR-03]**.

- An assessment must also be made regarding the risk of immediate disclosure of potentially personal training data by the AI application, given the various dimensions of the embedding, as is the case in generative models, for example.

**[DP-R-PD-ME-08] Federated learning**

Requirement: Do

- One way of making unwanted access to data more difficult is distributed learning or federated learning. This involves training models locally at different computer nodes so that the respective training data does not have to leave its local position. The separately created models are then combined to form a global model. If the ML model was built by federated learning (or a variation of it), this should be documented. In particular, it is necessary to describe to what extent the distributed learning of the AI component makes it more difficult to expose personal data.

### 9.1.3.3  Embedding

**[DP-R-PD-ME-09] Unintentional release of information**

Requirement: Do

- Through targeted querying of the AI application, it may be possible to reconstruct data, especially training data of the ML model, directly or through side channels. For example, through alternating queries of the AI application and adjusting the input, an input data point could be constructed that achieves a particular output. The form of the input data point could in turn provide an inference to the training data or to relationships learned through the ML model. If, in accordance with **[DP-R-PD-RI-04]**, there is a risk of personal data being exposed in this way, it is necessary to document what measures have been taken to prevent or impede the unintentional leakage of information through queries of the AI application or via side channels. For this purpose, reference can be made to data pre-processing measures taken, such as **[DP-R-PD-ME-01]** to **[DP-R-PD-ME-04]**, as well as to documentation from the **Risk area: integrity and availability (IA)** of the safety and security dimension, for example on the restriction of query options in **[S-R-IA-ME-10]** in accordance with **[S-R-IA-CR-03]**.

### 9.1.3.4  Measures for operation

**[DP-R-PD-ME-10] Storage and deletion**

Requirements: Do | Pr

- The technical implementation and the storage location of potentially personal data (training, input, output and usage data) must be documented as well as the measures taken to protect stored personal data from cyberattacks. If applicable, reference can be made here to measures from the **Risk area: integrity and availability (IA)** of the safety and security dimension, such as measures for ensuring confidentiality (see **[S-R-IA-ME-03]**) and for securing data (see **[S-R-IA-ME-04]**) and the storage location (see **[S-R-IA-ME-05]**).
- The technical procedures for deleting potentially personal data must also be described. In particular, it is necessary to explain how to deal with the fact that individuals are entitled to revoke their consent to the processing of their data. If operational processes have been established in relation to this, they must be documented.

**[DP-R-PD-ME-11] Ability to provide information about personal data**

Requirements: Do | Pr

- It must be documented how it is ensured that data subjects and users can obtain information about the data used by them or collected about them.
- How data subjects and users can find out what decisions the AI application has made regarding them or their queries must be documented. In some circumstances, reference can be made to the documentation from the **Dimension: Transparency (TR)** and/or **Dimension: Autonomy and Control (AC)**.
- A breach of confidentiality of data may not be recoverable, depending on its nature. However, procedures can be introduced to reduce damage, for example to notify data subjects. If these types of processes have been established, they should be documented.

## 9.1.4 Overall assessment

**[DP-R-PD-OA-01] Evaluating anonymization**

Requirement: Do

- It must be explained to what extent the anonymization measures taken and documented result in the anonymized data not allowing any undesired/unauthorized creation of a personal reference, or if this is only possible with a great deal of effort, which would typically not be proportional to the expected benefit.

**[DP-R-PD-OA-02] Declaration of data protection conformity**

Requirement: Do

- It must be documented that the use of the potentially personal data described in **[DP-R-PD-RI-01]** to **[DP-R-PD-RI-04]** is compliant with the GDPR and the BDSG (German Data Protection Law). Reference must be made to points including the risk assessment, the consent and protection mechanisms adopted and the legitimate interest for processing.
- In addition, a data protection impact assessment in line with Art. 35 Para. 1 of the GDPR must be made available, or clear justification must be provided that such an assessment is not necessary due to the measures taken (e.g., to anonymized data). If a data protection impact assessment has been prepared, the data protection officer must confirm its compliance with the GDPR.
- Lastly, a summary must be provided demonstrating that the objectives set in **[DP-R-PD-CR-01]** have been achieved.
- If not all requirements specified in **[DP-R-PD-CR-01]** are met, the deviations must be documented. This also applies to requirements that have only been partially met, e.g., where the criteria have not or not always been met.

## 9.2  Risk area: protection of business-relevant information (BI)

Aside from personal data, the digitalization of business processes is generating more and more business data that requires protection. Machine Learning models for various applications are being trained on powerful computers using large volumes of data: from anomaly detection in condition monitoring and predictive maintenance to recommendations for machine settings and autonomous vehicles, cooperative robots and smart controls. Currently, learning from these big data sets mainly happens in the cloud, i.e., on a central big data platform where historical data is continuously supplemented by data sets newly acquired from operation. However, this type of application is neither technically desirable nor legally possible in numerous areas: While data protection is the primary challenge to overcome in healthcare, the information contained in business data, for example about devices and machine manufacturers, carries the risk of internal and business secrets being exposed.

This means that non-personal data may also require protection, e.g., because it should remain secret on competitive grounds or because its use is contractually regulated. In the following, the latter will simply be referred to as licensed data, and, in short, concerns all types of data to which third parties have rights. In particular, this can also be publicly accessible data. For the sake of simplicity, the rights of third parties are referred to below as "licensing terms". In the following, business-relevant information refers to all data that relates to the operator or the operator's business and/or business contacts and that should be protected from access or unregulated viewing by third parties. This specifically includes data containing business secrets and, where applicable, licensed data. In particular, business-relevant information includes all data that, if disclosed, would negatively affect the company's competitiveness or damage the security or integrity of the operator. The model in the AI component itself can also be business-relevant information if, for example, it is a unique selling point of the operator and thus gives the operator a competitive advantage.

### 9.2.1  Risk analysis and objectives

The risk analysis for this risk area should determine the extent to which the AI application being assessed processes business-relevant information and which potential threats of the risk area are possible in the specific application context. Based on this, objectives for protecting business-relevant information must be defined.

**[DP-R-BI-RI-01] Risk analysis of training data**

Requirement: Do

- In addition to and, if applicable, with reference to **[DP-R-PD-RI-01]**, it must be documented whether and which elements of the training data used contain business-relevant information or, in particular, are licensed data.

**[DP-R-BI-RI-02] Risk analysis of model characteristics**

Requirement: Do

- It must be documented whether and which properties of the ML model (e.g., the type of model, model parameters) or of the AI component constitute business-relevant information or, in particular, are licensed data.

**[DP-R-BI-RI-03] Risk analysis of input and usage data**

Requirement: Do

- In addition to and, if applicable, referring to **[DP-R-PD-RI-02]**, it must be documented whether and which of the input and usage data used contains business-relevant information or, in particular, is licensed data.

**[DP-R-BI-RI-04] Model results and side channels**

Requirement: Do

- In addition to and, if applicable, with reference to **[DP-R-PD-RI-04]**, the sensitivity of the results of the AI application must be analyzed with regard to the violation of licensing terms or the undesired visibility of business-relevant information. In addition to simply analyzing the output of the model, this also includes analyzing the possibility of linking the output with background information. Furthermore, the AI application must be examined with regard to possible side channels.

**[DP-R-BI-RI-05] Risk analysis of the overall AI application**

Requirement: Do

- **Risk analysis:** In view of the business-related data described in **[DP-R-BI-RI-01]** to **[DP-R-BI-RI-04]**, it is necessary to analyze what risks exist with regard to unauthorized access by third parties to business-relevant information and, in particular, with regard to the violation of licensing terms, both during the development and during the use of the AI application. Thus, possible confidentiality threats must be specified i.e., plausible causes or possibilities of attack are examined for the particular application context, through which the sensitive data described could be accessed without authorization. The probability of these threats occurring must be estimated. It is also necessary to analyze the potential damage that could result from unauthorized access to business-relevant information and, in particular, licensed data.
- **Objectives:** Based on the risk analysis, the objectives for protecting business-relevant information and the licensing terms are formulated. This must at least include compliance with the licensing terms or other usage terms associated with the data.

### 9.2.2 Criteria for achieving objectives

**[DP-R-BI-CR-01] Quantification of risk**

Requirement: Do

- Criteria must be defined and documented to assess the risk of licensing terms being violated or business-relevant information not being adequately protected. Specifying different criteria for different data sets requires detailed justification for each data set. The criteria should preferably be quantitative in nature, but qualitative criteria can also be used. The following points should be addressed and discussed as a minimum when selecting the criteria:
  – Extent and type of publication or unwanted access to data
  – Volume and scope of unintentionally disclosed data
  – Criticality of the unintentionally disclosed data for the operator's business relations and competitiveness
  – Costs (e.g., in case of violation of licensing terms)
  – Criticality of the unintentionally disclosed data with respect to the manipulability or attackability of the AI application
  – Effort to consult background information
  – If relevant to the type of business-related data:
    · Group sizes when using anonymization methods (e.g., k-anonymity, l-diversity or t-closeness)
    · Group sizes in statistical aggregation
    · Privacy budgets (differential privacy) or limits on query options
  The choice of criteria must be justified in a transparent way. In particular, criteria that are not from the above list must be explained.
- Target intervals must also be defined for the selected quantitative criteria, which, if met, produce an acceptable level of risk.
- It must be demonstrated that the selected criteria and, if applicable, target intervals adequately reflect the objectives defined in **[DP-R-BI-RI-05]**.

### 9.2.3 Measures

### 9.2.3.1 Data

**[DP-R-BI-ME-01] Perturbation of modeling data**

Requirement: Do

- In addition to **[DP-R-PD-ME-03]**, it is necessary to document the ways in which data is changed (perturbed) by adding intentional random distortions during modeling to prevent or impede the extraction of business-relevant information and, in particular, licensed data. Possible perturbation methods include:
  – Adding additive or multiplicative random noise. The type (e.g., white or uniform noise) and volume (e.g., amplitude and standard deviation) of the noise must be specified.
  – Random shuffling of attribute values.
  If other methods are used, justification must be provided as to why they are suitable as a way of making the extraction of business-relevant information and, in particular, licensed data more difficult.

**[DP-R-BI-ME-02] Aggregations and generalization of data for modeling**

Requirement: Do

- In addition to **[DP-R-PD-ME-04]**, it is necessary to document the links/aggregations and generalizations of licensed data and business-relevant information created for modeling. Furthermore, the aggregations and generalizations must be evaluated in terms of the extent to which they make the unwanted visibility of this data more difficult.

**[DP-R-BI-ME-03] Anonymization**

Requirement: Do

- If applicable and appropriate for the type of business-relevant information, it is also possible to use anonymization mechanisms for licensed data or business-relevant information in addition to the measures in **[DP-R-PD-ME-01]**. If these types of measures are taken, it is necessary to explain which anonymization method has been applied to which business-related data set. In addition, the effectiveness of the method must be evaluated, including in relation to potentially available background information.

**[DP-R-BI-ME-04] Pseudonymization**

Requirement: Do

- If applicable and appropriate for the type of business-relevant information, it is also possible to use pseudonymization mechanisms for licensed data or business-relevant information in addition to the measures in **[DP-R-PD-ME-02]**. If these types of measures have been taken for business-related data, it is necessary to document the choice of method (such as hashing) as well as the data sets processed with it.
- Pseudonymization is not the same as anonymization and does not usually provide sufficient protection with regard to re-identifiability, for example of pseudonymized companies. Therefore, it must be demonstrated to what extent the procedures used for pseudonymization are effective in combination with other measures taken (e.g., from the **Risk area: integrity and availability (IA)**) and, if applicable, what gaps exist.

**[DP-R-BI-ME-05] Data obfuscation**

Requirement: Do

- In mitigation and/or in addition to **[DP-R-BI-ME-01]**, the content of stored data can be concealed. For example, birth dates could be stored as a real number in the normalized interval (-1, +1) rather than as years (taking into account the exact date). This can help ensure that if the data is lost, its semantic context is not recognized, thus making deductions more difficult without falsifying connections in the data set. It must be documented whether, to what extent and in what form data concealment was performed.

### 9.2.3.2 AI component

#### [DP-R-BI-ME-06] AI application purpose

Requirement: Do

- In the case of licensed data, it must be demonstrated and documented that the AI application only uses this data in accordance with the license granted. If necessary, reference can be made to the documentation from the **Dimension: Reliability (RE)**.
- If the AI application processes business-related data, it must be demonstrated and documented that the processing of the business-relevant information by the AI application is indeed necessary to ensure its functionality and suitability.

#### [DP-R-BI-ME-07] Novelty of outputs

Requirement: Do

- If the training data according to **[DP-R-BI-RI-01]** contains business-relevant information or, in particular, licensed data that requires protection, it is necessary to document, in line with **[DP-R-PD-ME-07]**, which measures have been taken to prevent the training data from being directly disclosed by outputs of the AI application. If applicable, direct reference can be made to **[DP-R-PD-ME-07]**, or to measures taken for data pre-processing such as **[DP-R-BI-ME-01]** to **[DP-R-BI-ME-04]**, as well as to documentation from the **Risk area: integrity and availability (IA)** of the safety and security dimension, for example on the restriction of query options (see **[S-R-IA-ME-10]**).
- An assessment must also be made regarding the risk of immediate disclosure of business-related training data by the AI application, given the various dimensions of the embedding as is the case in generative models, for example.

#### [DP-R-BI-ME-08] Federated learning

Requirement: Do

- In addition to **[DP-R-PD-ME-08]**, it is necessary to document the extent to which distributed learning of the AI component makes it more difficult or impossible to expose business-relevant information.

#### [DP-R-BI-ME-09] Weight signatures

Requirement: Do

- If learned weights within the AI component are an asset worth protecting, for example in the sense of copyright law, they can be given a digital signature[86], similar to a watermark. This ensures that authorship can always be proven through the model. It must be documented whether, to what extent and in what form the weights of the ML model were signed.

### 9.2.3.3 Embedding

#### [DP-R-BI-ME-10] Unintentional information leakage

Requirement: Do

- In accordance with **[DP-R-BI-RI-04]**, if there is a risk that business-relevant information and, in particular, licensed data may be exposed by means of targeted queries of the AI application, it must be documented, in line with **[DP-R-PD-ME-09]**, which measures have been taken to prevent or impede the unintentional leakage

---

of information. For this purpose, direct reference can be made to **[DP-R-PD-ME-09]**, or to data pre-processing measures taken, such as **[DP-R-BI-ME-01]** to **[DP-R-BI-ME-04]**, as well as to documentation from the **Risk area: integrity and availability (IA)** of the safety and security dimension, for example on the restriction of query options (see **[S-R-IA-ME-10]**).

**[DP-R-BI-ME-11] Preventing model extraction**

Requirement: Do

- It is necessary to demonstrate the extent to which the output of the AI application only contains/displays the results required for its use.
  **Example:** It is not usually necessary for the AI application to output the entire softmax vector; instead, it is sufficient for users to know which class achieves the highest softmax value.
- It is necessary to document what information regarding the technical properties of the AI application or its AI component is publicly available. In addition, it must be explained that this information does not go beyond what is necessary to inform subjects or users (see **Dimension: Transparency (TR)**).
- It must be demonstrated to what extent the sensitive model characteristics defined in **[DP-R-BI-RI-02]** are protected against reconstruction in view of the freely accessible information about the AI application and the allocated query options (see **[S-R-IA-ME-10]**). If required, the possible contradictions or trade-offs with regard to providing information to subjects and users (see **Dimension: Transparency (TR)**) must be addressed.

### 9.2.3.4 Measures for operation

**[DP-R-BI-ME-08] Storage and deletion**

Requirement: Do

- In addition to **[DP-R-PD-ME-10]**, it is necessary to document the technical implementation and the location of the storage of licensed data or business-relevant information. If possible, direct reference can be made to **[DP-R-PD-ME-10]**, or to measures from the **Risk area: integrity and availability (IA)** of the safety and security dimension, such as **[S-R-IA-ME-04]** and **[S-R-IA-ME-05]**.
- In addition, the technical procedures used to delete data when the license expires must be documented.
- It is also necessary to explain what measures have been taken to protect the licensed data or business-relevant information from cyberattacks. In this case, reference can be made to measures from the **Risk area: integrity and availability (IA)**, for example **[S-R-IA-ME-03]**.

### 9.2.4 Overall assessment

**[DP-R-BI-OA] Overall assessment**

Requirement: Do

- Documentation should be available that conclusively and collectively justifies that the residual risk with regard to the confidentiality of business-relevant information in the context of the AI application is acceptable in accordance with **[DP-R-BI-CR-01]**. In particular, justification must be provided that licensed data that is processed or collected in the AI application, for example for training or as input, is only used according to the licensing terms.
- If not all requirements specified in **[DP-R-BI-CR-01]** are met, the deviations must be documented.
  This also applies to requirements that have only been partially met, e.g., where the criteria have not or not always been met.

## 9.3 Risk area: control of dynamics (CD)

The purpose of the control of dynamics risk area is to ensure that data protection is maintained during the operation of the AI application. Changed external circumstances may require action or adjustments even after the AI application has been put into operation. For example, new technologies or newly available background information may significantly increase the risk of individuals being re-identified in a data set or the risk of unwanted visibility of sensitive business data. In addition, a change in the framework conditions, such as the legal situation or licensing terms, can present operators and/or developers with new challenges. Last but not least, the behavior of users and subjects also influences the operating conditions of the AI application. For example, whether an individual consents to the processing of their data depends on various factors including if they trust the operator to sufficiently protect their data. This trust can be strengthened by external circumstances but it can also be weakened by public scandals.

### 9.3.1 Risk analysis and objectives

**[DP-R-CD-RI-01] Risk analysis and objectives**

Requirement: Do

- **Risk analysis:** It must be documented whether and to what extent the AI application processes new incoming data during operation, of which category (personal/business-relevant/licensed) the new incoming data is and what requirements are placed on the data. A description must also be provided of the processes or mechanisms in place to control and protect the new incoming data. The damage caused if the requirements for the new incoming data are not met must be examined.
  Furthermore, it is necessary to analyze the framework conditions for data processing by the AI application (e.g., laws, internal specifications) and the external factors on which the protective measures are based (e.g., available background knowledge, competitive situation). In particular, it is necessary to estimate with what probability the framework conditions or relevant external factors would change in the course of operation and what damage this would inflict.
- **Objectives:** The objective regarding data protection during operation of the AI application is set based on the identified risks. This must at least include the fulfillment of the criteria specified in **[DP-R-PD-CR-01]** and **[DP-R-BI-CR-01]**.

### 9.3.2 Criteria for achieving objectives

**[DP-R-CD-CR-01] Quantification of risk**

Requirement: Do

- Criteria must be defined and documented to assess the control of operational dynamics in view of new incoming data as well as possible changes in the framework conditions. This should at least include compliance with the criteria specified in **[DP-R-PD-CR-01]** and **[DP-R-BI-CR-01]**.
- An explanation must be provided that the criteria specified conform to the objectives defined in **[DP-R-CD-RI-01]**.

### 9.3.3 Measures

The measures listed below apply to the data, AI component, embedding and operation in equal measure.

### 9.3.3.1 Data

### 9.3.3.2 AI component

### 9.3.3.3 Embedding

### 9.3.3.4 Measures for operation

**[DP-R-CD-ME-01] Consent, complaints, deletion of personal data**

Requirements: Do | Pr

- Documentation should be available that explains how the data protection requirements regarding consent for a specific purpose, revocation of consent, complaints in the event of suspected non-compliance and deletion are implemented in the handling of personal data in the course of operating the AI application. If operational processes have been established for this purpose, they must be described in detail.

**[DP-R-CD-ME-02] Future development relating to personal data**

Requirement: Do

- It is necessary to investigate, assess and document how the privacy risk will develop in the context of the AI application considering the collection of further data (training, input, usage data) and with respect to background knowledge that will be (generally) available in the future.

**[DP-R-CD-ME-03] Consent, complaints, deletion of licensed data**

Requirements: Do | Pr

- Documentation should be available that explains how compliance with licensing terms is achieved in the course of operating the AI application when dealing with licensed data. If operational processes have been established for this purpose, they must be described in detail.

**[DP-R-CD-ME-04] Future development relating to business-relevant information**

Requirement: Do

- It is necessary to investigate, assess and document how the context-specific significance of business-related information will develop in the future. More specifically, it must be analyzed whether it is foreseeable that certain data processed by the AI application or the model itself will be classified as business-relevant information in the future or whether this categorization will no longer apply to this data in the future.

### 9.3.4 Overall assessment

**[DP-R-CD-OA] Overall assessment**

Requirement: Do

- Considering the measures taken, it is necessary to demonstrate that the residual risk regarding data protection in the course of operating the AI application is acceptable according to **[DP-R-CD-CR-01]**.
- If not all requirements specified in **[DP-R-CD-CR-01]** are met, the deviations must be documented. This also applies to requirements that have only been partially met, e.g., where the criteria have not or not always been met.

## Summary

**[DP-S] Summary of the dimension**

Requirement: Do

- If there is a medium or high protection requirement for this dimension, documentation must be prepared for the remaining residual risks. First of all, the residual risks from the various risk areas in this dimension are summarized. Subsequently, and taking into account the protection requirement, the identified residual risks are collectively assessed as negligible, non-negligible (but acceptable) or unacceptable. This analysis should specifically assess the impact of measures from the security dimension in terms of whether they help mitigate or control data protection risks. The result of the analysis must be explained.
- If risks or measures under this dimension have been identified as having potentially negative effects on other dimensions, such as transparency, they must be documented.
- A conclusion must be made about the dimension that includes the assessment of residual risks.

# 10. Cross-dimensional Assessment of Trustworthiness (AT)

The discussion of the individual risk areas in the previous chapters always includes a concluding assessment of the measures taken (overall assessment), which argues that these measures are sufficient to meet the quality criteria defined on the basis of the risk analysis. However, as explained in the summary of the individual dimensions, there may be conflicting objectives between different quality dimensions. The purpose of this chapter is to illustrate how these conflicting objectives should be dealt with.

Conflicting objectives between the dimensions can result, for example, from a lack of feasibility of opposing requirements or because meeting the requirements of one dimension would increase the risks related to another dimension. The complexity of possible quality requirements can be demonstrated in detail using the example of an AI application for assessing creditworthiness. A trade-off must be made as early on as the point at which the features are selected, i.e., the characteristics contained in the input data on which the AI component operates. This is because requirements regarding data minimization and fairness – given that sensitive personality traits should not influence decision on creditworthiness in any way – could conflict with the objective of high accuracy, which is usually increased by providing as much information (features) as possible about the individual. Another conflicting objective regarding the performance could arise from the choice of fairness concept. For example, implementing statistical parity when faced with an "unfair" data set goes against a perfect prediction (compared to the unfair data set). In addition, conflicting objectives usually arise from the choice of model. For example, models that provide reliable results about the creditworthiness of potential customers may not be interpretable by experts. In this sensitive application context and after careful consideration, it may be possible to accept a decline in performance provided that an interpretable model is used. Last but not least, the intervention options during operation and the amount of information provided to employees or customers via the system, for example, must also be discussed in this example. The essential requirement here of human supervision and autonomy may possibly conflict with security in the sense that options for attacking or manipulating the AI application could be opened up or made easier.

It is important to consider all key stakeholder interests in order to achieve a sustainable balance between the existing conflicting objectives as well as the associated residual risks. In particular, risks and their effects should be considered from two different perspectives, as shown in the following. In the six dimensions, risks are mainly examined in terms of potential effects on users, affected persons or the (immediate) environment. However, risks such as faulty or even harmful behavior of an AI application also have an impact on the organization operating it. For example, an AI application for credit lending involves the risk of discrimination, whereby the personal rights of customers are violated as well as the reputation of the credit institution involved is damaged. This example demonstrates that AI risks must be considered in an organization's decision-making process. Organizations that use or operate AI applications should establish AI governance for this purpose. They should also form organizational structures[87] that manage roles and responsibilities regarding AI (risk) management. In terms of implementation, this includes having an appropriate process to balance conflicting objectives and potential residual risks. In particular, an authority is needed within an organization to confirm the outcome of the evaluation process and assume responsibility for the associated residual risks.

---

[87] For a detailed review of these types of issues, see the Fraunhofer IAIS study "AI Management Systems" (available now), which discusses requirements for organizations dealing with AI in terms of governance, management and technical and organizational measures, including the current standardization work of ISO/IEC JTC 1/SC 42 "Artificial Intelligence".

For this purpose, the High-Level Expert Group on AI suggests an "AI Ethics Review Board"[88] to discuss responsibilities and ethical practices with respect to the use of AI, as well as processes to perform ongoing evaluation of the system.

Thus, the following requirement results from the preceding discussion:

**[AT] Cross-dimensional assessment of the trustworthiness of the AI application**

Requirement: Do

- The AI application must not be deemed trustworthy if the conclusion was made in a dimension with a medium or high protection requirement that unacceptable residual risks exist.
- If no unacceptable, but still non-negligible residual risks were identified, it is necessary to investigate the extent to which these are related to potential conflicting objectives between the dimensions. In particular, it is necessary to discuss the extent to which residual risks in one dimension may be unavoidable in order to mitigate risks in another dimension. If it is argued that a residual risk cannot or should not be mitigated due to a conflicting objective, it is necessary to assess and justify the chosen prioritization in relation to the trade-off in question. In particular, the justification should take into account the protection requirement of the dimensions being examined.
  – If it cannot be plausibly justified that the existing residual risks are unavoidable due to the existence of conflicting objectives, the AI application should not be classified as trustworthy.
  – If it can be plausibly demonstrated that all existing residual risks must be accepted due to barely avoidable conflicting objectives, and the chosen prioritization has been explained in relation to the existing trade-offs, it is possible to judge the AI application as trustworthy despite non-negligible residual risks. The judgment of whether the AI application is trustworthy must be explained in detail.
- The AI application is to be deemed trustworthy if it was concluded in each dimension with a medium or high protection requirement that the residual risks are negligible.

---

88  High-Level Expert Group on AI (HLEG) (July 2020). The Assessment List for Trustworthy Artificial Intelligence (ALTAI). Issued by the European Commission. https://digital-strategy.ec.europa.eu/en/library/assessment-list-trustworthy-artificial-intelligence-altai-self-assessment (last accessed: 06/21/2021)

# Publishing Notes

**Published by**
Fraunhofer Institute for Intelligent
Analysis and Information Systems IAIS
Schloss Birlinghoven 1
53757 Sankt Augustin, Germany

**Editorial team**
Silke Loh
Evelyn Stolberg
Daria Tomala

**Graphics and layout**
Achim Kapusta
Pascal Ochel

**Photo acknowledgments**
Cover image: Alex – stock.adobe.com
P. 7, photo of Prof. Dr. Andreas Pinkwart: ©MWIDE NRW/E. Lichtenscheid

**Date**
January 2023

**1st edition**
© Fraunhofer Institute for Intelligent Analysis and Information Systems IAIS, Sankt Augustin 2023

First edition in German published July 2021. First edition of the English translation published January 2023.

German version: https://www.iais.fraunhofer.de/de/forschung/kuenstliche-intelligenz/ki-pruefkatalog.html
English version: https://www.iais.fraunhofer.de/en/research/artificial-intelligence/ai-assessment-catalog.html

In cooperation with

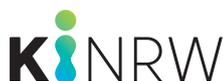

KI.NRW is the central point of contact for Artificial Intelligence in North Rhine-Westphalia. The competence platform supports the transformation of the state into a nationwide leading location for applied AI. The aim is to accelerate the transfer of AI from cutting-edge research to industry and to promote social dialogue about AI. In doing so, KI.NRW places people and their ethical principles at the center of the design of Artificial Intelligence.
**www.ki.nrw/en**

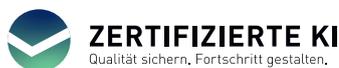

The flagship project "Zertifizierte KI" (Certified AI) powered by KI.NRW is developing assessment criteria, methods and tools for AI systems in order to enable quality assessments carried out by a third body. The consortium consists of Fraunhofer IAIS, BSI, DIN and other research partners.
**www.certified-ai.com**